\documentclass[fleqn,usenatbib]{mnras}
\usepackage{newtxtext,newtxmath}

\usepackage[T1]{fontenc}
\usepackage{ae,aecompl}

\usepackage{graphicx}	
\usepackage{amsmath}	
\usepackage{amssymb}	

\usepackage{longtable}  
\usepackage{multicol}
\usepackage{xtab,booktabs}   

\usepackage{threeparttable}  

\title[Short GRB afterglows and AT2017gfo]{A comparison between short 
GRB afterglows and kilonova AT2017gfo: shedding light on kilonovae properties}

\author[A. Rossi et al.]{A. Rossi$^{1,2}$
G. Stratta$^{1,3}$, 
E. Maiorano$^{1}$, 
D. Spighi$^{1}$, 
N. Masetti$^{1,4}$,
E. Palazzi$^{1}$, 
\newauthor
A. Gardini$^{5}$, 
A. Melandri$^{6}$,
L. Nicastro$^{1}$, 
E. Pian$^{1}$,
M. Branchesi$^{7}$,
M. Dadina$^{1}$,
\newauthor 
V. Testa$^{2}$,
S. Brocato$^{2,8}$ ,
S. Benetti$^{9}$,
R. Ciolfi$^{9,10}$,
S. Covino$^{6}$,
V. D'Elia$^{11,2}$,
\newauthor 
A. Grado$^{12}$,
L. Izzo$^{5}$,
A. Perego$^{13}$,
S. Piranomonte$^{2}$,
R. Salvaterra$^{14}$,
J. Selsing$^{15}$,
\newauthor
L. Tomasella$^{9}$,
S. Yang$^{9}$,
D. Vergani$^{1}$,
L. Amati$^{1}$,
J. B. Stephen$^{1}$,
\newauthor
on behalf of the Gravitational Wave Inaf TeAm (GRAWITA),
\\
$^{1}$INAF - Osservatorio di Astrofisica e Scienza dello Spazio, via Piero Gobetti 93/3, 40129 Bologna, Italy\\
$^{2}$INAF - Osservatorio Astronomico di Roma, via Frascati 33, 00040 Monte Porzio Catone, Italy\\
$^{3}$ INFN-Firenze, via Sansone 1, I-50019, Firenze, Italy \\
$^{4}$Departamento de Ciencias F\'{\i}sicas, Universidad Andr\'es Bello, Fern\'andez Concha 700, Las Condes, Santiago, Chile\\
$^{5}$Instituto de Astrof\'{\i}sica de Andaluc\'{\i}a (IAA-CSIC), Glorieta de la Astronom\'{\i}a s/n, E-18008 Granada, Spain\\
$^{6}$INAF - Osservatorio Astronomico di Brera, Via E. Bianchi 46, I-23807, Merate (LC), Italy\\
$^{7}$Gran Sasso Science Institute, Viale F. Crispi 7, I-67100 L'Aquila, Italy \\
$^{8}$ INAF - Osservatorio Astronomico d'Abruzzo, Via M. Maggini snc, I-64100 Teramo, Italy \\
$^{9}$ INAF - Osservatorio Astronomico di Padova, Vicolo dell'Osservatorio 5, I-35122 Padova, Italy \\
$^{10}$ INFN-TIFPA, Trento Institute for Fundamental Physics and Applications, via Sommarive 14, I-38123 Trento, Italy\\ 
$^{11}$	ASI-Science Data Center, via del Politecnico snc, 00133 Rome, Italy \\
$^{12}$ INAF - Osservatorio Astronomico di Capodimonte, salita Moiariello 16, I-80131 Napoli, Italy\\
$^{13}$ Università degli Studi di Milano-Bicocca, Piazza dell'Ateneo Nuovo, 1 - 20126, Milano, Italy\\
$^{14}$ INAF, Istituto di Astrofisica Spaziale e Fisica Cosmica di Milano, via E. Bassini 15, I-20133 Milano, Italy\\
$^{15}$ Dark Cosmology Centre, Niels Bohr Institute, Juliane Maries Vej 30,  2100 Copenhagen \O, Denmark \\
}

\date{Accepted XXX. Received YYY; in original form ZZZ}

\pubyear{2019}
\begin{document}
\label{firstpage}
\pagerange{\pageref{firstpage}--\pageref{lastpage}}
\maketitle

\begin{abstract}

Multi-messenger astronomy received a great boost following the discovery of kilonova AT2017gfo, the optical counterpart of the gravitational wave source GW170817 associated with the short gamma-ray burst GRB 170817A. AT2017gfo was the first kilonova that could be extensively monitored in time 
both photometrically and spectroscopically. Previously, only few candidates have been observed against the glare of short GRB afterglows. In this work, 
we aim to search the fingerprints of AT2017gfo-like kilonova emissions in the optical/NIR light curves of 39 short GRBs with known redshift.
For the first time, our results allow us to study separately the range of luminosity of the 
 blue and red components of AT2017gfo-like kilonovae in short GRBs. In particular, {\bf the red component is similar in luminosity to AT2017gfo,} while the blue kilonova can be more than 10 times brighter.
{\bf Finally, we find further evidence to support all the claimed kilonova detections and we exclude an AT2017gfo-like kilonova in GRBs 050509B and 061201.}

\end{abstract}



\section{Introduction}
\label{sec:intro}

 Gamma-ray bursts (GRBs) are divided in two populations consisting of long and short GRBs \citep[e.g.,][]{Kouveliotou1993a}.
Long GRBs (i.e. GRBs with a burst duration longer than $\sim2$s) have been conclusively linked to the explosive deaths of massive stars \citep[e.g.,][]{Hjorth2003b}. For a long time only indirect evidence associated short GRBs to the merging of compact objects, however a watershed occurred after the simultaneous detection of the gravitational wave (GW)
source GW170817 \citep{Abbott2017a} by aLIGO/AdVirgo \citep[][]{Ligo2015a,Virgo2015a} and the short GRB 170817A \citep{Goldstein2017a,Savchenko2017a,Abbott2017c}. 
Their identification
with the same astrophysical source has provided the first direct evidence that at least a fraction of short GRBs is associated with the merging of two neutron stars (NSs). At the same time, 
the discovery of the optical counterpart of GW170817, AT2017gfo \citep{Coulter2017a}, and its identification with the elusive  ``kilonova'' (KN) emission \citep[e.g.,][]{Li1998a,Metzger2010a}, has indirectly told us that these poorly sampled astrophysical phenomena can potentially be detected as a possible additional component to the optical and near-infrared (NIR) afterglow of (nearby) short GRBs in the temporal window that goes from about a few hours to a few weeks after the onset of the burst \citep[e.g.,][]{Kasen2015a,Barnes2016a,Fernandez2016a,Metzger2017a}.     

AT2017gfo was discovered during its 
brightening phase at $\sim11$ hours after the gravitational wave event \citep{Coulter2017a} and was followed up by several groups both photometrically and spectroscopically in the optical and NIR bands \citep[][]{Andreoni2017a,Arcavi2017a,Covino2017a,Chornock2017a,Drout2017a,Evans2017a,Pian2017a,Smartt2017a,Tanvir2017a,Drout2017a,Kasliwal2017b,Troja2017a}. 
This enormous observational effort, well summarised in \citet[][]{Abbott2017b}, allowed several of these groups to recognise 
a thermal emission in the data, with a black-body temperature evolving from 
$\sim7300$ K at $\sim0.6$ days \citep{Evans2017a} to 
$\sim5000$ K at 1.5 days after the GW event \citep[][]{Pian2017a,Smartt2017a}. 
At $\sim6$ days, the maximum moved to the longer wavelengths peaking in the J band, indicating rapid cooling.
This behaviour was markedly different not only from an afterglow but also from a supernova event. Instead, both the light curve evolution and the early ($<15$ days) spectra nicely matched the expected kilonova modelling, i.e. a thermal emission powered by the radioactive decay of elements formed 
via r-process nucleosynthesis in the ejecta of the NS-NS merger \citep[e.g.,][]{Kasen2017a,Metzger2018a}. 
In particular, the observations are consistent with a kilonova characterised by a blue, rapidly decaying, component and a red, more slowly evolving, component.
Moreover, \citet{Covino2017a} reported a low degree of linear polarization
of the optical blue component which is consistent with a symmetric geometry of the emitting region and with low inclination of the merger system. 

At the same position of AT2017gfo, a non-thermal emission consistent with a GRB afterglow was first identified in the X-ray and radio bands no more than one week after the GW event \citep[][]{Troja2017a,Hallinan2017a}, and only months later in the optical 
\citep{Lyman2018a,Margutti2018a,Rossi2018gcn,Piro2019a} due to Sun observational constraints. Afterwards, the multi-wavelength follow-up continued for up to one year after the GW event \citep{Davanzo2018a,Troja2018a,Piro2019a}. 
The observations show an achromatic slow rising flux up to $\sim150$ days after the explosion followed by a 
decay and are interpreted as emission from a structured jet expanding in the ISM and 
observed off-axis from a viewing angle of $\sim20$ deg with respect to the jet axis
 \citep[][]{Kathirgamaraju2018a,Kathirgamaraju2018b,Mooley2018a,Troja2018a,Ghirlanda2019a,Salafia2019a}.
 This is consistent with the inclination of the system derived combining the GW signal and the distance of the source {\bf \citep{Abbott2017c,Mandel2018a}}.
Such a scenario predicts a late-time rising afterglow, in contrast with the on-axis case (i.e. when the viewing angle is along, or very close to, the jet axis). The afterglow observations and their consistency with the off-axis model further confirm that all the early (i.e. $<$1 month after the GW event) optical/NIR data of AT2017gfo are not contaminated by the afterglow emission, as the latter is initially much fainter as it was already noticed \citep[e.g.,][]{Pian2017a}.

The precise nature of the different ejection mechanisms and of the different ejecta components is still under debate 
\citep[][]{Tanaka2018a,Kasen2017a,Metzger2018a,Perego2017a,Radice2018a}. 
Numerical simulations show that, during the merger of two NSs, a small fraction ($\sim$0.05 $M_{\odot}$ or less) of the total mass is ejected into space with a latitude-dependent pattern of density, velocity and opacity. Specifically, it is thought that along the polar regions the ejecta have lower velocities and opacities 
\citep[the \emph{blue} kilonova component;][]{Kasen2017a}
with respect to the equatorial region if a NS remnant is formed after the merger. If a {\bf black hole (BH)} is promptly formed, the ejecta are mostly concentrated on the equatorial plane and have high velocity and large opacities \citep[the \emph{red} kilonova component; ][]{Kasen2017a}. 
The analysis of the complete set of data of AT2017gfo has clearly demonstrated that {\bf early} ultraviolet and optical observations are of key importance to disentangle the different thermal contributions that are present in the observed emission \citep[e.g.,][]{Pian2017a,Arcavi2018a,Villar2017b,Cowperthwaite2017a,Bulla2018a} and at least two, possibly three, different emitting components have been identified \citep[e.g.,][]{Perego2017a}.

The most plausible evidence of a kilonova before AT2017gfo is that observed as an emerging component {\bf in the light curve of} the NIR afterglow of the short GRB 130603B at z=0.356 \citep{Tanvir2013a,Berger2013b}.
Other possible kilonova signatures were found in the optical counterpart {\bf light curves} of GRBs 050709 at z=0.161 \citep{Jin2016a}, 060614 at z=0.125 \citep{Yang2015a,Jin2015a}, 080503 \citep{Perley2009a,Gao2017a} at unknwon redshift, 
 150101B at z=0.134 \citep{Troja2018a}, and {\bf both in the NIR and (perhaps less clearly) in the optical counterpart light curve of 160821B at z=0.16 \citep{Jin2018a,Kasliwal2017a,Troja2019a,Lamb2019a}}.
\cite{Gao2017a} found three other possible kilonova candidates associated to GRBs 050724, 070714B and 061006. However, their peak luminosity at $\sim$1 day after the burst are more than one order of magnitude brighter than the typical predicted values of the kilonova associated with GRBs 050709 and 130603B.
It should also be noted that in all these possible kilonova identifications (but GRB 150101B), the emission was preceded by a bright GRB {\bf afterglow} indicating an on-axis configuration, thus suggesting that the kilonova emission may exceed the afterglow luminosity even for on-axis GRBs.

After the discovery of AT2017gfo, \cite{Gompertz2018a}
compared the optical/NIR light curves of AT2017gfo with those of all 23 short GRBs with redshift below 0.5. They were able to firmly exclude the presence of an AT2017gfo-like component in three GRBs (050509B, 061201 and 080905A). 
At the same time, they confirmed that AT2017gfo was much fainter than the {\bf claimed} kilonova candidates 
\citep[see also][]{Fong2017a}. These results suggest that kilonovae may display very different luminosity, colours and timescale {\bf evolutions}.

To further investigate the possible range of kilonova luminosity, we compare the optical/NIR light curves of all short GRB with known redshift up to June 2019 with those of AT2017gfo.
This paper is organised as follows. In \S~\ref{sec:data} we describe the AT2017gfo data used in this work and the short GRB sample selection. In \S~\ref{sec:met} we describe the methods we used to 
compare AT2017gfo with other short GRB optical/NIR counterparts. Section~\ref{sec:res} then illustrates the results about the most compelling short GRBs. 
These results are then discussed in \S~\ref{sec:dis}. Finally, our conclusions are given in \S \ref{sec:con}.

Throughout this work, we adopt the notation according to which the flux density of a counterpart is described as $F_\nu (t) \propto t^{-\alpha} \nu^{-\beta}$ 
and we use a $\Lambda$CDM world model with $\Omega_M = 0.308$, $\Omega_{\Lambda} = 0.692$, and $H_0 = 67.8$ km s$^{-1}$ Mpc$^{-1}$ \citep{Planck2016a}.

\begin{table}
\centering
\caption{The 39 short GRBs used in this work. 
}
\begin{threeparttable}
\begin{tabular}{c c c c c c}
\toprule
GRB              & z        &$P_{cc}(<\delta R)$&  References       & Accurate \\
                 &          &\tnote{a}                   &                   & redshift \\
\midrule 
\small     
050509B\tnote{b} & 0.225    & $ 5\times10^{-3}  $      &  1-2              &  {\bf y}   \\
050709\tnote{b}	 & 0.161    & $ 3\times10^{-3}  $      &  1                &  {\bf y}   \\
050724\tnote{b}  & 0.258    & $ 2\times10^{-5}  $      &  1                &  {\bf y}   \\
051221A          & 0.546    & $ 5\times10^{-5}  $      &  1                &  {\bf y}	 \\    
060502B          & 0.287    & $ 0.03            $      &  1-3              &  n	 \\
060614\tnote{c}  & 0.125\tnote{d}    &       &  1-4              &  {\bf y}	 \\
060801           & 1.13     & $0.02             $      &  1-3              &  n	 \\
061006\tnote{b}	 & 0.438    & $4\times10^{-4}   $      &  1                &  {\bf y}	 \\
061201\tnote{b}  & 0.111    & $ 0.08            $      &  1-5-6            &  n	 \\
061210\tnote{b}  & 0.41     & $ 0.02            $      &  1-3              &  {\bf y}	 \\
061217           & 0.827    & $ 0.24            $      &  1-5              &  n	 \\
070429B          & 0.902    & $ 3\times10^{-3}  $      &  1                &  {\bf y}	 \\
070714B\tnote{b} & 0.923    & $ 5\times10^{-3}  $      &  1                &  {\bf y}	 \\
070724A\tnote{b} & 0.456    & $ 8\times10^{-4}  $      &  1                &  {\bf y}	 \\    
070729           & 0.8      & $  0.05           $      &  1-5-7            &  {\bf y}   \\     
070809           & 0.473    & $ 0.03            $      &  1-4-6            &  n  	 \\  
071227\tnote{b}  & 0.381    & $ 0.01            $      &  1                &  {\bf y}	 \\   
080905A\tnote{b} & 0.122    & $0.01             $      &  1                &  n	 \\   
090510\tnote{b}  & 0.903    & $ 8\times10^{-3}  $      &  1-5              &  {\bf y}	 \\   
090515           & 0.403    & $ 0.15            $      &  1-4              &  n	 \\    
100117A          & 0.915    & $ 7\times10^{-5}  $      &  1                &  {\bf y}	 \\   
100206A          & 0.407    & $ 1\times10^{-3}  $      &  1-8              &  {\bf y}	 \\   
100625A          & 0.452    & $ 0.04            $      &  1                &  {\bf y}	 \\   
100816A\tnote{c} & 0.805\tnote{d}    & $ --     $      &  9                &  {\bf y}	 \\   
101219A          & 0.718    & $ 0.06            $      &  1                &  {\bf y}	 \\   
111117A          & 2.211    & $ 0.02            $      & 1-10-11           &  {\bf y}	 \\   
120804A          & 1.3      & $ 210-4           $      &  12               &  {\bf y}	 \\
130603B\tnote{b} & 0.356\tnote{d}  & --                &  13               &  {\bf y}	 \\
131004A          & 0.717\tnote{d}  & --                &  14-15            &  {\bf y}	 \\
140903A\tnote{b} & 0.351    & $3\times10^{-4}   $      &  16               &  {\bf y}	 \\
141212A          & 0.596    & $ 0.03            $      &  17-18-This work  &  n	         \\
150101B          & 0.134    & $ 4.8\times10^{-4}$      &  19               &  {\bf y}	 \\
150120A          & 0.46     & $ 0.02            $      &  20-21-This work  &  n	        \\
150423A\tnote{c} & 1.394\tnote{d}    & see text        &  22-23               &  n	 \\
150424A\tnote{c} & 0.3      & $ 0.02            $      &  24-25-26-This work  &  n	       \\
160410A\tnote{c} & 1.717\tnote{d}    &  --             &  22               &  {\bf y}	 \\
160624A\tnote{c} & 0.483    & $ 0.01            $      &  27-This work     &  {\bf y}   \\
160821B\tnote{c} & 0.16     & $ 0.02            $      &  28-29            &  {\bf y}	 \\
170428A\tnote{c} & 0.454    & $ 4\times10^{-3}  $      &  30-This work     &  {\bf y}	 \\
\bottomrule
\end{tabular}
\begin{tablenotes}\footnotesize
\item[a] probability of chance coincidence \citep{Bloom2002a}.
\item[b] light curve updated with respect to \citet{Fong2015a} with new data. See Table \ref{tab:newgrbs}.
\item[c] not in \cite{Fong2015a}, photometry in Table \ref{tab:newgrbs}. 
\item[d] redshift measured from the afterglow spectrum.
In these cases association with a host galaxy was not necessary and therefore not reported. 
{\bf References for the probability of chance association and redshift}: 
(1)  \cite{Fong2013a};
(2)  \cite{Bloom2006a};
(3)  \cite{Berger2007a};
(4) \citep{Price2006a};
(5)  \cite{Berger2010a}; 
(6)  \cite{Stratta2007a};
(7)  \cite{Fong2013b};
(8)  \cite{Perley2012a};
(9) \citep{Tanvir2010a};
(10)  \cite{Margutti2012a};
(11) \cite{Selsing2018a}; 
(12)  \cite{Berger2013a};
(13) \cite{deUgartePostigo2014a};
(14) \cite{Chornock2013a};
(15) \cite{Delia2013a};
(16)  \cite{Troja2016a};
(17) \cite{Malesani2014a};
(18) \cite{Chornock2014a};
(19)  \cite{Fong2016a};
(20)  \cite{Chornock2015a};
(21) \cite{Perley2015b}; 
(22) \citep{Selsing2019a};
(23) \citep{Malesani2015a};
(24) \citep{Castro-Tirado2015a};
(25) \citet{Jin2018a};
(26) \citet{Tanvir2015a};
(27) \citep{Cucchiara2016a};
(28) \citep{Levan2016a};
(29) \cite{Troja2019a}; 
(30) \citep{Izzo2017a}. 
\end{tablenotes}
\end{threeparttable}
\label{tab:allnew}
\end{table}

\section{Data}
\label{sec:data}

In this section we describe spectroscopic and photometric data of AT2017gfo, and the optical/NIR photometric data of short GRBs with known redshift that we compiled and used in this work.

\subsection{AT2017gfo data \label{sec:KN170817}}

The follow-up with VLT/X-Shooter of AT2017gfo is not only the first spectroscopic observation of a kilonova, but it also provided the first temporal sampling of this new class of sources.
The 10 spectra, described in \citet{Pian2017a} and \citet{Smartt2017a}, were taken between $\sim$1.5 and $\sim$10.5 days after the GW trigger and have a coverage from UV to NIR bands. 
We did not consider Gemini-S/GMOS and VLT/FORS spectroscopic observations which are limited only to the optical window.
All but two epochs are obtained from \citet{Pian2017a}. The two epochs at $\sim$2.5 and $\sim$4.5 after the GW trigger are from \citet{Smartt2017a}\footnote{They are limited to $\sim22000$~\AA~due to the presence of K-band blocking filter.} and 
 have been taken from the last version available on \emph{WISeREP} \citep{Yaron2012a}. 

At epochs earlier than the first spectrum (i.e. <1.5 days after the GW trigger), we have collected photometric observations from the works of \citet{Tanvir2017a}, \citet{Drout2017a}, \citet{Evans2017a}, \citet{Covino2017a}, \citet{Coulter2017a}, \citet{Troja2017a}, \citet{Pian2017a}, \citet{Cowperthwaite2017a}; see also the Kilonova Project: \citet{Guillochon2017a}.
We interpolated the photometric light curves using a cubic spline 
to build the spectral energy distribution (SEDs) at 
three epochs. The first epoch at $\sim$0.5 days after the trigger roughly corresponds to the first optical/NIR observations, the second epoch at 0.66 days is the first one with UV data, and the third epoch at $\sim$1 day after the trigger lays between the first photometric and the first X-Shooter observations (Fig.\ref{fig:knfit}). 
Note that we did not use all data available in the literature, 
because these data show great variation in values, even though the single data points have in most cases very small uncertainties. This can be ascribed to different calibration
and the problematic removal of light from the underlying host galaxy. Therefore, we decided to use 
only photometric data from large telescopes 
and from the restricted number of works given above.

All data have been corrected for the Galactic absorption using the interstellar extinction curve derived by \citet{Cardelli1989}, the dust maps of \citet{SchlaflyFinkbeiner2011a}, and an optical total-to-selective extinction ratio $R_V=3.1$. All observations have been converted to flux densities $F_{\nu}$ 
using transmission curves or instrument-specific conversion factors when available, or the standard conversions following \citet{Blanton2007a}.

\subsection{The short GRB data sample}

Our starting sample of short GRBs is that presented by 
\citet{Fong2015a} which includes 87 short GRBs with optical and NIR counterparts observed between November 2004 and March 2015.
We considered only the 33 events that have 
a redshift determination.
We extended this sample by including 6 short GRBs with known redshift, detected between March 2015 and December 2018. 

In addition, we took into account many works that show that the short/hard versus long/soft division does not map directly onto what would be expected from the two classes of progenitors \citep[e.g.,][]{Kann2011a}. 
For instance, \cite{Bromberg2012a} showed that the 2~s duration commonly used to separate collapsars and non-collapsars is inconsistent with the duration distributions of Swift and Fermi GRBs and only holds for old BATSE GRBs. For this reason we included the two peculiar long GRBs 060614 and 100816A, because their spectral hardness and negligible spectral lags are typical of short GRBs (see also \citealt{Bernardini2015a}). 
With respect to the \cite{Fong2015a} sample we removed GRB 140622A because only very early upper limits exist (i.e. $<0.1$ hours after trigger) that could not be compared with AT2017gfo observations that started 0.5 days after the trigger.
We also excluded GRB 090426 which, although having a duration shorter than 2 seconds, has features similar to collapsar events (soft spectra, dwarf blue host, very luminous afterglow,  \citealt{Antonelli2009a,Nicuesa2011a,Nicuesa2012a}).
In addition, we have also updated the light curves of the whole short GRB sample by adding photometric measurements that were not included in the original \cite{Fong2015a} data set (see Tab.~\ref{tab:newgrbs}). Finally, we have updated the redshift of GRB 111117A with the more refined measure of z=2.211 {\bf \citep{Selsing2018a,Sakamoto2013a}}.
Note that, contrary to \citet{Gompertz2018a}, we decided to not include GRB 051210 because, according to the most recent literature, only a lower limit on the redshift exists
\citep[$z>1.4$, see][]{Berger2007a,Fong2015a}.
The final sample thus consists of 39 short GRBs within the redshift range $0.1 \leq z \leq 2.2$ and is summarised in Table \ref{tab:allnew}.

In all cases we pay particular attention to not include photometry that was dominated by the host according to the literature from which we obtained the data. In the case the origin of the emission was not specified 
or was not certain in the literature, then we considered only data that showed to be fading. However, in all cases we have not considered necessary to model the light curves to search for a constant component, i.e., the host.

\subsection{On the redshift accuracy \label{sec:redshift}}

Accurate and reliable redshift determination through optical/NIR spectroscopy of short GRB afterglows have been obtained only in {\bf three cases (GRBs 100816A, 130603B, 160410A). 
In case of GRBs 060614 and 131004A the redshift is measured from emission lines of the host superposed from the afterglow spectrum\footnote{In case of GRB 131004A possible weak absorption lines at this same redshift are also present \citep{Chornock2013a}.}.} 
In all other 34 cases, the redshifts have been obtained through spectroscopy of the associated host galaxies. 
{
\bf
To assess the probability that the burst originated from a host candidate, we have collected or calculated the probability of chance coincidence, $P_{cc}(<\delta R)$
\citep{Bloom2002a}, at a given angular separation ($\delta R$), and apparent magnitude ($m$) for galaxies candidates.  

In table \ref{tab:allnew} we indicate redshifts, probabilities, and their references. 
 In few cases the probabilities are not negligible (larger than 1\%), but given the lack of any other possible galaxy with similarly low chance association, they are considered as good association. 
 They are GRBs 061210, 070729, 100625A, 101219A, 111117A. In 9 cases no clear association can be made with a galaxy in the field 
 (GRBs 060502B, 060801, 061201, 061217, 070809, 090515, 141212A, 150120A, 150424A).
Therefore, they do not have well-defined 
redshift measurements.
}

{\bf In almost all other cases the association with the host is well defined following the criteria of \citet{Bloom2002a} and it was possible to measure a redshift (table \ref{tab:allnew}).
Only two cases deserve more caution: GRBs 080905A and 150423A.} In the first case, \cite{Davanzo2014a} find that its properties are not consistent with the E$_{peak}$-L$_{iso}$ \citep{Yonetoku2004a}, and the E$_{peak}$-E$_{iso}$ relations \citep{Amati2002a}. They conclude that either GRB 080905A is really a peculiar sub-luminous (and sub-energetic) burst, or the associated host galaxy is just a foreground source, and the distance is underestimated.
{\bf In the case of GRB 150423A we adopted the redshift of $z=1.394$ measured by \citet{Malesani2014a} and \citet{Selsing2019a}. However,  as noted in  \citet{Malesani2014a} the redshift is based only on a tentative  detection of an absorption doublet in the faint afterglow continuum, and identified as \ion{Mg}{II} at $z = 1.394$.
Our independent analysis of the reduced spectrum, have not permitted us to confirm the presence of the absorption lines.} We note that \citet{Perley2015a} reports the presence of a galaxy at redshift of $z=0.456$ with $r=23.3$ \citep{Varela2015a} and $4\farcs$ away from the afterglow position (with $P_{cc}=0.13$) and thus leaving the distance measurement for this burst still uncertain.  

Finally, we note that in the case of GRB 061201 we used the redshift of $0.111$  of the nearest galaxy \citep{Stratta2007a}, used also by \citet{Fong2015a}, which is different from $z=0.084$ used by \citet{Gompertz2018a} that is the redshfit of the galaxy cluster within which this GRB happened. {\bf In summary, we estimate that for 28 short GRBs out of 39 events, the associated redshift is highly reliable.}

\begin{figure}
\begin{center}
\includegraphics[width=0.45\textwidth,angle=0]{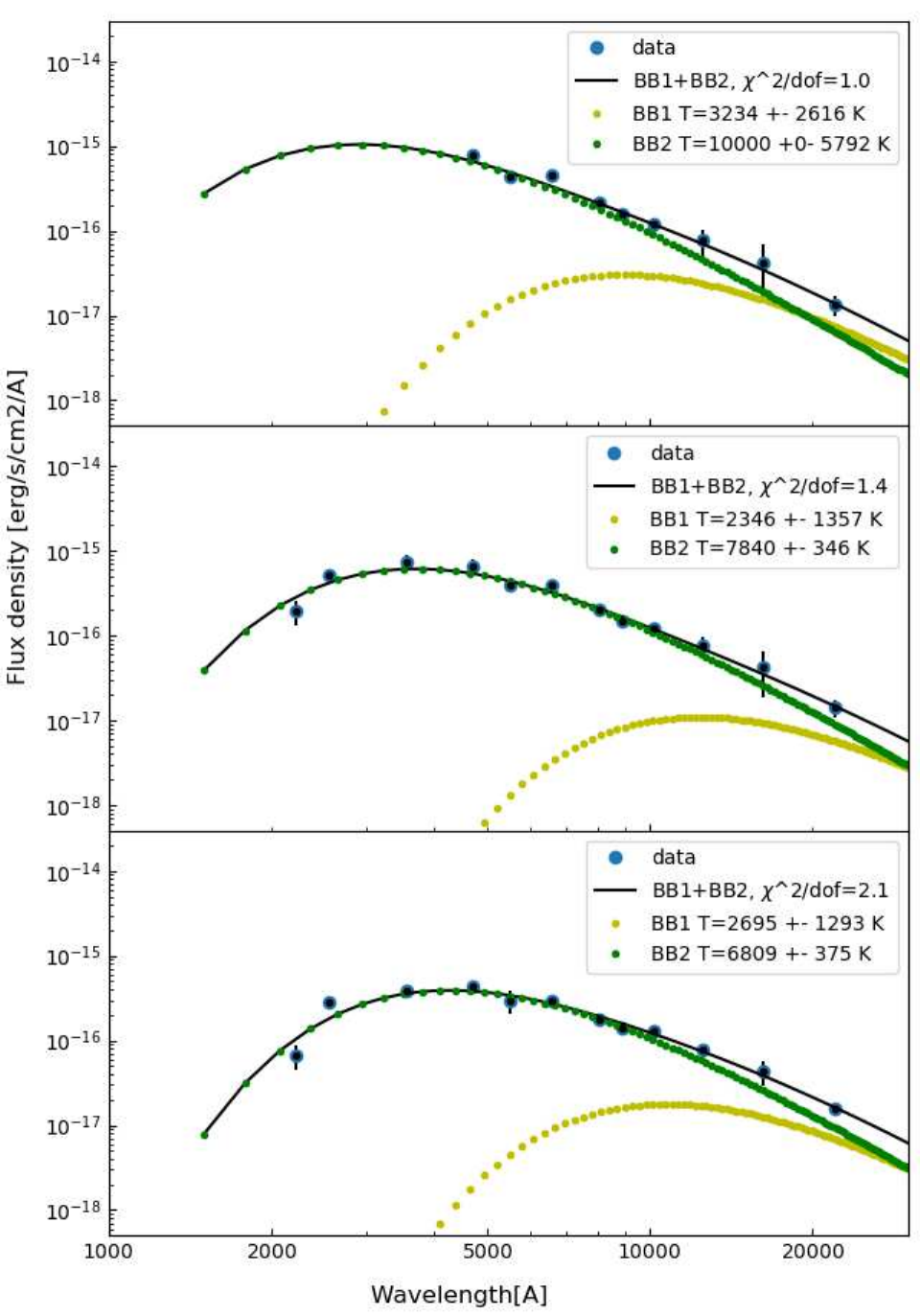}
\caption{Photometry at 0.5, 0.66 and 1 days after trigger ({blue dots with error-bars}) of AT2017gfo modelled with a double black-body model ({in} black), with single black-body components in green and yellow; see \S~\ref{sec:models}). 
}
\label{fig:knfit}
\end{center}
\end{figure}

\begin{figure*}
\includegraphics[width=0.80\textwidth,angle=0]{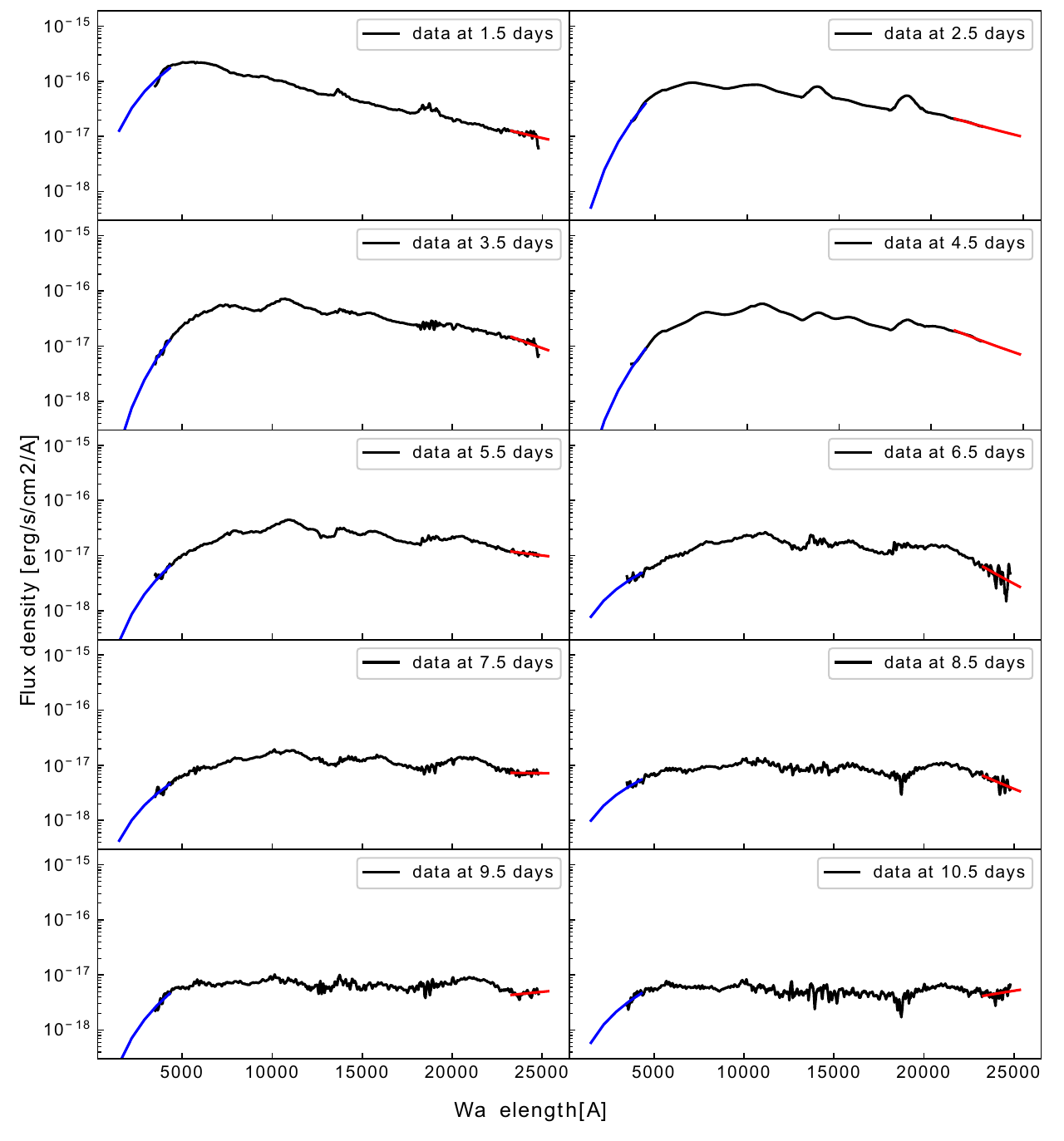}
\caption{X-Shooter spectra between 1.5 and 10.5 days after the trigger ({black}) of AT2017gfo. The blue and NIR tails have been modelled with a power-law $F(\nu) \propto \nu^a$ (blue and red lines; see \S~\ref{sec:models}). 
}
\label{fig:knfit2}
\end{figure*}

\begin{figure}
\begin{center}
\includegraphics[width=0.48\textwidth,angle=0]{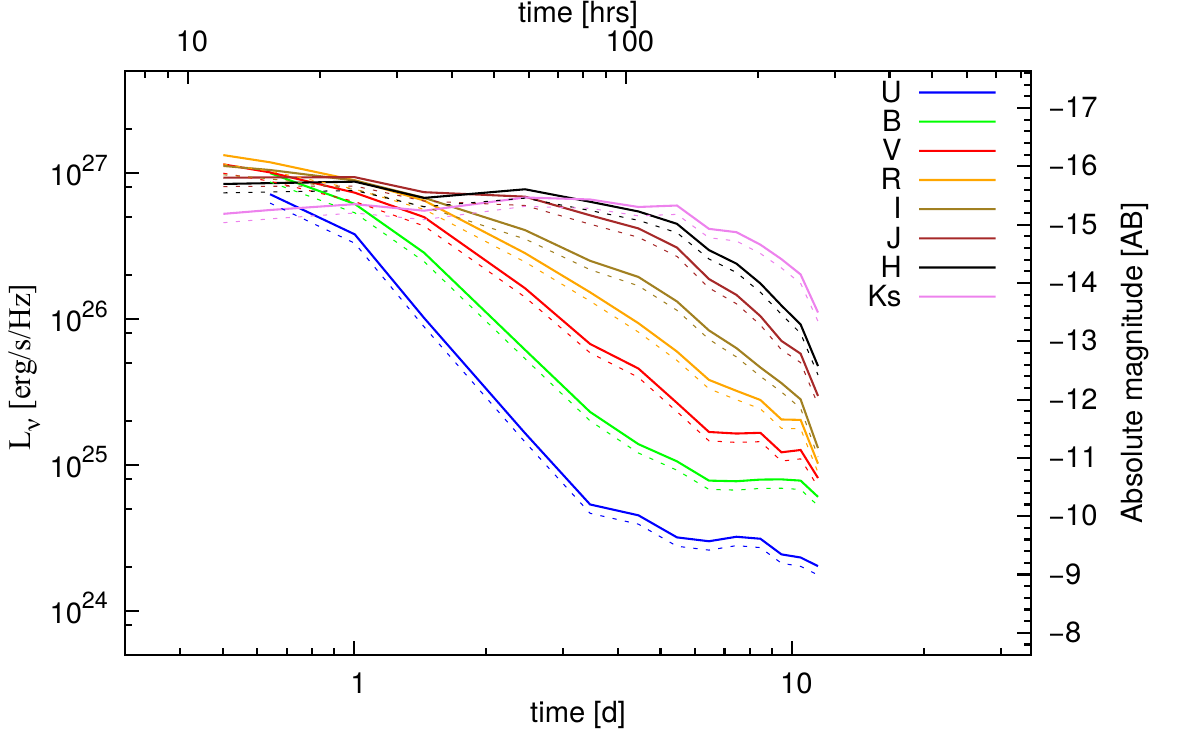}
\caption{Rest frame light-curves of AT2017gfo in few selected filters. The $U$ and $Ks$-band light curves  were obtained by extrapolating photometry and spectroscopy to obtain a full coverage of the filter transmissions (see \S~\ref{sec:models} and Table~\ref{tab:lckn}). 
The dashed-line curves have been obtained assuming a distance of $40.7$ Mpc as found by \citet{Cantiello2018a}. }
\label{fig:lckn}
\end{center}
\end{figure}

\section{Data analysis}
\label{sec:met}

In order to compare AT2017gfo with GRB optical counterparts, 
we computed the AT2017gfo luminosity in the GRB rest-frame filters. 
This approach allows us to use
the exceptionally high quality data set 
of AT2017gfo which provides
much better spectral accuracy and coverage than 
that of typical GRB afterglows, enabling a more precise flux estimate in the redshifted frequencies. 
We thus first built a set of rest-frame AT2017gfo spectra at different epochs, 
which hereafter we will refer as
kilonova spectral templates. We then convolved these spectra with the optical/NIR filters scaled to the GRB rest frames and proceeded with the luminosity comparison.

This procedure is similar to that used by \citet{Gompertz2018a}.
However, the use of X-Shooter spectra allows us to simplify the approach.
First of all, except for the black-body modelling of the photometry for the first 3 epochs, we do not interpolate the SED to obtain the photometry in a specific filter, because the X-Shooter spectroscopy guarantees full spectral coverage. Secondly, we do not need to make any special assumption for the bluer part of the spectra of AT2017gfo, because the X-Shooter spectra extends down to $3500$~\AA (including half of the $U$-band). 
With our method we can give robust constraints to the rest-frame optical data ($<6000$~\AA) up to redshift $\sim$0.7.
We also note that, in our case, we do not use an analytical function to model the light curves, since we have better temporal sampling. Instead, we interpolate the photometry derived from the spectral templates when we need to compute the luminosity ratio (Table~\ref{tab:ratiolum}).
Below we explain all the steps of this analysis in more detail.

\subsection{Kilonova spectral templates \label{sec:models}}

After including the three SEDs built at 0.5, 0.66 and 1 days after the trigger, we have a total of 13 epochs that we use to build the kilonova spectral templates. 


We modelled the SEDs (at 0.5, 0.66 and 1 days after the trigger) following the current theoretical interpretation of AT2017gfo, where the observed emission is the combination of at least two different black body components\footnote{We note that a more sophisticated model has been used by other authors \citep[][]{Villar2017b,Cowperthwaite2017a} who found some evidence of an intermediate black-body component. However, given that our goal is only to model the extremes of the spectral interval, we considered that an intermediate component is too sophisticated for the aim of this work.}. Results are plotted in Fig. \ref{fig:knfit}. 
We note that even in the early phase when matter is extremely dense, opaque and hot, strong line blanketing can be at work and the absorption in the UV may not be negligible and thus the real temperature of the blue component can be higher.
In the first epoch at 0.5 days after the trigger, the peak of the black-body is at much bluer wavelengths than the available photometry and cannot be constrained, therefore, in this case, we have only extrapolated the model to the NIR, after imposing the black-body temperatures to be $\leq$10000 K. For this reason, the U and B band KN template light curves start at 0.66 and not at 0.5 days after the trigger, like the other bands.

In order to maximise the possibilities of comparison with GRB counterparts down to the UV and up to the NIR regimes, 
we then modelled the UV and NIR extremes of the 
spectra, leaving untouched the rest of the spectral interval already covered. At the time of the X-shooter spectra (>1.5 days) it is likely that 
the ejected matter becomes more transparent and absorption features starts to dominate the spectra. Therefore, modelling
the data with one or more black-body components without considering absorption is not possible. However, given that we are only interested in expanding the spectral interval of the templates in the case of the X-Shooter spectra we have modelled the data with two power-laws, one below 5000~\AA~and one above 21000~\AA~(see Fig. \ref{fig:knfit2}).
The best fit models have been used to extrapolate the AT2017gfo flux down to 1,500~\AA ~and up to 26,000~\AA. 
Finally, we have computed 
the best fit spectral models 
in the kilonova rest-frame.  
For AT2017gfo we have adopted the redshift $z_{KN}=0.0098$ \citep{Hjorth2017a} that, with the assumed PLANCK cosmology, corresponds to a luminosity distance of $D_L=43.7$ Mpc.

The result of this procedure is a set of spectral templates, covering the UV to NIR range, computed at different epochs between 0.5 and 10 days after the GW trigger. 
We then used these templates to produce rest-frame light curves of AT2017gfo for all the GRB filters as explained in the next section (Table~\ref{tab:lckn}). 
In Figure~\ref{fig:lckn} we show a sample of light curves for few selected optical/NIR filters, computed also assuming a luminosity distance of $40.7$ Mpc, as found by \citet{Cantiello2018a}.
 {\bf It is clear that the peak of the optical emission lies in the first day after the trigger, while the NIR emission is almost constant during the first 6 days and dominates the emission after 2 days.}

\subsection{Comparison with short GRBs \label{sec:sgrb}}

To proceed with the AT2017gfo - short GRBs comparison, each short GRB flux $F_{\nu}$ measured at the time $t_{GRB}$ was converted to a luminosity and, for each filter, a rest-frame light curve was built.  
In order to compare the AT2017gfo and GRB luminosity in the same frequency, we proceeded as follows. 
For each GRB of our sample, we have a set of 
filters used for the observations. Given a GRB at redshift $z_{GRB}$, for each filter $X$ centred at the observed frequency $\nu_X$, we computed an effective rest-frame filter $X_{eff}$ centred at $\nu_{X,eff}=\nu_X\times(1+z_{GRB})$. 
By integrating the AT2017gfo luminosity spectra taken at different epochs ($t$) over the effective rest-frame filter \footnote{The effective rest-frame filter was obtained by multiplying the filter response matrix by $(1+z_{GRB})$}\label{footnote_3} 
$X_{eff}$, we were able to build 
a AT2017gfo luminosity light curve $L_{X,eff}(t)$ in the rest frame filter $X_{eff}$, i.e., the same in which the GRB was observed.

With this procedure, 
we built a set of AT2017gfo luminosity light curves in the same set of filters used to observe a given GRB. In this way we could proceed in a straightforward manner to the comparison of the luminosity of AT2017gfo and the GRB afterglow in each filter. 
Note that using the distance of 40.7 Mpc found by \cite{Cantiello2018a} 
AT2017gfo would become fainter of a factor $\sim$1.15 and therefore the luminosity ratios would change. However, this would not change 
qualitatively our conclusions.

\section{Results}%
\label{sec:res}

In the following we present the results obtained 
from the comparison in each filter of the AT2017gfo luminosity and those of the optical and NIR counterparts of our selected sample of short GRBs. 

The GRB counterparts and AT2017gfo light curves in different bands are plotted as luminosity (left side) and apparent magnitude (right side) versus rest-frame time in Figs.\ref{fig:constraints}, \ref{fig:lcplot}, \ref{fig:lcplot3}, \ref{fig:noconstraints}. 
Note that several filters with similar wavelength have been grouped into a single one in the plots for visualisation purposes, and that the filters quoted in each plot are the "effective" ones, i.e. the observed filters shifted to the GRB rest frame (\S\ref{sec:sgrb}). 

In order to avoid any model-dependent temporal extrapolation, in this work we limited the comparison to the short GRBs observations that fall in the sampled AT2017gfo temporal window ($0.5$-$10.5$ days in rest-frame). For this reason, for 13 short GRBs the comparison with AT2017gfo was not possible.
However, we still show their light curves as illustrative examples of the magnitude range of AT2017gfo-like emission in comparison with those of the observed GRB counterparts (see Fig. \ref{fig:noconstraints}). 
In the case of GRB 100206A, albeit not covered by the KN templates, the gap is negligible and the
observations are clearly fainter than the kilonova template (see below \S~\ref{sec:constraints}). In Table \ref{tab:ratiolum} we quote the luminosity as well as the ratios between GRB counterpart and AT2017gfo luminosity in the spectral bands and at the time of the observations at which such comparison was possible.

\begin{figure*}
\centering
\includegraphics[scale=0.71]{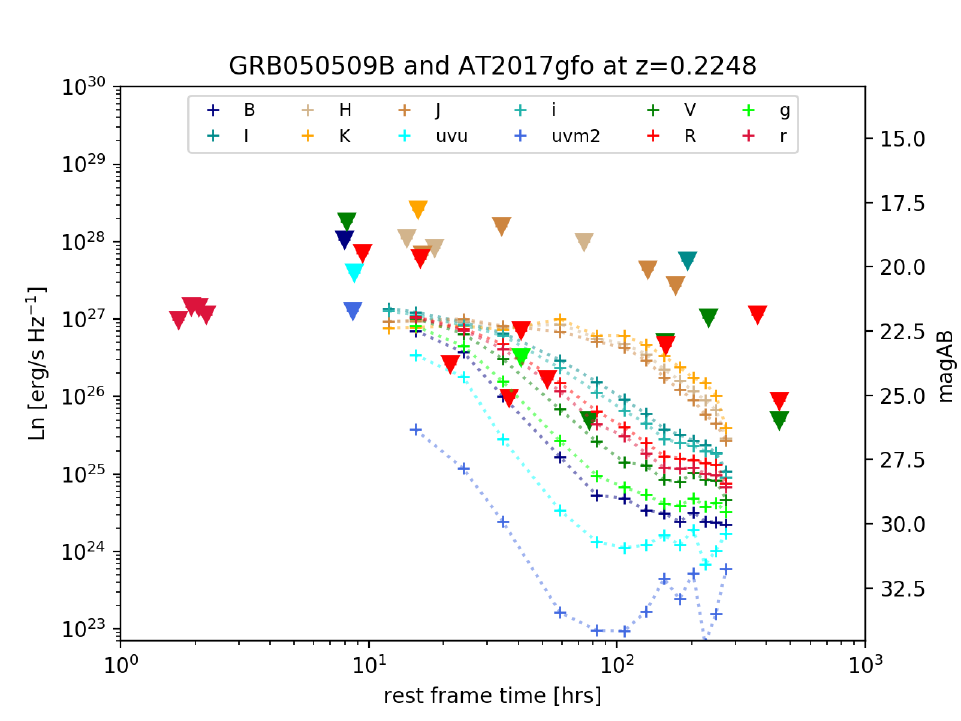}
\includegraphics[scale=0.71]{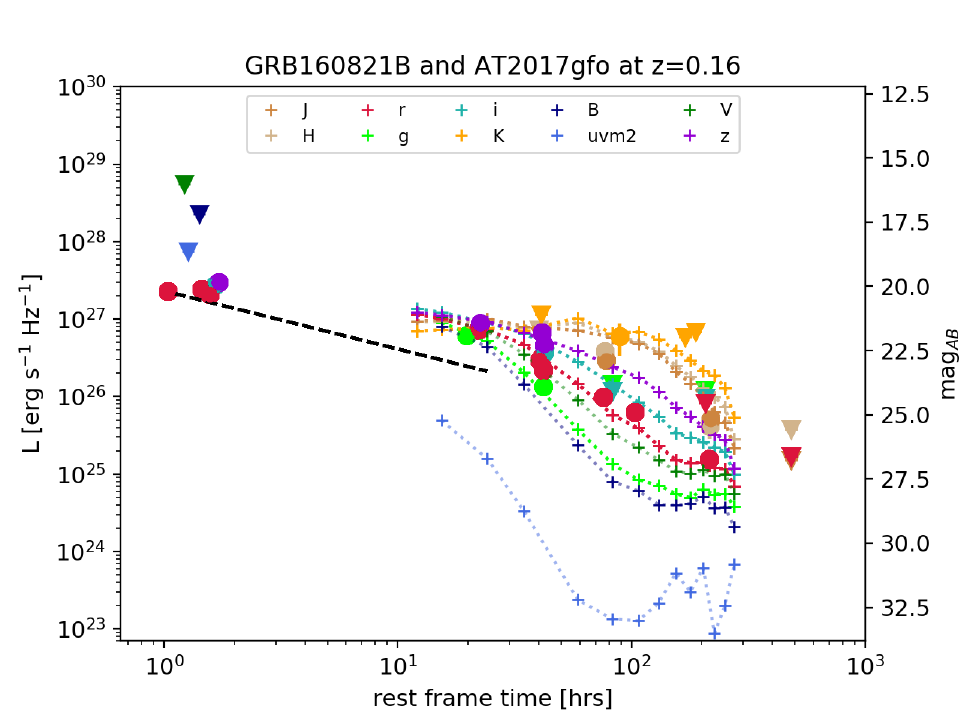}
\caption{{\bf Two examples of short GRBs for which the optical counterpart luminosity (circle for a detection and triangle for an upper limit) are fainter than AT2017gfo luminosity (dotted lines with crosses) in at least one effective rest-frame filter (see \S\ref{sec:sgrb} for effective rest-frame filter definition). See figure \ref{fig:constraints} for the full sample of short GRBs with similar properties. 
Note that GRB 160821B ({\it right}) shows evidence of a temporal decay index lower than the shallowest index predicted by the fireball model (i.e. $\alpha=0.75$, see \S \ref{sec:flatag}, black dashed line). 
If this anomalous shallow decay is due to an emerging kilonova emission, its NIR luminosity is a factor of 1.2-2 fainter than AT2017gfo at epochs later than 3 days after the merger time (see Tab.2).
}
}
\label{fig:constraints_example}
\end{figure*}

On the basis of the luminosity ratios and on the temporal behaviour of the  GRB counterpart luminosity, we built three main groups as described below (see \S \ref{sec:constraints}, \S \ref{sec:flatag}, and \S \ref{sec:130603B}). 
{\bf 
The first group includes 7 GRBs with a counterpart fainter than AT2017gfo in at least one filter. 
The light curves of these GRBs are shown in Figure \ref{fig:constraints} and two examples are plotted in Figure \ref{fig:constraints_example}.
The 19 short GRB counterparts brighter than AT2017gfo are plotted in Figure \ref{fig:lcplot} and Figure \ref{fig:lcplot3}. 
}
Moreover, we distinguish between the \emph{blue} and \emph{red} components, depending on whether the rest-frame effective wavelength is below or above $9000$~\AA, respectively.
In Table \ref{tab:ressum1} we summarise the short GRBs that stand out for their properties.

\begin{table}
\caption{
{\bf Summary of the GRB to kilonova luminosity ratios in the effective rest-frame filters (\S\ref{sec:sgrb}) of short GRBs that stand out for their properties (full Table in \ref{tab:ratiolum}). The sample is divided between events fainter (first part of the table) and brighter (second part of the table) than AT2017gfo. GRBs with accurate redshift (\S 2.3) are quoted in bold. }
}
\small
\begin{threeparttable}
\renewcommand{\tabcolsep}{2mm}
\begin{tabular}{p{3.2em} p{1.7em} p{2.2em}  p{1.4em} c p{1.1em}  p{1.2em}}
\toprule
GRB      & time\tnote{a}  & Band\tnote{b}           &$\lambda_{eff}$\tnote{b} & $L$-ratio\tnote{c} &  ID\tnote{d} & Com.\tnote{e} \\
         & (hrs) & (eff)          &(nm)            &${GRB}/{KN}$ &    &      \\
\midrule                                               
{\bf 050509B}  &  21.2 & R$_{/(1+z)}$  &    523    &$<$0.3  & blue  & --\\
         &  36.6 & R$_{/(1+z)}$  &    523    &$<$0.2  & blue  & \\
         &  52.2 & R$_{/(1+z)}$  &    523    &$<$0.7  & blue  & \\
{\bf 050709}   &  50.8 & I$_{/(1+z)}$  &    681    &0.8     & blue & S,KN \\
061201   &  29.8 & I$_{/(1+z)}$  &   712     & $<$0.3 & blue & --\\
         &  73.3 & I$_{/(1+z)}$  &   712     & $<$0.7 & blue &  \\
080905A\tnote{f}  &  12.8 & R$_{/(1+z)}$  &   572     & 0.3 & blue&  --\\
090515   &  17.8 & r$_{/(1+z)}$ &    446     & 0.4    & blue &S \\
{\bf 160821B}  & 22.1 & r$_{/(1+z)}$ & 539 & 0.9   & blue & S,KN \\
         &  42.0 & r$_{/(1+z)}$ & 539 & 0.6   & blue &  \\
         &  78.1 & J$_{/(1+z)}$  &  1081     & 0.5    & red & \\
         & 102.8 & K$_{/(1+z)}$  &   1863    & 0.9    & red  & \\
{\bf 100206A} &$\sim$10& i$_{/(1+z)}$  &   544     & $<1$    & blue &--\\
\midrule  
{\bf 050724} &27.7   & I$_{/(1+z)}$   & 628      & 6.0    & blue &  S,MKN \\
         &66.1   & I$_{/(1+z)}$   & 628      & 2.4    & blue & \\
         &27.8   & R$_{/(1+z)}$   &510       & 9.1    & blue & \\
{\bf 060614} &  12.8 & R$_{/(1+z)}$    &  570    & 16.9   & blue & S,KN\\
         &143.8 & R$_{/(1+z)}$ &  570 & 8.6  & blue & \\
{\bf 070714B} & 54.9 & R$_{/(1+z)}$ & 333 & 263 & blue & {MKN}\\
{ 070809} & 23.8 & R$_{/(1+z)}$ & 435& 3.4 & blue & {KN}\\   
{\bf 130603B}  &  28.3 & r$_{/(1+z)}$   &    462   &  1.3   & blue & KN  \\
         & 164.7 & H$_{/(1+z)}$    &   1218  &  3.0   & red  &   \\
{\bf 150101B}  &  35.1 & r$_{/(1+z)}$   &    552   &  2.1   & blue &KN \\
150424A&16.4&r$_{/(1+z)}$&  481     &  15.2  &blue  &  S \\
                &12.4&J$_{/(1+z)}$&  964     &  37.2  &red   &  \\
\bottomrule
\end{tabular}
\begin{tablenotes}\footnotesize
\item[a] rest-frame time
\item[b] rest-frame effective band and effective wavelength (\S\ref{sec:sgrb})
\item[c] GRB optical counterpart to AT2017gfo luminosity ratio 
\item[d] This column indicates if the effective wavelength is below (blue) or above (red) $9000$~\AA. 
\item[e] 'S'= evidence of shallow decay from this work; 'KN'= evidence of kilonova from the literature; ; 'MKN'= evidence of magnetar-powered KN from \cite{Gao2017a}
\item[f] In the case of GRB 080905A, the decay index of the optical/NIR light curve is $\alpha=0.4\pm1.3$ between 7 and 16 hours after the trigger (rest-frame), i.e., not enough to be constrained.
\end{tablenotes}
\end{threeparttable}
\label{tab:ressum1}
\end{table}

\subsection{Short GRBs with optical counterpart fainter than AT2017gfo}
\label{sec:constraints}

We find that in seven cases (namely GRBs 050509B, 050709, 061201, 080905A, 090515, 100206A and 160821B) 
the luminosity of the optical counterparts is smaller than that of AT2017gfo in at least one filter. 
This is also true in the case of GRB 100206A although the photometric monitoring ending before the temporal window of the light curve of AT2017gfo.  

In {\bf the first part of} 
Table \ref{tab:ressum1}
we report the rest-frame time after the GRB event 
(together with the effective rest-frame filters and wavelengths; see \S\ref{sec:sgrb}) 
in which we find that 
AT2017gfo was fainter than the optical/NIR counterpart of the GRB
by a factor quoted in the fifth column as the luminosity ratio.
We note that for 2 of these 7 short GRBs (namely GRB 050709 and GRB 160821B) a kilonova emission has been invoked in the literature \citep{Jin2016a,Jin2018a,Kasliwal2017b,Troja2019a,Lamb2019a}. These GRBs have been labelled in the last column of Table \ref{tab:ressum1} with "KN". 
{\bf The cases with evidence of magnetar-powered KN \citep[see][]{Gao2017a} have been labelled with "MKN".}
Moreover, for 3 GRBs we find evidence of a shallow decay not consistent with the standard fireball model (see next Section). The latter ones are labelled with "S" in Table \ref{tab:ressum1}.
For all the short GRBs belonging to this group we could probe the \emph{blue} kilonova component and constrain its luminosity within a range of 0.2-1 times the AT2017gfo luminosity. {\bf In the NIR, only the red counterpart of GRB 160821B is fainter (0.4-0.9 times) than At2017gfo (Tab.2)}.

\begin{table}
\centering
\caption{Summary of the GRBs with evidence of an anomalous shallow decay and/or with a claimed kilonova in the literature. GRBs with well defined redshift (\S 2.3) are quoted in bold and represent our golden sample (\S \ref{sec:gold}). 
}
\small
\begin{threeparttable}
\begin{tabular}{p{3.3em} p{2.3em} p{4.1em}  p{1.8em} p{1.5em}  l}
\toprule
GRB & Band\tnote{a} & $\alpha$\tnote{b} & $t_{start}$\tnote{c} & $t_{end}$\tnote{c} & Com.\tnote{d}\\
    & (eff)     &  & (hrs)       & (hrs)     &  \\ 
\midrule
{\bf 050709}             & I$_{/(1+z)}$ & $0.6\pm0.2$ & 50.7   & 115.8 &  KN, S\\
                         & I$_{/(1+z)}$ & $1.2\pm0.1$ & 115.8  & 202.1 &  \\
{\bf 050724} & R$_{/(1+z)}$ & $<0$        & 4.4    & 9.4   &  MKN, S\\
	                     & R$_{/(1+z)}$ & $1.5\pm0.1$ & 9.4    & 27.8  &  \\
{\bf 060614}             & R$_{/(1+z)}$ & $0.2\pm0.1$ & 1.0   & 10.0   &  KN, S\\
090515                  & r$_{/(1+z)}$ & $0.1\pm0.1$& 1.2   & 17.8   &  S\\
150423A                 & r$_{/(1+z)}$ & $0.4\pm0.2$  & 0.4   &  1.7   &  S\\
150424A                 & R$_{/(1+z)}$ & $0.09\pm0.03$& 1.2   & 10.1   &  S\\
{\bf 160821B}           & R$_{/(1+z)}$ & $0.0\pm0.1$ & 1.0   & 1.6    &  KN,S\\
                        & R$_{/(1+z)}$ & $0.40\pm0.03$  & 1.6   & 22.1   &  \\
{\bf 070714B} & R$_{/(1+z)}$ & $1.0\pm0.3$  & 12.3 & 54.9   &  MKN\\ 
{\bf 130603B} & H$_{/(1+z)}$ & $1.6\pm0.1$ & 10.7   & 164.7  & KN\\
	                   & R$_{/(1+z)}$ & $2.7\pm0.1$ & 10.9   & 28.3   & \\
{\bf 150101B} & r$_{/(1+z)}$ & $1.0\pm0.3$ & 35.1   & 56.1   &  KN\\
\bottomrule
\end{tabular}
\begin{tablenotes}\footnotesize
\item[a] rest-frame effective band (\S\ref{sec:sgrb})
\item[b] decay index ($F_{\nu}(t)\propto t^{-\alpha}$)
\item[c] rest-frame time interval within wich $\alpha$ was computed
\item[d] 'S'= evidence of shallow decay from this work; 'KN'= evidence of kilonova from the literature; 'MKN'= evidence of magnetar-powered KN from \cite{Gao2017a}
\end{tablenotes}
\end{threeparttable}
\label{tab:decay}
\end{table}

\subsection{Short GRB counterparts with shallow decay}
\label{sec:flatag}

A kilonova is expected to show a shallow evolution close to its maximum brightness. Therefore, it can be distinguished from the standard afterglow decay, which at the typical observing time (i.e., $>$min after the burst) has a constant power-law decay \citep[e.g.,][]{Sari1998a,Sari1999a,Zhang2004a,Zhang2006c}. 

Kilonova peak brightness is typically estimated around a few days after the merger assuming simple one-component modelling and fiducial values 
\citep[e.g.,][]{Metzger2010a}. However, according to more sophisticated models (e.g. \citealt{Radice2018a}), the kilonova maximum brightness 
can be as early as $<$0.1 days, depending on the nature of the central remnant (i.e. if a black hole or a neutron star is generated, see their Fig. 28). Thus, we considered {\bf as possible} evidence of a kilonova 
a shallow evolution in the optical counterpart of the short GRBs that can happen from a 
few hours after the burst up to a few days.
{\bf 
Our simple method of kilonova identification is effective only near the kilonova peak epoch.
In fact, it is not suited to find events where the kilonova is dominating but its decay is too steep to be distinguished from an afterglow (for example see the case of GRB 150101B in \S\ref{sec:130603B}).
A shallow decay simultaneous to a X-ray plateau 
can possibly indicate the presence of a MKN \citep[see ][]{Gao2017a}, although other explanations are possible \citep[e.g.,][]{Mangano2007a}. Therefore, we also describe the simultaneous behaviour of the X-ray light curve. 
}

Under the reasonable assumption of a slow cooling regime for the electrons producing the observed afterglow radiation \citep[see][]{Sari1998a,Sari1999a} the predicted shallowest flux decay power law index is $\alpha=3(p-1)/4$ where $p$ is the power law index of the electron energy distribution.
In this context, using a minimal electron index $p=2$ in the slow cooling regime, we considered a decay to be anomalously shallow when $\alpha<0.75$. 
We computed the decay index $\alpha$ for all GRBs for which two or more observations were available. 
We note that the flattening in the GRBs 071027 and 061006 is due to the contribution from the host
\citep{Davanzo2009a}, and thus they are not considered here.

\begin{figure}
\includegraphics[width=0.48\textwidth,angle=0]{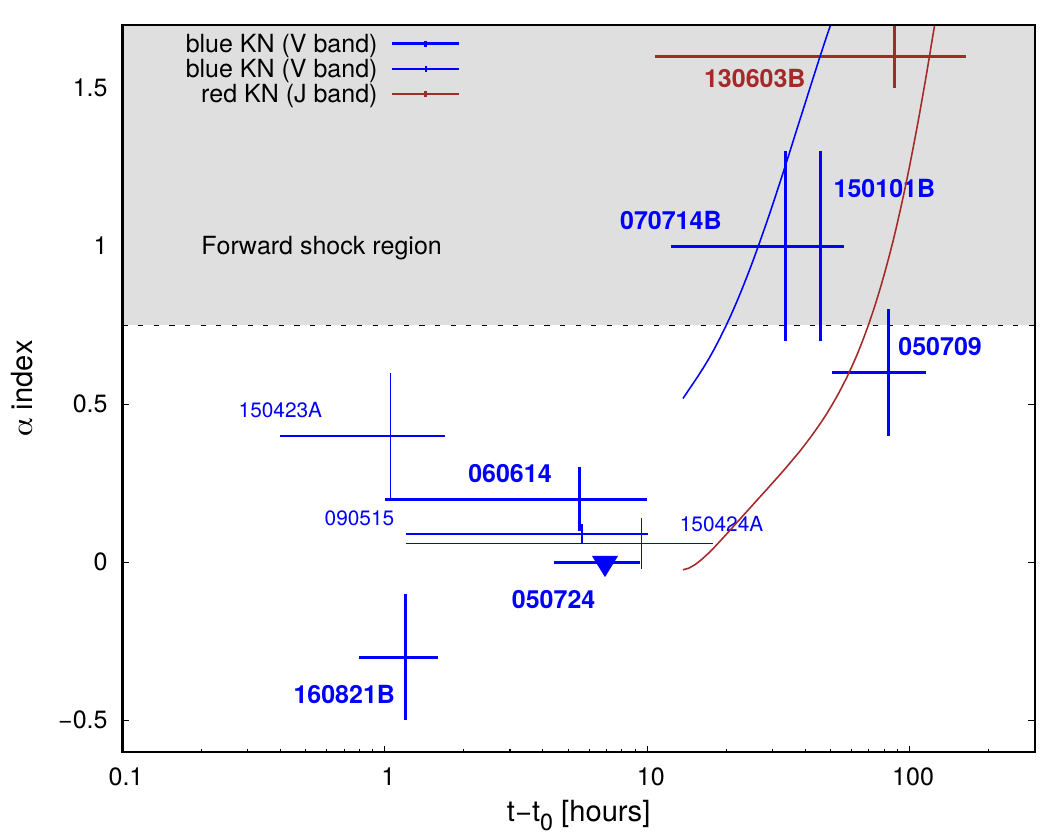}
\caption{The lowest decay indices of the GRB optical counterparts with shallow decay. See Table \ref{tab:decay}. The short GRBs labels are coloured following the band in which the decay index was computed.
Upper limits are indicated as downward triangles. 
{\bf GRBs in the golden sample are highlighted in bold.}
The horizontal line indicates the lowest possible decay index predicted by the afterglow theory ($\alpha=0.75$; see \S \ref{sec:flatag}). 
The curves are the smoothed splines of the decay indices of AT2017gfo in the V and $J$-bands, from top to bottom, computed in the 13 epochs. These are derived by computing the decay between two epochs in the template light curves shown in \S \ref{fig:lckn} with a time step of 0.5 days.
}
\label{fig:alpha}
\end{figure}

In {\bf 7} cases we have found a suspicious shallow decay that can indicate the presence of a kilonova emission dominating over the afterglow component. {\bf We quote them in Table \ref{tab:ressum1}.} 
Note that in all cases, 
we could measure a shallow decay only in the optical filters, i.e. in the regime of the blue component.
The results are summarised in Table~\ref{tab:decay}.
In Figure~\ref{fig:alpha} we compare the estimated decay indexes with those of AT2017gfo, computed assuming a power-law evolution between {\bf consecutive} template epochs (Fig.  \ref{fig:lckn}).
Note that the decay index of the kilonova is always smaller in the $J$-band than in optical bands thus {\bf reflecting the expected}   smoother evolution in the red band with respect to the blue one.

In the following, we discuss in more detail these bursts. Note that {\bf 3} of them (GRBs 050709, 090515, 160821B) are also part of the first group, i.e., those GRB optical counterparts that are fainter than AT2017gfo.
{\bf According to AT2017gfo templates (Fig.\ref{fig:lckn}), the blue component dominates during the first 24 hours, therefore}
we first list those short GRBs for which 
we measured a shallow decay {\bf before 1 day. We include also those short GRBs for which we measured a shallow decay 
at epochs earlier than the first AT2017gfo observation (i.e. before $\sim11$ hours).} 

\begin{itemize}



\item 
{\bf GRB 050724.} Its early $R$-band light curve is rising between 5.6 and 11.8 hours after the trigger {\bf (4.4 and 9.4 hours rest-frame)}, with $\alpha \sim-0.5$.
{\bf  Simultaneous X-ray and radio data show a similar trend. 
\cite{Gao2017a} interpreted this behaviour as evidence for a kilonova powered by a magnetar. We find that, if it were an emerging kilonova, its blue component is brighter than AT2017gfo up a factor 9 and 6 in $R$ and $I$ bands at $\sim28$ hours (rest-frame), respectively. This factor 
decreases to $\sim2.4$ at $\sim66$ hours.
}

\item {\bf GRB 060614A.}
In this case the $R$-band light curve shows a clear shallow decay phase
{\bf between 1.5 and 11 hours (1 and 10 hours in rest-frame) simultaneous with an X-ray plateau feature,}
with a brightening peaking at $\sim4$ hours after the trigger. 
In this case, the detected luminosity are a factor of $\sim$16 larger than the $R$-band luminosity of AT2017gfo at $\sim$12.8~hours after the trigger. 
{\bf \citet[][]{Jin2015a,Yang2015a} found a kilonova component for this burst at more than $\sim$3 days after the trigger from an optical excess in the afterglow. At this time we find no clear evidence of an optical shallow decay (see \S~\ref{sec:peculiar})}.

\item {\bf GRB 090515.} In this case the $r$-band light curve has a decay index $\alpha\sim0.1$ between 1.7 and 25 hours after the trigger {\bf (1.2 and 17.8 hours rest-frame)}. Moreover, a very late ($\sim10^3$~hrs) deep upper limit confirms that no emission from an underlying host is affecting the early data. 
{\bf  The X-ray afterglow is very weak and no data is available for a comparison after the first hour.}
If a kilonova is emerging from the afterglow emission, it is fainter than AT2017gfo by a factor of 2.3 at $\sim18$ hours after the trigger.
Note that this case was already presented in \S 4.1.

\item 
{\bf GRB 150423A.}
This burst shows a shallow decay behaviour in the i- and z-bands, with index $\alpha =0.4\pm0.2$ in the r-band before 4 hours after the trigger {\bf (1.7 hours in rest-frame)}.
The temporal mismatch with the AT2017gfo light curves prevents us from performing a more quantitative comparison.
{\bf We note that during the same time interval the X-ray light curve is much steeper ($\alpha_X\sim0.96\pm0.07$), thus suggesting a different origin with respect to the optical counterpart.
}

\item 
{\bf GRB 150424A.}
In this case the r-band light curve has an atypical shallow decay with index $\alpha\sim0.1$ between 1.6 and 13 hours after the trigger
{\bf (1.2 and 10.1 hours in rest-frame), simultaneous with an X-ray plateau feature}. 
{\bf If a kilonova is the dominant component, its blue component is brighter than AT2017gfo by a factor of $\sim15$ at $\sim16.4$ hours after the trigger (rest-frame), i.e. at the end of the shallow phase. At the same time in the $J,H$ bands the light curve is brighter than AT2017gfo by a factor $\sim30$ and falls to $\sim2.3$ times brighter at $\sim$124 hours after the trigger.} 

\item {\bf GRB 160821B.} For this event, the presence of a kilonova emission was claimed by \citet{Troja2019a}.  
In this case, the $R$-band counterpart shows a brightening at $\sim1$ hour after the trigger ($\alpha\sim-0.3$), which then steepens to $\alpha~\sim0.9$ between 1 and 86 hours after the trigger {\bf (1.6 and 74.5 hours in rest-frame). A very weak X-ray afterglow possibly show evidence of a plateau lasting $\sim1$ day after the burst}.
If an underlying \emph{blue} kilonova component is peaking 
at $\sim1$ day after the trigger, the kilonova is fainter than AT2017gfo by a factor in the range 1.1-2.5 (see Tab.2). As also noted by \citet{Troja2019a} the late $J$ and $H$-band light curves have a behaviour similar to that of KN170817, but have a decay index larger than one \citep[see also][]{Lamb2019a}. This case was already presented in \S 4.1.

\end{itemize}
Below we list those short GRBs for which we measured a shallow decay at more than 24 hours after the trigger.
\begin{itemize} 

\item {\bf GRB 050709A.} For this burst the presence of a kilonova was claimed by \cite{Jin2016a}. 
In this case the $R$-band light curve is very shallow with $\alpha\sim0.7$ between 59 and 134 hours after the trigger {\bf (51 and 116 hours rest-frame) and no straightforward comparison with X-ray data was possible due to the lack of enough statistics in the data \citep{Fox2005a}}. From the comparison with AT2017gfo, if the claimed kilonova signature is real, then its luminosity is comparable within the uncertainties in the $K$, $I$ and $V$-bands, although afterglow contamination is still possible.
This case was already presented in the previous section. 



\end{itemize}

\subsection{Short GRBs with known kilonova candidates}
\label{sec:130603B}
In few cases
a kilonova component has been found in the optical/NIR GRB counterpart {\bf light curves} and published in the literature. Two of them (GRBs 050709 and 160821B) have an optical counterpart fainter than AT2017gfo and are already part of the first group in Table~\ref{tab:ressum1}. {\bf Others
(namely GRBs 050724, 060614, 070714B, 070809, 130603B and 150101B) have an optical/NIR counterpart brighter than AT2017gfo and we report them in the second part of Table \ref{tab:ressum1}.}

{\bf \citet{Gao2017a} found evidence of kilonova in GRBs 050724 (see \S\ref{sec:flatag}), 061006 and 070714B by modelling the X-rays and optical light curves with a MKN model. However, as we noted in \S \ref{sec:flatag}, 
\citet{Davanzo2009a} found that the optical light curve of 061006 is  dominated by the host galaxy, thus we have included only GRB 050724 and GRB 070714B in our analysis (see Tab. \ref{tab:decay}).
 In case of 070714B, we cannot identify a shallow phase, but \citet{Gao2017a} found a MKN peaking at $\sim2$ days (observer frame; $\sim$1 day in rest-frame). 
During the time when this component is dominating there are two photometric epochs, one at $\sim11$ hours (just earlier than the first AT2017gfo photometric measurement) and one at $\sim55$ hours. During the first epoch the blue counterpart is clearly more than 10 times brighter than AT2017gfo, and the ratio increases to $\sim260$ times at the second epoch.
} 

{\bf In the case of GRB 070809, recently \citet{Jin2019a} discovered a kilonova in its optical light curve. Unfortunately, its redshift is not well determined and therefore it cannot be used to constrain the optical luminosity of the kilonova with confidence.
}

For GRBs 130603B and 150101B 
a kilonova component was claimed in the literature \citep{Tanvir2013a}. 
{\bf The kilonova was detected with an observation taken $\sim7$ days after the burst in the $H$-band \citep{Tanvir2013a}. 
At that time, its luminosity in the same band is $\sim3$ times brighter than AT2017gfo (see Fig. \ref{fig:lcplot3}).}

In the case of GRB 150101B the r-band light curve has a decay index $\alpha\sim1$ between 35 and 56 hours after the trigger which is above the shallow decay limit.
\citet{Troja2018b} show that this light curve is compatible with the late evolution of the blue component of AT2017gfo, well after the peak emission.  
We find that if a blue kilonova is the dominant component, it is brighter than AT2017gfo by a factor $\sim2$ at $\sim35$ hours after the trigger. Moreover, the deep and late upper limits confirm that no emission from an underlying host is affecting the early data. 


\begin{figure*}
\begin{center}
\includegraphics[width=0.48\textwidth,angle=0]{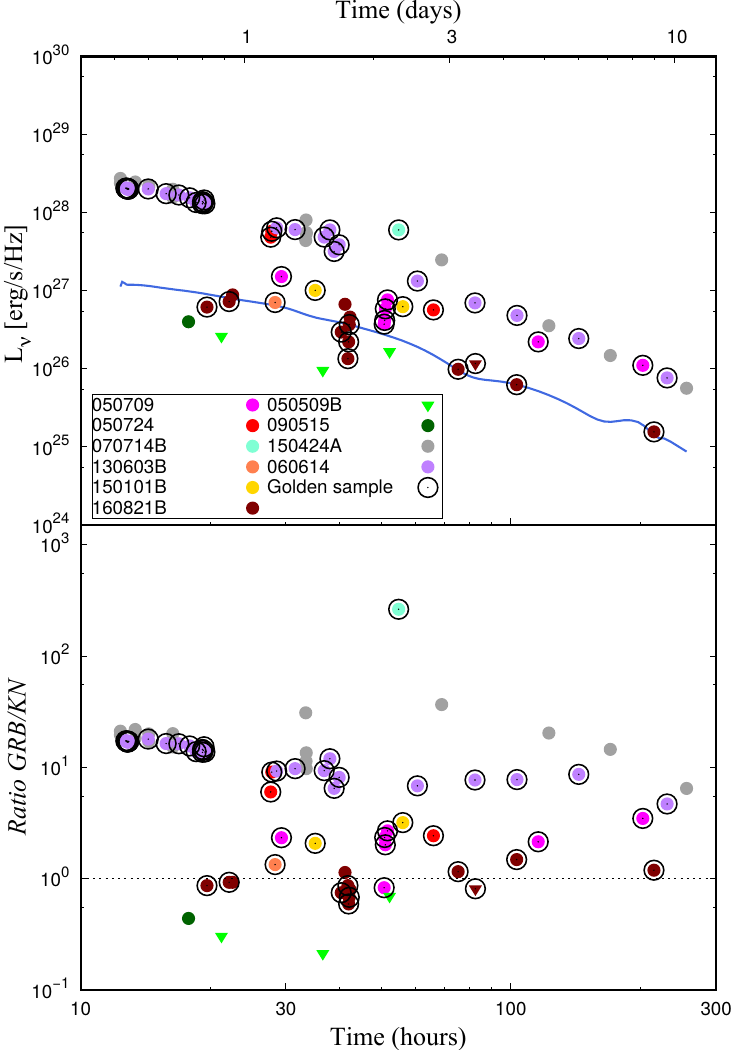}
\includegraphics[width=0.48\textwidth,angle=0]{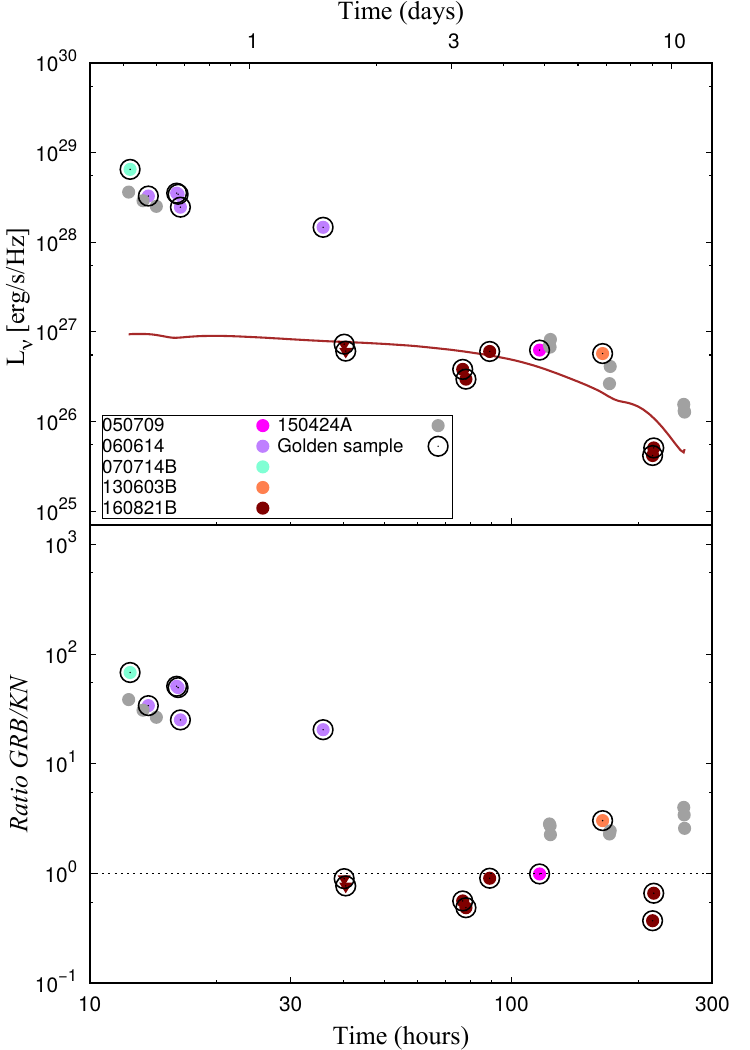}
\caption{luminosity ({\it top}) and luminosity ratios ({\it bottom}) of GRB and AT2017gfo versus time from merger. Data are taken from Table \ref{tab:ratiolum} after shifting the effective wavelengths of the filter bands to the rest frame value. The whole analysed sample is plotted in blue ({\it left}) and red ({\it right}) where the colour code indicates two different spectral bands (blue: $<$900 nm, red: $>900$ nm). Upper limits are indicated as downward triangles {\bf and only when below AT2017gfo luminosity}. 
Short GRBs with anomalous shallow decay are highlighted with different colours, including those without a well defined redshift. Bursts which belong to our golden sample are highlighted with a black circle (see \ref{sec:gold}).
The blue and red solid lines in the top panels indicate the AT2017gfo luminosity at 800 nm and 1600 nm wavelength, respectively.
}
\label{fig:ratiotime}
\end{center}
\end{figure*}

\section{Discussion}
\label{sec:dis}

In the following we first discuss those GRBs that are fainter than AT2017gfo,
without considering those with uncertain redshift.
Afterwards, we will discuss the cases with 
shallow decay or {\bf with} a claimed kilonova, {\bf by defining a golden sample of GRBs with accurate redshifts and by using} their luminosity to constrain that of AT2017gfo-like blue and red kilonova components.

\subsection{Upper limits to AT2017gfo-like kilonovae} \label{sec:ul}

In the previous section (\S\ref{sec:constraints}) we show that in seven cases an AT2017gfo-like emission could have been detected since its expected luminosity is well above the observed optical counterpart luminosity 
(see Tab. \ref{tab:ressum1} and Tab. \ref{tab:ratiolum}). 
{\bf In particular, we note that in two cases (GRBs 050509B and 061201) 
if a kilonova was present, it should be less luminous than AT2017gfo up to more than a factor of 5 for the blue component \citep[see also ][]{Gompertz2018a}.}
In the following, we will consider only those with accurate redshift determination (see \S\ref{sec:redshift}), which are {\bf GRB 050509B}, 050709, 100206A and 160821B.
The counterpart luminosity of these {\bf four GRBs} enable us to robustly set constraining upper limits to a possible underlying kilonova component that is fainter than AT2017gfo {\bf \citep[see also][]{Dichiara2019a}}.

In Figure \ref{fig:ratiotime} we compare their optical and NIR counterpart luminosity 
with the blue and red component of AT2017gfo (the spectral bands of the blue and red component have been defined in \S\ref{sec:res}). 
In the blue component spectral band, 
{\bf the strongest constrain is given by GRB 050509B that is more than 5 times fainter than AT2017gfo 
$\sim35$ hours after the trigger.}
GRBs 100206A and 050709 are marginally fainter and still comparable in luminosity to AT2017gfo at $12$ and $\sim50$ hours after the trigger, respectively.

For the red component, GRB 160821B was fainter than AT2017gfo, although its blue component has similar luminosity. The deepest and earliest constraint is a factor $\sim2$ fainter than AT2017gfo at $\sim1000$nm at $\sim78$ hours after the burst, close to the actual NIR peak of AT2017gfo, indicating that in this case the red kilonova is at least partially suppressed.
{\bf Intriguingly, in the bottom right panel in figure \ref{fig:ratiotime} we show that there are no NIR upper limits 
below AT2017gfo luminosity after $\sim50$ hours
(see also \S\ref{sec:peculiar}). 
The same is not true in the left panel, where upper limits exist below AT2017gfo luminosity level, thus possibly suggesting a larger range of luminosity for the blue counterpart with respect to the red one. 
}

\subsection{Golden sample of GRBs with kilonova candidates}
\label{sec:gold} 

Past evidence of kilonova emission was found in the GRB optical counterparts of {\bf 6} short GRBs (050709A, 060614A, 080503, 130603B, 150101B, and 160821B; see \S \ref{sec:intro} and references therein). {\bf In addition to those, there are the magnetar-powered kilonovae identified by \citet{Gao2017a} (GRBs 050724, 070714B).}
{\bf 
In all cases but GRB 080503 the redshift is well defined.
Therefore, we define a golden sample that includes all 7 GRBs with kilonova candidates claimed in the literature that have accurate redshift.} 
 \subsubsection{Interesting extreme events}
 \label{sec:peculiar}

{\bf 
We define "extreme" events those cases in our golden sample that are more than 10 times brighter or fainter than AT2017gfo either in the blue and/or in the red bands.
We find two cases 
with a bright blue counterpart, namely GRBs 060614 and 070714B. 
}
{\bf 
Concerning GRB 060614, in this work we show that its optical light curve has a shallow decay until 10 hours after the trigger (rest-frame; see \S\ref{sec:flatag}) and the blue counterpart is 17 times brighter than AT2017gfo at 13 hours (i.e. at the end of the shallow decay).
We note that a kilonova has been found by \citet[][]{Yang2015a} but dominating 3 days after the GRB, with a peak in the infrared \citep[][]{Jin2015a}. 
In \citet{Mangano2007a} the early optical behaviour of GRB 060614 is explained as the counterpart of the plateau observed in X-rays.
 However, the similarity of the optical and X-ray light curves of GRB 060614 with GRB 050724 (see also Fig.\ref{fig:lcplot}), for which a MKN was claimed by \citet{Gao2017a}, may support the blue component interpretation of the early emission for both cases. 
}

{\bf 
The blue counterpart of GRB 070714B is between 10 and $\sim260$ times brighter than AT2017gfo between 11 and 55 hours (\S\ref{sec:130603B}). During this time
\citet{Gao2017a} propose that a MKN is dominating.
They show that the peak bolometric luminosity of MKNs is $\gtrsim10$ times more luminous than other kilonovae like the one associated to 050709, which we find more similar to AT2017gfo.
This is also the case for the MKN associated GRB 050724, which instead is 
more similar to AT2017gfo in our analysis (see \S\ref{sec:gold}). However, the proposed MKN peaks at $\sim0.5$ days, i.e. too early for a comparison with AT2017gfo templates, and afterwards it decays rapidly. Therefore, our analysis cannot constrain the peak bolometric luminosity.}

{\bf 
In the NIR band, all the kilonovae 
detected (GRBs 050709, 130603B, 160821B, \S\ref{sec:130603B}),
are no more than 3 times brighter or fainter than AT2017gfo
during the time where the red kilonova dominates \citep[after 1 day, see ][]{Jin2016a,Tanvir2013a,Lamb2019a}.
Moreover, as we noted already in \S~\ref{sec:ul}, there are no upper limits in the NIR comparable to AT2017gfo luminosity after $\sim$2 days. In other words, in all cases when observations comparable to AT2017gfo NIR emission exist the kilonova counterpart has been detected. This suggests that all red kilonova detected so far have similar luminosity, although we are aware that the numbers are not high enough for a meaningful statistic.
}

\subsection{Interesting events without accurate redshifts \label{sec:interes}}

The redshifts of GRBs 090515 and 150424A are not well defined (see \S~\ref{sec:redshift}), 
{\bf nevertheless they can be 
useful to constrain the luminosity of AT2017gfo-like kilonovae since
both show evidence of shallow decay 
that can suggest the presence of a kilonova 
component. 
In particular, GRB 090515 is the most interesting 
because the blue component have a well constrained luminosity $2.3$ times fainter than AT2017gfo
(Fig.~\ref{fig:alpha}). }
If its redshift is correct, {\bf together with 050509B} it would provide the strongest and earliest constraints to the blue component of an AT2017gfo-like kilonova (Fig.~\ref{fig:ratiotime}). 

GRB 150424A has a shallow decay but its blue component is $\sim$15-40 times brighter than AT2017gfo in the optical (Fig.~\ref{fig:ratiotime}).
{\bf During the same time interval the X-ray light curve shows evidence of a plateau feature. 
This could be a MKN similar to GRBs 050724, 070714B and 060614 (\S~\ref{sec:peculiar}), although
\citet{Knust2017a} found that energy injection from a down spinning magnetar can explain X-rays, optical, and NIR data without invoking a kilonova.} 

{\bf A more detailed analysis is needed to separate the kilonova and afterglow components modelling together both optical and X-ray data \citep[see e.g.,][]{Jin2019a,Gao2017a,Lamb2019a}. Unfortunately, in case of GRB 090515 the X-ray data is weak and detected only during the first hour.
} 

\subsection{Interpreting the large range of luminosity}

AT2017gfo is the only kilonova that has been very well sampled and studied so far, but it is a one-of-its-kind example and other kilonovae may differ for their evolution and colours. 
Providing a theoretical explanation for kilonovae tens times brighter than AT2017gfo is beyond the scope of this paper. However, in the following we describe the possible causes to the large range in luminosity that we have found.

According to the most accredited model \citep[e.g.,][]{Mooley2018b,Ghirlanda2019a}, AT2017gfo was observed $\sim$15 degrees off-axis while here we are comparing it with likely on-axis events \citep[see also][]{Bulla2018a,Mandel2018a}. 
A kilonova luminosity gradient at a given wavelength is expected between 
the polar and the equatorial direction of the binary plane system, and
many parameters including its magnitude depend on the fate of the central remnant \cite[e.g.,][]{Kasen2017a,Radice2018a}. 
According to recent  numerical computations (see e.g., figure 24 in \citealt{Radice2018a}), 
in the case of a binary NS system (BNS) promptly forming a BH 
the result is an overall decrease of luminosity by less than a factor 2 (i.e., $\sim0.5$ mag) {\bf in the polar direction with respect to the equatorial one. Accounting for an off-axis inclination like AT2017gfo, then the polar emission should be less than 2 times fainter}.
This is not enough to explain the low luminosity ratio of those GRBs with optical counterpart fainter than AT2017gfo (Tab. \ref{tab:ressum1}).
In the case of an hyper massive NS (HMNS) or a stable NS being formed, then the polar luminosity should increase in the rest-frame $g$ and $z$ bands by a factor of less than 1.5 {\bf with respect to the equatorial direction} (i.e., a decrement of $\Delta g \leq 0.4$ mag and $\Delta z \leq 0.2$ mag). {\bf Again,} this factor is not high enough to explain the measured large luminosity ratios for the peculiar events we describe in \S~\ref{sec:peculiar}.
We conclude that, in the prompt BH formation case we cannot explain the measured luminosity gradient for any viewing angle and even assuming a central {\bf HMNS formation before the collapse to a BH}, the viewing angle correction factors are not large enough to recover the observed luminosity gradients (Tab.~\ref{tab:ressum1}). 

A possible solution to explain the {\bf faint} emission of the kilonova associated with the 7 GRBs for which the  optical counterpart was fainter than AT2017gfo may invoke not only a different viewing angle 
but also a different progenitor, i.e. NS-BH instead of BNS, where larger opacities are expected with respect to a NS-NS merger case \citep[][]{Kasen2015a,Metzger2017a,Barbieri2019a}. Although in the most dramatic cases lower masses and velocities of the ejecta can play an important role \citep[e.g.,][]{Dichiara2019a}, any further investigation is beyond the scope of this paper.

{\bf On the other side, a possible explanation for the largest  luminosity ratios may invoke the presence of a long-lived NS remnant that can alter the kilonova luminosity.}
In this case, its spin-down emission could illuminate the ejecta on timescales much longer (up to hours or even more) than the typical timescale of baryon wind ejection and neutrino irradiation (less than few seconds), effectively increasing the ejecta kinetic and thermal energy and thus potentially altering the brightness of the corresponding KN \citep[e.g,][]{MetzgerPiro2014a,Gao2017a}.

A highly magnetised millisecond pulsar (a magnetar) has been previously proposed to explain the plateaus observed in the X-ray light curves of GRBs, where the magnetar loses energy via dipole radiation and thus provides the energy to sustain the X-ray plateau phase \citep[][]{ZhangMeszaros2001a,Yu2013a,MetzgerPiro2014a,Siegel2016a,Siegel2016b}. 
Note that an X-ray plateau was found in the light curves of GRBs 060614A
\citep[][]{Mangano2007a,Stratta2018a}, and 150424 \citep{Knust2017a} that show a blue component brighter than AT2017gfo.
Therefore, it is possible that what we observed in these cases was a BNS merger exploding as a short GRB with a bright X-ray plateau and an luminosity-enhanced blue kilonova, leaving a magnetar as the final remnant of the merger, 
{\bf similarly to what proposed by \citet{Gao2017a} for GRB 050724.}

\begin{figure}
\begin{center}
\includegraphics[width=0.5\textwidth,angle=0]{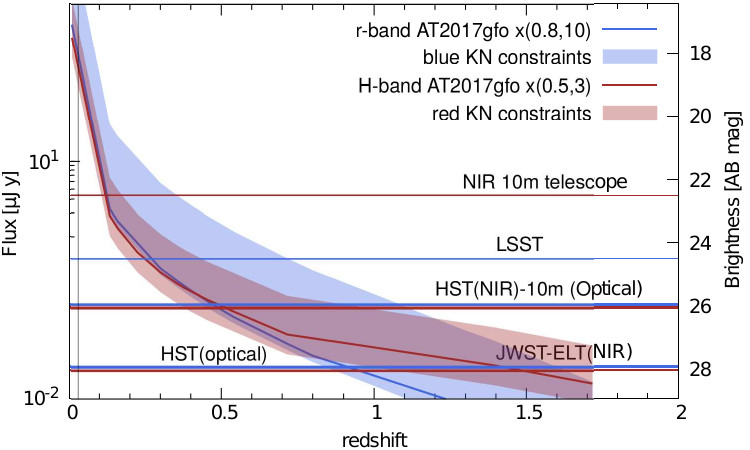}
\caption{Peak brightness of AT2017gfo in the $r$ (green) and $H$ (red) bands at different redshifts, within the constraints we derived for an AT2017gfo-like kilonova from the golden sample. The redshift range is limited at $z\sim2$ at which ET will be able to observe a GW signal from a merging BNS. The vertical line is the aLIGO/AVirgo detection limits for a BNS event. 
The horizontal lines are different detection limits for different class of telescopes with an exposure time of 10 minutes.
}
\label{fig:gwz}
\end{center}
\end{figure}

\subsection{Future perspectives for high redshift events}
\label{sec:hiz}

In the following, we want to investigate up to which redshift a kilonova can be followed-up, considering the current and future optical and NIR facilities.
In doing so we do not consider the challenge to search and identify a kilonova within the error boxes given by the GW detectors \citep[ e.g.,][]{Brocato2018a}.  

In Fig. \ref{fig:gwz} we show the maximum brightness of AT2017gfo in 
the observed $r$-band (at 12 hours in the rest frame) and $H$-band (at 58 hours in the rest frame) up to the redshift at which
the future Einstein Telescope (ET; \citealt{ET2012a}) will be able to observe a GW signal from a merging BNS ($z\sim2$).
In light of the results from the section~\ref{sec:gold}, and conservatively assuming that the blue kilonova is brighter up to 10 times AT2017gfo, we can constrain the peak luminosity of the blue kilonova between 0.8 and 10 times that of AT2017gfo and for the red component between 0.5 and 3 times. These constrain identify blue and red coulored regions in figure \ref{fig:gwz}. 
We put a lower limit to the 3-$\sigma$ detection with the current largest ground-based and orbiting telescopes dedicated to the characterization of the source: e.g., VLT, LBT ($r = 26$, $H = 23$ mag in the AB system), and the Hubble space telescope (HST) along with the forthcoming LSST (\citealt{LSST2019}; $r=25$) and ELT \citep{Spyromilio2008a} ground-based telescopes and the JWST \citep{Gardner2006} space telescope (H $\sim28$ mag, AB system) assuming 10 min exposure time \citep[see also][]{Maiorano2018a}.
An AT2017gfo-like kilonova would be detectable up to redshift 0.5 in the optical and 0.2 in the NIR by ground-based very large telescopes.
The JWST will be able to detect AT2017gfo at redshift larger than one.

From figure \ref{fig:gwz}, we note that a AT2017gfo-like kilonova would be brighter in NIR bands at redshift larger than $\sim0.5$, but only the JWST or the ELT would be able to detect this emission. 
Note that the current largest telescopes are able to detect the brightest
AT2017gfo-like blue kilonovae above $z=1$, a distance at which only HST is able to detect the brightest red kilonovae.
 The situation will improve when, thanks to JWST and ELT, we will be  able to detect a kilonova up to $z\sim0.7-1.6$ for the blue component and $z\sim1-2$ for the red  component (Fig. \ref{fig:gwz}).
This shows that follow-up of GRB/kilonovae with large-size ground-based telescopes and space observatories at redshifts beyond that of AT2017gfo is possible, although in most cases
it can be difficult to distinguish the GRB afterglow from the kilonova component.

{\bf Again, we stress that the real challenge will be to search and identify a kilonova within the error boxes given by the GW detectors. Distant GW sources (z>0.5) will be discovered only with interferometers of third generation as ET and will be localised within several thousands square degrees with a single interferometer and within few tens of square degrees with three detector network \citep[e.g.][]{Chan2018a}. Therefore, only the association with a GRB will permit to localize high-redshift kilonovae with enough accuracy. This can be provided by future space-based GRB dedicated missions as for example THESEUS \citep{Amati2018a,Stratta2018b,Rossi2018a}.
}

\begin{figure*}
\begin{center}
\includegraphics[width=0.6\textwidth,angle=0]{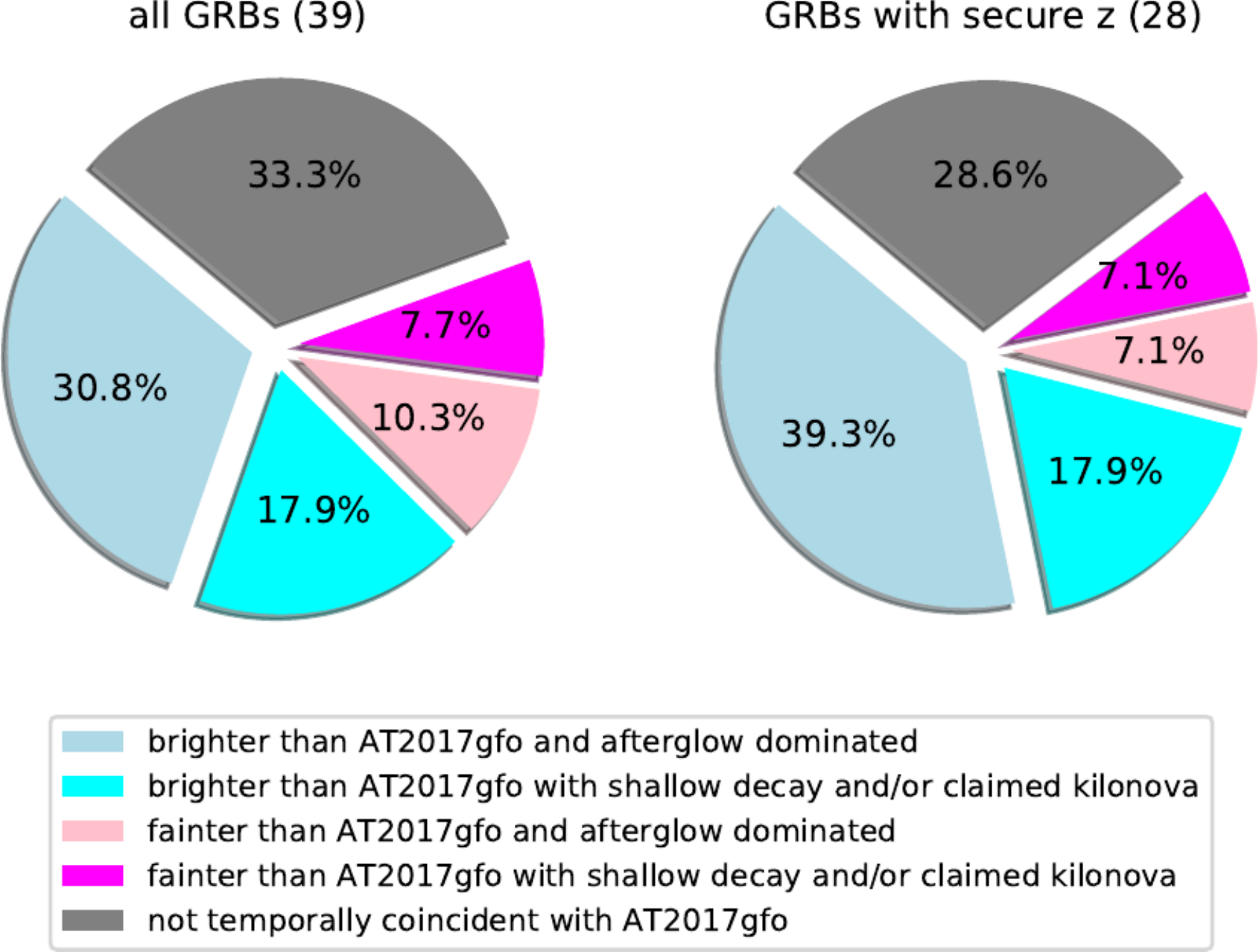}
\caption{Pie-chart summary of analysed short GRBs with optical counterpart brighter, or fainter in at least one filter, than AT2017gfo, and with claimed evidence of kilonovae and/or evidence of shallow decay. We also show all short GRBs with no evidence of kilonova or shallow decay, that we call \emph{afterglow dominated}, and those with no data in the temporal window where AT2017gfo was sampled, that we call \emph{not temporally coincident with AT2017gfo}.}
\label{fig:pie}
\end{center}
\end{figure*}

\section{Summary and conclusions}
\label{sec:con}

{\color{red}
}

 The discovery of GW170817 and GRB 170817A has provided the first direct evidence of the association of at least a fraction of short GRBs with binary NS merging systems. It also provided the most compelling evidence that kilonova emission may be an additional component in sGRB optical/NIR afterglows. 
Motivated by this discovery, we have searched for AT2017gfo-like kilonova emissions in the optical/NIR light curves of 39 short GRBs with known redshift, using optical and NIR rest-frame light curves obtained from the spectroscopic and photometric data-set of AT2017gfo. 

In addition to past works, due to the large spectral coverage of our data sample, we were able to confirm the presence of a significant kilonova luminosity gradient for both the blue and red components. Our main results and conclusions are summarized in Figure \ref{fig:pie} and below: 
\begin{itemize}

\item 
{\it We find robust evidence that not all short GRBs are associated with a AT2017gfo-like kilonova.} {\bf Indeed, we find 7 events in which the GRB optical counterpart is less luminous than AT2017gfo in at least one filter (pink slices in Fig. \ref{fig:pie}). For these cases, if an AT2017gfo-like kilonova were present, it should had been detected. In particular for two GRBs with accurate redshift} 
(050509B and 061201), the optical counterpart luminosity is fainter than AT2017gfo 
by a factor up to $>5$ for the blue component (see Figure \ref{fig:ratiotime} and Tab.2). 

\item 
{\it We find evidence for a significant kilonova luminosity gradient for the blue component. } 
In 7\% of the cases with well-defined redshift (GRBs 050709 and 160821B), the kilonova luminosity is fainter than AT2017gfo (violet slice in Fig.~\ref{fig:pie}) while 18\% is brighter (bright cyan slice), providing evidence for a luminosity range
of $\sim[0.6-17]$ times the  AT2017gfo luminosity for the blue component
{\bf and more than 200 times in the case of the claimed magnetar-powered kilonova of GRB 070714B.}
(see Fig. \ref{fig:ratiotime} and Tab.2). 
These percentages become 8\% (fainter) and 20\% (brighter) if we consider all the sample (i.e., also those GRBs with not well-defined redshift).
As noted by others, a different observer angle is not sufficient to explain the measured luminosity range 
\citep[e.g.,][]{Gompertz2018a}, and the central remnant can play a role
\citep[e.g.,][]{Metzger2018a,Ascenzi2019a}. In particular, it is possible that if a magnetar forms after the merger even for a short time, it can inject energy in the blue kilonova emission \citep[e.g.,][]{Gao2017a}. 

\item 
{\bf 
{\it We find evidence for a similar kilonova luminosity for all kilonovae detected in the NIR (the red component).} 
In three cases (GRBs 050709, 130603B, 160821B) the kilonova is detected in the NIR  after $\sim$2 days, and it is less than a factor $\sim[0.5-3]$ times the AT2017gfo luminosity. Although the numbers are small, this suggests that the red component is similar in luminosity to AT2017gfo.
}


\end{itemize}
{\bf By taking into account a conservative range of blue luminosity for the kilonova}, we estimate up to which redshift the kilonova peak brightness can be detected with current and future facilities. We find that for example with the ELT and JWST we will be able to follow-up a kilonova with redshift $z\sim1-2$ 
(Fig. \ref{fig:gwz}).
{\bf
The precise sky localization for the kilonova follow-up will be provided by the associated GRB and afterglow that will be detected by future space-based GRB dedicated missions as for example THESEUS.}


\section*{Acknowledgements}

The authors thank the anonymous referee for a very constructive report.
We thank D. Radice for useful discussion about the kilonova models,
and G. Raimondo for valuable suggestions.
A. Rossi acknowledges support from Premiale LBT 2013. 
We acknowledge support from the PRIN-INAF SKA-CTA 2016 project
``Toward the SKA and CTA era: discovery, localization, and physics of transient sources''.
 This work made use of the Weizmann interactive supernova data repository - http://wiserep.weizmann.ac.il.
This research has made use of the NASA/IPAC Infrared Science Archive, which is operated by the Jet Propulsion Laboratory, California Institute of Technology, under contract with the National Aeronautics and Space Administration.




\bibliographystyle{mnras}
\bibliography{biblio} 

\begin{thebibliography}{}
\makeatletter
\relax
\def\mn@urlcharsother{\let\do\@makeother \do\$\do\&\do\#\do\^\do\_\do\%\do\~}
\def\mn@doi{\begingroup\mn@urlcharsother \@ifnextchar [ {\mn@doi@}
  {\mn@doi@[]}}
\def\mn@doi@[#1]#2{\def\@tempa{#1}\ifx\@tempa\@empty \href
  {http://dx.doi.org/#2} {doi:#2}\else \href {http://dx.doi.org/#2} {#1}\fi
  \endgroup}
\def\mn@eprint#1#2{\mn@eprint@#1:#2::\@nil}
\def\mn@eprint@arXiv#1{\href {http://arxiv.org/abs/#1} {{\tt arXiv:#1}}}
\def\mn@eprint@dblp#1{\href {http://dblp.uni-trier.de/rec/bibtex/#1.xml}
  {dblp:#1}}
\def\mn@eprint@#1:#2:#3:#4\@nil{\def\@tempa {#1}\def\@tempb {#2}\def\@tempc
  {#3}\ifx \@tempc \@empty \let \@tempc \@tempb \let \@tempb \@tempa \fi \ifx
  \@tempb \@empty \def\@tempb {arXiv}\fi \@ifundefined
  {mn@eprint@\@tempb}{\@tempb:\@tempc}{\expandafter \expandafter \csname
  mn@eprint@\@tempb\endcsname \expandafter{\@tempc}}}

\bibitem[\protect\citeauthoryear{{Abbott} et~al.,}{{Abbott}
  et~al.}{2017a}]{Abbott2017a}
{Abbott} B.~P.,  et~al., 2017a, \mn@doi [Physical Review Letters]
  {10.1103/PhysRevLett.119.161101}, \href
  {http://adsabs.harvard.edu/abs/2017PhRvL.119p1101A} {119, 161101}

\bibitem[\protect\citeauthoryear{{Abbott} et~al.,}{{Abbott}
  et~al.}{2017b}]{Abbott2017b}
{Abbott} B.~P.,  et~al., 2017b, \mn@doi [\apjl] {10.3847/2041-8213/aa91c9},
  \href {http://adsabs.harvard.edu/abs/2017ApJ...848L..12A} {848, L12}

\bibitem[\protect\citeauthoryear{{Abbott} et~al.,}{{Abbott}
  et~al.}{2017c}]{Abbott2017c}
{Abbott} B.~P.,  et~al., 2017c, \mn@doi [\apjl] {10.3847/2041-8213/aa920c},
  \href {http://adsabs.harvard.edu/abs/2017ApJ...848L..13A} {848, L13}

\bibitem[\protect\citeauthoryear{{Acernese} et~al.,}{{Acernese}
  et~al.}{2015}]{Virgo2015a}
{Acernese} F.,  et~al., 2015, \mn@doi [Classical and Quantum Gravity]
  {10.1088/0264-9381/32/2/024001}, \href
  {http://adsabs.harvard.edu/abs/2015CQGra..32b4001A} {32, 024001}

\bibitem[\protect\citeauthoryear{{Amati} et~al.,}{{Amati}
  et~al.}{2002}]{Amati2002a}
{Amati} L.,  et~al., 2002, \mn@doi [\aap] {10.1051/0004-6361:20020722}, \href
  {http://adsabs.harvard.edu/abs/2002A%26A...390...81A} {390, 81}

\bibitem[\protect\citeauthoryear{{Amati} et~al.,}{{Amati}
  et~al.}{2018}]{Amati2018a}
{Amati} L.,  et~al., 2018, \mn@doi [Advances in Space Research]
  {10.1016/j.asr.2018.03.010}, \href
  {https://ui.adsabs.harvard.edu/abs/2018AdSpR..62..191A} {62, 191}

\bibitem[\protect\citeauthoryear{{Andreoni} et~al.,}{{Andreoni}
  et~al.}{2017}]{Andreoni2017a}
{Andreoni} I.,  et~al., 2017, \mn@doi [\pasa] {10.1017/pasa.2017.65}, \href
  {http://adsabs.harvard.edu/abs/2017PASA...34...69A} {34, e069}

\bibitem[\protect\citeauthoryear{{Antonelli} et~al.,}{{Antonelli}
  et~al.}{2009}]{Antonelli2009a}
{Antonelli} L.~A.,  et~al., 2009, \mn@doi [\aap] {10.1051/0004-6361/200913062},
  \href {http://adsabs.harvard.edu/abs/2009A%26A...507L..45A} {507, L45}

\bibitem[\protect\citeauthoryear{{Arcavi}}{{Arcavi}}{2018}]{Arcavi2018a}
{Arcavi} I.,  2018, \mn@doi [\apjl] {10.3847/2041-8213/aab267}, \href
  {http://adsabs.harvard.edu/abs/2018ApJ...855L..23A} {855, L23}

\bibitem[\protect\citeauthoryear{{Arcavi} et~al.,}{{Arcavi}
  et~al.}{2017}]{Arcavi2017a}
{Arcavi} I.,  et~al., 2017, \mn@doi [\nat] {10.1038/nature24291}, \href
  {http://adsabs.harvard.edu/abs/2017Natur.551...64A} {551, 64}

\bibitem[\protect\citeauthoryear{{Ascenzi} et~al.,}{{Ascenzi}
  et~al.}{2019}]{Ascenzi2019a}
{Ascenzi} S.,  et~al., 2019, \mn@doi [\mnras] {10.1093/mnras/stz891}, \href
  {https://ui.adsabs.harvard.edu/abs/2019MNRAS.486..672A} {486, 672}

\bibitem[\protect\citeauthoryear{{Barbieri}, {Salafia}, {Perego}, {Colpi}  \&
  {Ghirlanda}}{{Barbieri} et~al.}{2019}]{Barbieri2019a}
{Barbieri} C.,  {Salafia} O.~S.,  {Perego} A.,  {Colpi} M.,   {Ghirlanda} G.,
  2019, \mn@doi [\aap] {10.1051/0004-6361/201935443}, \href
  {https://ui.adsabs.harvard.edu/abs/2019A&A...625A.152B} {625, A152}

\bibitem[\protect\citeauthoryear{{Barnes}, {Kasen}, {Wu}  \&
  {Mart{\'{\i}}nez-Pinedo}}{{Barnes} et~al.}{2016}]{Barnes2016a}
{Barnes} J.,  {Kasen} D.,  {Wu} M.-R.,   {Mart{\'{\i}}nez-Pinedo} G.,  2016,
  \mn@doi [\apj] {10.3847/0004-637X/829/2/110}, \href
  {http://adsabs.harvard.edu/abs/2016ApJ...829..110B} {829, 110}

\bibitem[\protect\citeauthoryear{{Berger}}{{Berger}}{2010}]{Berger2010a}
{Berger} E.,  2010, \mn@doi [\apj] {10.1088/0004-637X/722/2/1946}, \href
  {https://ui.adsabs.harvard.edu/abs/2010ApJ...722.1946B} {722, 1946}

\bibitem[\protect\citeauthoryear{{Berger} et~al.,}{{Berger}
  et~al.}{2007}]{Berger2007a}
{Berger} E.,  et~al., 2007, \mn@doi [\apj] {10.1086/518762}, \href
  {http://adsabs.harvard.edu/abs/2007ApJ...664.1000B} {664, 1000}

\bibitem[\protect\citeauthoryear{{Berger} et~al.,}{{Berger}
  et~al.}{2013a}]{Berger2013a}
{Berger} E.,  et~al., 2013a, \mn@doi [\apj] {10.1088/0004-637X/765/2/121},
  \href {https://ui.adsabs.harvard.edu/abs/2013ApJ...765..121B} {765, 121}

\bibitem[\protect\citeauthoryear{{Berger}, {Fong}  \& {Chornock}}{{Berger}
  et~al.}{2013b}]{Berger2013b}
{Berger} E.,  {Fong} W.,   {Chornock} R.,  2013b, \mn@doi [\apjl]
  {10.1088/2041-8205/774/2/L23}, \href
  {https://ui.adsabs.harvard.edu/abs/2013ApJ...774L..23B} {774, L23}

\bibitem[\protect\citeauthoryear{{Bernardini} et~al.,}{{Bernardini}
  et~al.}{2015}]{Bernardini2015a}
{Bernardini} M.~G.,  et~al., 2015, \mn@doi [\mnras] {10.1093/mnras/stu2153},
  \href {http://adsabs.harvard.edu/abs/2015MNRAS.446.1129B} {446, 1129}

\bibitem[\protect\citeauthoryear{{Blanton} \& {Roweis}}{{Blanton} \&
  {Roweis}}{2007}]{Blanton2007a}
{Blanton} M.~R.,  {Roweis} S.,  2007, \mn@doi [\aj] {10.1086/510127}, \href
  {http://adsabs.harvard.edu/abs/2007AJ....133..734B} {133, 734}

\bibitem[\protect\citeauthoryear{{Bloom}, {Kulkarni}  \& {Djorgovski}}{{Bloom}
  et~al.}{2002}]{Bloom2002a}
{Bloom} J.~S.,  {Kulkarni} S.~R.,   {Djorgovski} S.~G.,  2002, \mn@doi [\aj]
  {10.1086/338893}, \href
  {https://ui.adsabs.harvard.edu/abs/2002AJ....123.1111B} {123, 1111}

\bibitem[\protect\citeauthoryear{{Bloom} et~al.,}{{Bloom}
  et~al.}{2006}]{Bloom2006a}
{Bloom} J.~S.,  et~al., 2006, \mn@doi [\apj] {10.1086/498107}, \href
  {http://adsabs.harvard.edu/abs/2006ApJ...638..354B} {638, 354}

\bibitem[\protect\citeauthoryear{{Brocato} et~al.,}{{Brocato}
  et~al.}{2018}]{Brocato2018a}
{Brocato} E.,  et~al., 2018, \mn@doi [\mnras] {10.1093/mnras/stx2730}, \href
  {http://adsabs.harvard.edu/abs/2018MNRAS.474..411B} {474, 411}

\bibitem[\protect\citeauthoryear{{Bromberg}, {Nakar}, {Piran}  \&
  {Sari}}{{Bromberg} et~al.}{2012}]{Bromberg2012a}
{Bromberg} O.,  {Nakar} E.,  {Piran} T.,   {Sari} R.,  2012, \mn@doi [\apj]
  {10.1088/0004-637X/749/2/110}, \href
  {http://adsabs.harvard.edu/abs/2012ApJ...749..110B} {749, 110}

\bibitem[\protect\citeauthoryear{{Bulla} et~al.,}{{Bulla}
  et~al.}{2018}]{Bulla2018a}
{Bulla} M.,  et~al., 2018, \mn@doi [Nature Astronomy]
  {10.1038/s41550-018-0593-y}, \href
  {http://adsabs.harvard.edu/abs/2018NatAs.tmp..164B} {}

\bibitem[\protect\citeauthoryear{{Cantiello} et~al.,}{{Cantiello}
  et~al.}{2018}]{Cantiello2018a}
{Cantiello} M.,  et~al., 2018, \mn@doi [\apjl] {10.3847/2041-8213/aaad64},
  \href {http://adsabs.harvard.edu/abs/2018ApJ...854L..31C} {854, L31}

\bibitem[\protect\citeauthoryear{{Cardelli}, {Clayton}  \& {Mathis}}{{Cardelli}
  et~al.}{1989}]{Cardelli1989}
{Cardelli} J.~A.,  {Clayton} G.~C.,   {Mathis} J.~S.,  1989, \mn@doi [\apj]
  {10.1086/167900}, \href {http://adsabs.harvard.edu/abs/1989ApJ...345..245C}
  {345, 245}

\bibitem[\protect\citeauthoryear{{Castro-Tirado} et~al.,}{{Castro-Tirado}
  et~al.}{2005}]{CastroTirado2005a}
{Castro-Tirado} A.~J.,  et~al., 2005, \mn@doi [\aap]
  {10.1051/0004-6361:200500147}, \href
  {http://adsabs.harvard.edu/abs/2005A%26A...439L..15C} {439, L15}

\bibitem[\protect\citeauthoryear{{Castro-Tirado}, {Sanchez-Ramirez}, {Lombardi}
   \& {Rivero}}{{Castro-Tirado} et~al.}{2015}]{Castro-Tirado2015a}
{Castro-Tirado} A.~J.,  {Sanchez-Ramirez} R.,  {Lombardi} G.,   {Rivero} M.~A.,
   2015, GRB Coordinates Network, Circular Service, No.~17758, \#1 (2015),
  \href {http://adsabs.harvard.edu/abs/2015GCN.17758....1C} {17758}

\bibitem[\protect\citeauthoryear{{Chan}, {Messenger}, {Heng}  \&
  {Hendry}}{{Chan} et~al.}{2018}]{Chan2018a}
{Chan} M.~L.,  {Messenger} C.,  {Heng} I.~S.,   {Hendry} M.,  2018, \mn@doi
  [\prd] {10.1103/PhysRevD.97.123014}, \href
  {https://ui.adsabs.harvard.edu/abs/2018PhRvD..97l3014C} {97, 123014}

\bibitem[\protect\citeauthoryear{{Chornock} \& {Fong}}{{Chornock} \&
  {Fong}}{2015}]{Chornock2015a}
{Chornock} R.,  {Fong} W.,  2015, GRB Coordinates Network, \href
  {https://ui.adsabs.harvard.edu/abs/2015GCN.17358....1C} {17358, 1}

\bibitem[\protect\citeauthoryear{{Chornock}, {Lunnan}  \& {Berger}}{{Chornock}
  et~al.}{2013}]{Chornock2013a}
{Chornock} R.,  {Lunnan} R.,   {Berger} E.,  2013, GRB Coordinates Network,
  15307, 1

\bibitem[\protect\citeauthoryear{{Chornock}, {Fong}  \& {Fox}}{{Chornock}
  et~al.}{2014}]{Chornock2014a}
{Chornock} R.,  {Fong} W.,   {Fox} D.~B.,  2014, GRB Coordinates Network, \href
  {https://ui.adsabs.harvard.edu/abs/2014GCN.17177....1C} {17177, 1}

\bibitem[\protect\citeauthoryear{{Chornock} et~al.,}{{Chornock}
  et~al.}{2017}]{Chornock2017a}
{Chornock} R.,  et~al., 2017, \mn@doi [\apjl] {10.3847/2041-8213/aa905c}, \href
  {http://adsabs.harvard.edu/abs/2017ApJ...848L..19C} {848, L19}

\bibitem[\protect\citeauthoryear{{Coulter} et~al.,}{{Coulter}
  et~al.}{2017}]{Coulter2017a}
{Coulter} D.~A.,  et~al., 2017, \mn@doi [Science] {10.1126/science.aap9811},
  \href {http://adsabs.harvard.edu/abs/2017Sci...358.1556C} {358, 1556}

\bibitem[\protect\citeauthoryear{{Covino} et~al.,}{{Covino}
  et~al.}{2017}]{Covino2017a}
{Covino} S.,  et~al., 2017, \mn@doi [Nature Astronomy]
  {10.1038/s41550-017-0285-z}, \href
  {http://adsabs.harvard.edu/abs/2017NatAs...1..791C} {1, 791}

\bibitem[\protect\citeauthoryear{{Cowperthwaite} et~al.,}{{Cowperthwaite}
  et~al.}{2017}]{Cowperthwaite2017a}
{Cowperthwaite} P.~S.,  et~al., 2017, \mn@doi [\apjl]
  {10.3847/2041-8213/aa8fc7}, \href
  {http://adsabs.harvard.edu/abs/2017ApJ...848L..17C} {848, L17}

\bibitem[\protect\citeauthoryear{{Cucchiara} \& {Levan}}{{Cucchiara} \&
  {Levan}}{2016}]{Cucchiara2016a}
{Cucchiara} A.,  {Levan} A.~J.,  2016, GRB Coordinates Network, Circular
  Service, No.~19565, \#1 (2016), \href
  {http://adsabs.harvard.edu/abs/2016GCN.19565....1C} {19565}

\bibitem[\protect\citeauthoryear{{D'Avanzo} et~al.,}{{D'Avanzo}
  et~al.}{2009}]{Davanzo2009a}
{D'Avanzo} P.,  et~al., 2009, \mn@doi [\aap] {10.1051/0004-6361/200811294},
  \href {http://adsabs.harvard.edu/abs/2009A%26A...498..711D} {498, 711}

\bibitem[\protect\citeauthoryear{{D'Avanzo} et~al.,}{{D'Avanzo}
  et~al.}{2014}]{Davanzo2014a}
{D'Avanzo} P.,  et~al., 2014, \mn@doi [\mnras] {10.1093/mnras/stu994}, \href
  {http://adsabs.harvard.edu/abs/2014MNRAS.442.2342D} {442, 2342}

\bibitem[\protect\citeauthoryear{{D'Avanzo} et~al.,}{{D'Avanzo}
  et~al.}{2018}]{Davanzo2018a}
{D'Avanzo} P.,  et~al., 2018, \mn@doi [\aap] {10.1051/0004-6361/201832664},
  \href {http://adsabs.harvard.edu/abs/2018A%26A...613L...1D} {613, L1}

\bibitem[\protect\citeauthoryear{{D'Elia}, {D'Avanzo}, {Malesani}, {di
  Fabrizio}  \& {Tessicini}}{{D'Elia} et~al.}{2013}]{Delia2013a}
{D'Elia} V.,  {D'Avanzo} P.,  {Malesani} D.,  {di Fabrizio} L.,   {Tessicini}
  G.,  2013, GRB Coordinates Network, 15310, 1

\bibitem[\protect\citeauthoryear{{Della Valle} et~al.,}{{Della Valle}
  et~al.}{2006}]{DellaValle2006a}
{Della Valle} M.,  et~al., 2006, \mn@doi [\nat] {10.1038/nature05374}, \href
  {https://ui.adsabs.harvard.edu/abs/2006Natur.444.1050D} {444, 1050}

\bibitem[\protect\citeauthoryear{{Dichiara}, {Troja}, {O'Connor}, {Marshall},
  {Beniamini}, {Cannizzo}, {Lien}  \& {Sakamoto}}{{Dichiara}
  et~al.}{2019}]{Dichiara2019a}
{Dichiara} S.,  {Troja} E.,  {O'Connor} B.,  {Marshall} F.~E.,  {Beniamini} P.,
   {Cannizzo} J.~K.,  {Lien} A.~Y.,   {Sakamoto} T.,  2019, arXiv e-prints,
  \href {https://ui.adsabs.harvard.edu/abs/2019arXiv191208698D} {p.
  arXiv:1912.08698}

\bibitem[\protect\citeauthoryear{{Drout} et~al.,}{{Drout}
  et~al.}{2017}]{Drout2017a}
{Drout} M.~R.,  et~al., 2017, \mn@doi [Science] {10.1126/science.aaq0049},
  \href {http://adsabs.harvard.edu/abs/2017Sci...358.1570D} {358, 1570}

\bibitem[\protect\citeauthoryear{{Evans} et~al.,}{{Evans}
  et~al.}{2017}]{Evans2017a}
{Evans} P.~A.,  et~al., 2017, \mn@doi [Science] {10.1126/science.aap9580},
  \href {http://adsabs.harvard.edu/abs/2017Sci...358.1565E} {358, 1565}

\bibitem[\protect\citeauthoryear{{Fern{\'a}ndez} \& {Metzger}}{{Fern{\'a}ndez}
  \& {Metzger}}{2016}]{Fernandez2016a}
{Fern{\'a}ndez} R.,  {Metzger} B.~D.,  2016, \mn@doi [Annual Review of Nuclear
  and Particle Science] {10.1146/annurev-nucl-102115-044819}, \href
  {http://adsabs.harvard.edu/abs/2016ARNPS..66...23F} {66, 23}

\bibitem[\protect\citeauthoryear{{Fong} \& {Berger}}{{Fong} \&
  {Berger}}{2013}]{Fong2013b}
{Fong} W.,  {Berger} E.,  2013, \mn@doi [\apj] {10.1088/0004-637X/776/1/18},
  \href {https://ui.adsabs.harvard.edu/abs/2013ApJ...776...18F} {776, 18}

\bibitem[\protect\citeauthoryear{{Fong} et~al.,}{{Fong}
  et~al.}{2013}]{Fong2013a}
{Fong} W.,  et~al., 2013, \mn@doi [\apj] {10.1088/0004-637X/769/1/56}, \href
  {https://ui.adsabs.harvard.edu/abs/2013ApJ...769...56F} {769, 56}

\bibitem[\protect\citeauthoryear{{Fong}, {Berger}, {Margutti}  \&
  {Zauderer}}{{Fong} et~al.}{2015}]{Fong2015a}
{Fong} W.,  {Berger} E.,  {Margutti} R.,   {Zauderer} B.~A.,  2015, \mn@doi
  [\apj] {10.1088/0004-637X/815/2/102}, \href
  {http://adsabs.harvard.edu/abs/2015ApJ...815..102F} {815, 102}

\bibitem[\protect\citeauthoryear{{Fong} et~al.,}{{Fong}
  et~al.}{2016}]{Fong2016a}
{Fong} W.,  et~al., 2016, \mn@doi [\apj] {10.3847/1538-4357/833/2/151}, \href
  {http://adsabs.harvard.edu/abs/2016ApJ...833..151F} {833, 151}

\bibitem[\protect\citeauthoryear{{Fong} et~al.,}{{Fong}
  et~al.}{2017}]{Fong2017a}
{Fong} W.,  et~al., 2017, \mn@doi [\apjl] {10.3847/2041-8213/aa9018}, \href
  {http://adsabs.harvard.edu/abs/2017ApJ...848L..23F} {848, L23}

\bibitem[\protect\citeauthoryear{{Fox} et~al.,}{{Fox} et~al.}{2005}]{Fox2005a}
{Fox} D.~B.,  et~al., 2005, \mn@doi [\nat] {10.1038/nature04189}, \href
  {http://adsabs.harvard.edu/abs/2005Natur.437..845F} {437, 845}

\bibitem[\protect\citeauthoryear{{Fynbo} et~al.,}{{Fynbo}
  et~al.}{2006}]{Fynbo2006a}
{Fynbo} J. P.~U.,  et~al., 2006, \mn@doi [\nat] {10.1038/nature05375}, \href
  {https://ui.adsabs.harvard.edu/abs/2006Natur.444.1047F} {444, 1047}

\bibitem[\protect\citeauthoryear{{Gao}, {Zhang}, {L{\"u}}  \& {Li}}{{Gao}
  et~al.}{2017}]{Gao2017a}
{Gao} H.,  {Zhang} B.,  {L{\"u}} H.-J.,   {Li} Y.,  2017, \mn@doi [\apj]
  {10.3847/1538-4357/aa5be3}, \href
  {http://adsabs.harvard.edu/abs/2017ApJ...837...50G} {837, 50}

\bibitem[\protect\citeauthoryear{Gardner et~al.,}{Gardner
  et~al.}{2006}]{Gardner2006}
Gardner J.~P.,  et~al., 2006, \mn@doi [Space Science Reviews]
  {10.1007/s11214-006-8315-7}, 123, 485

\bibitem[\protect\citeauthoryear{{Ghirlanda} et~al.,}{{Ghirlanda}
  et~al.}{2019}]{Ghirlanda2019a}
{Ghirlanda} G.,  et~al., 2019, \mn@doi [Science] {10.1126/science.aau8815},
  \href {https://ui.adsabs.harvard.edu/abs/2019Sci...363..968G} {363, 968}

\bibitem[\protect\citeauthoryear{{Goldstein} et~al.,}{{Goldstein}
  et~al.}{2017}]{Goldstein2017a}
{Goldstein} A.,  et~al., 2017, \mn@doi [\apjl] {10.3847/2041-8213/aa8f41},
  \href {http://adsabs.harvard.edu/abs/2017ApJ...848L..14G} {848, L14}

\bibitem[\protect\citeauthoryear{{Gompertz} et~al.,}{{Gompertz}
  et~al.}{2018}]{Gompertz2018a}
{Gompertz} B.~P.,  et~al., 2018, \mn@doi [\apj] {10.3847/1538-4357/aac206},
  \href {http://adsabs.harvard.edu/abs/2018ApJ...860...62G} {860, 62}

\bibitem[\protect\citeauthoryear{{Guillochon}, {Parrent}, {Kelley}  \&
  {Margutti}}{{Guillochon} et~al.}{2017}]{Guillochon2017a}
{Guillochon} J.,  {Parrent} J.,  {Kelley} L.~Z.,   {Margutti} R.,  2017,
  \mn@doi [\apj] {10.3847/1538-4357/835/1/64}, \href
  {http://adsabs.harvard.edu/abs/2017ApJ...835...64G} {835, 64}

\bibitem[\protect\citeauthoryear{{Hallinan} et~al.,}{{Hallinan}
  et~al.}{2017}]{Hallinan2017a}
{Hallinan} G.,  et~al., 2017, \mn@doi [Science] {10.1126/science.aap9855},
  \href {https://ui.adsabs.harvard.edu/abs/2017Sci...358.1579H} {358, 1579}

\bibitem[\protect\citeauthoryear{{Hjorth} et~al.,}{{Hjorth}
  et~al.}{2003}]{Hjorth2003b}
{Hjorth} J.,  et~al., 2003, \nat, \href
  {http://adsabs.harvard.edu/abs/2003Natur.423..847H} {423, 847}

\bibitem[\protect\citeauthoryear{{Hjorth} et~al.,}{{Hjorth}
  et~al.}{2005}]{Hjorth2005a}
{Hjorth} J.,  et~al., 2005, \mn@doi [\apjl] {10.1086/491733}, \href
  {http://adsabs.harvard.edu/abs/2005ApJ...630L.117H} {630, L117}

\bibitem[\protect\citeauthoryear{{Hjorth} et~al.,}{{Hjorth}
  et~al.}{2017}]{Hjorth2017a}
{Hjorth} J.,  et~al., 2017, \mn@doi [\apjl] {10.3847/2041-8213/aa9110}, \href
  {http://adsabs.harvard.edu/abs/2017ApJ...848L..31H} {848, L31}

\bibitem[\protect\citeauthoryear{{Ivezi{\'c}} et~al.,}{{Ivezi{\'c}}
  et~al.}{2019}]{LSST2019}
{Ivezi{\'c}} {\v{Z}}.,  et~al., 2019, \mn@doi [\apj]
  {10.3847/1538-4357/ab042c}, \href
  {https://ui.adsabs.harvard.edu/abs/2019ApJ...873..111I} {873, 111}

\bibitem[\protect\citeauthoryear{{Izzo}, {Cano}, {de Ugarte Postigo}, {Kann},
  {Thoene}  \& {Geier}}{{Izzo} et~al.}{2017}]{Izzo2017a}
{Izzo} L.,  {Cano} Z.,  {de Ugarte Postigo} A.,  {Kann} D.~A.,  {Thoene} C.,
  {Geier} S.,  2017, GRB Coordinates Network, Circular Service, 21059

\bibitem[\protect\citeauthoryear{{Jin}, {Li}, {Cano}, {Covino}, {Fan}  \&
  {Wei}}{{Jin} et~al.}{2015}]{Jin2015a}
{Jin} Z.-P.,  {Li} X.,  {Cano} Z.,  {Covino} S.,  {Fan} Y.-Z.,   {Wei} D.-M.,
  2015, \mn@doi [\apjl] {10.1088/2041-8205/811/2/L22}, \href
  {http://adsabs.harvard.edu/abs/2015ApJ...811L..22J} {811, L22}

\bibitem[\protect\citeauthoryear{{Jin} et~al.,}{{Jin} et~al.}{2016}]{Jin2016a}
{Jin} Z.-P.,  et~al., 2016, \mn@doi [Nature Communications]
  {10.1038/ncomms12898}, \href
  {http://adsabs.harvard.edu/abs/2016NatCo...712898J} {7, 12898}

\bibitem[\protect\citeauthoryear{{Jin} et~al.,}{{Jin} et~al.}{2018}]{Jin2018a}
{Jin} Z.-P.,  et~al., 2018, \mn@doi [\apj] {10.3847/1538-4357/aab76d}, \href
  {http://adsabs.harvard.edu/abs/2018ApJ...857..128J} {857, 128}

\bibitem[\protect\citeauthoryear{{Jin}, {Covino}, {Liao}, {Li}, {D'Avanzo},
  {Fan}  \& {Wei}}{{Jin} et~al.}{2019}]{Jin2019a}
{Jin} Z.-P.,  {Covino} S.,  {Liao} N.-H.,  {Li} X.,  {D'Avanzo} P.,  {Fan}
  Y.-Z.,   {Wei} D.-M.,  2019, \mn@doi [Nature Astronomy]
  {10.1038/s41550-019-0892-y}, \href
  {https://ui.adsabs.harvard.edu/abs/2019NatAs.tmp..461J} {p.~461}

\bibitem[\protect\citeauthoryear{{Kann} et~al.,}{{Kann}
  et~al.}{2011}]{Kann2011a}
{Kann} D.~A.,  et~al., 2011, \mn@doi [\apj] {10.1088/0004-637X/734/2/96}, \href
  {http://adsabs.harvard.edu/abs/2011ApJ...734...96K} {734, 96}

\bibitem[\protect\citeauthoryear{{Kasen}, {Fern{\'a}ndez}  \&
  {Metzger}}{{Kasen} et~al.}{2015}]{Kasen2015a}
{Kasen} D.,  {Fern{\'a}ndez} R.,   {Metzger} B.~D.,  2015, \mn@doi [\mnras]
  {10.1093/mnras/stv721}, \href
  {http://adsabs.harvard.edu/abs/2015MNRAS.450.1777K} {450, 1777}

\bibitem[\protect\citeauthoryear{{Kasen}, {Metzger}, {Barnes}, {Quataert}  \&
  {Ramirez-Ruiz}}{{Kasen} et~al.}{2017}]{Kasen2017a}
{Kasen} D.,  {Metzger} B.,  {Barnes} J.,  {Quataert} E.,   {Ramirez-Ruiz} E.,
  2017, \mn@doi [\nat] {10.1038/nature24453}, \href
  {http://adsabs.harvard.edu/abs/2017Natur.551...80K} {551, 80}

\bibitem[\protect\citeauthoryear{{Kasliwal} et~al.,}{{Kasliwal}
  et~al.}{2017a}]{Kasliwal2017b}
{Kasliwal} M.~M.,  et~al., 2017a, \mn@doi [Science] {10.1126/science.aap9455},
  \href {http://adsabs.harvard.edu/abs/2017Sci...358.1559K} {358, 1559}

\bibitem[\protect\citeauthoryear{{Kasliwal}, {Korobkin}, {Lau}, {Wollaeger}  \&
  {Fryer}}{{Kasliwal} et~al.}{2017b}]{Kasliwal2017a}
{Kasliwal} M.~M.,  {Korobkin} O.,  {Lau} R.~M.,  {Wollaeger} R.,   {Fryer}
  C.~L.,  2017b, \mn@doi [\apjl] {10.3847/2041-8213/aa799d}, \href
  {http://adsabs.harvard.edu/abs/2017ApJ...843L..34K} {843, L34}

\bibitem[\protect\citeauthoryear{{Kathirgamaraju}, {Barniol Duran}  \&
  {Giannios}}{{Kathirgamaraju} et~al.}{2018}]{Kathirgamaraju2018a}
{Kathirgamaraju} A.,  {Barniol Duran} R.,   {Giannios} D.,  2018, \mn@doi
  [\mnras] {10.1093/mnrasl/slx175}, \href
  {https://ui.adsabs.harvard.edu/\#abs/2018MNRAS.473L.121K} {473, L121}

\bibitem[\protect\citeauthoryear{{Kathirgamaraju}, {Tchekhovskoy}, {Giannios}
  \& {Barniol Duran}}{{Kathirgamaraju} et~al.}{2019}]{Kathirgamaraju2018b}
{Kathirgamaraju} A.,  {Tchekhovskoy} A.,  {Giannios} D.,   {Barniol Duran} R.,
  2019, \mn@doi [\mnras] {10.1093/mnrasl/slz012}, \href
  {https://ui.adsabs.harvard.edu/abs/2019MNRAS.484L..98K} {484, L98}

\bibitem[\protect\citeauthoryear{{Knust} et~al.,}{{Knust}
  et~al.}{2017}]{Knust2017a}
{Knust} F.,  et~al., 2017, \mn@doi [\aap] {10.1051/0004-6361/201730578}, \href
  {http://adsabs.harvard.edu/abs/2017A%26A...607A..84K} {607, A84}

\bibitem[\protect\citeauthoryear{{Kocevski} et~al.,}{{Kocevski}
  et~al.}{2010}]{Kocevski2010a}
{Kocevski} D.,  et~al., 2010, \mn@doi [\mnras]
  {10.1111/j.1365-2966.2010.16327.x}, \href
  {http://adsabs.harvard.edu/abs/2010MNRAS.404..963K} {404, 963}

\bibitem[\protect\citeauthoryear{{Kouveliotou}, {Meegan}, {Fishman}, {Bhat},
  {Briggs}, {Koshut}, {Paciesas}  \& {Pendleton}}{{Kouveliotou}
  et~al.}{1993}]{Kouveliotou1993a}
{Kouveliotou} C.,  {Meegan} C.~A.,  {Fishman} G.~J.,  {Bhat} N.~P.,  {Briggs}
  M.~S.,  {Koshut} T.~M.,  {Paciesas} W.~S.,   {Pendleton} G.~N.,  1993,
  \mn@doi [\apjl] {10.1086/186969}, \href
  {http://adsabs.harvard.edu/abs/1993ApJ...413L.101K} {413, L101}

\bibitem[\protect\citeauthoryear{{LIGO Scientific Collaboration} et~al.,}{{LIGO
  Scientific Collaboration} et~al.}{2015}]{Ligo2015a}
{LIGO Scientific Collaboration} et~al., 2015, \mn@doi [Classical and Quantum
  Gravity] {10.1088/0264-9381/32/7/074001}, \href
  {http://adsabs.harvard.edu/abs/2015CQGra..32g4001L} {32, 074001}

\bibitem[\protect\citeauthoryear{{Lamb} et~al.,}{{Lamb}
  et~al.}{2019}]{Lamb2019a}
{Lamb} G.~P.,  et~al., 2019, \mn@doi [\apj] {10.3847/1538-4357/ab38bb}, \href
  {https://ui.adsabs.harvard.edu/abs/2019ApJ...883...48L} {883, 48}

\bibitem[\protect\citeauthoryear{{Levan}, {Wiersema}, {Tanvir}, {Malesani},
  {Xu}  \& {de Ugarte Postigo}}{{Levan} et~al.}{2016}]{Levan2016a}
{Levan} A.~J.,  {Wiersema} K.,  {Tanvir} N.~R.,  {Malesani} D.,  {Xu} D.,   {de
  Ugarte Postigo} A.,  2016, GRB Coordinates Network, Circular Service,
  No.~19846, \#1 (2016), \href
  {http://adsabs.harvard.edu/abs/2016GCN.19846....1L} {19846}

\bibitem[\protect\citeauthoryear{{Li} \& {Paczy{\'n}ski}}{{Li} \&
  {Paczy{\'n}ski}}{1998}]{Li1998a}
{Li} L.-X.,  {Paczy{\'n}ski} B.,  1998, \mn@doi [\apjl] {10.1086/311680}, \href
  {http://adsabs.harvard.edu/abs/1998ApJ...507L..59L} {507, L59}

\bibitem[\protect\citeauthoryear{{Lyman} et~al.,}{{Lyman}
  et~al.}{2018}]{Lyman2018a}
{Lyman} J.~D.,  et~al., 2018, \mn@doi [Nature Astronomy]
  {10.1038/s41550-018-0511-3}, \href
  {http://adsabs.harvard.edu/abs/2018NatAs...2..751L} {2, 751}

\bibitem[\protect\citeauthoryear{{Maiorano}, {Amati}, {Rossi}, {Stratta},
  {Palazzi}  \& {Nicastro}}{{Maiorano} et~al.}{2018}]{Maiorano2018a}
{Maiorano} E.,  {Amati} L.,  {Rossi} A.,  {Stratta} G.,  {Palazzi} E.,
  {Nicastro} L.,  2018, \memsai, \href
  {https://ui.adsabs.harvard.edu/abs/2018MmSAI..89..181M} {89, 181}

\bibitem[\protect\citeauthoryear{{Malesani} et~al.,}{{Malesani}
  et~al.}{2007}]{Malesani2007a}
{Malesani} D.,  et~al., 2007, \mn@doi [\aap] {10.1051/0004-6361:20077868},
  \href {http://adsabs.harvard.edu/abs/2007A%26A...473...77M} {473, 77}

\bibitem[\protect\citeauthoryear{{Malesani}, {D'Avanzo}, {D'Elia}, {Vergani},
  {Andreuzzi}, {Garcia}, {Escudero}  \& {Bonomo}}{{Malesani}
  et~al.}{2014}]{Malesani2014a}
{Malesani} D.,  {D'Avanzo} P.,  {D'Elia} V.,  {Vergani} S.~D.,  {Andreuzzi} G.,
   {Garcia} A.,  {Escudero} G.,   {Bonomo} A.,  2014, GRB Coordinates Network,
  \href {https://ui.adsabs.harvard.edu/abs/2014GCN.17170....1M} {17170, 1}

\bibitem[\protect\citeauthoryear{{Malesani} et~al.,}{{Malesani}
  et~al.}{2015}]{Malesani2015a}
{Malesani} D.,  et~al., 2015, GRB Coordinates Network, \href
  {https://ui.adsabs.harvard.edu/abs/2015GCN.17755....1M} {17755, 1}

\bibitem[\protect\citeauthoryear{Mandel}{Mandel}{2018}]{Mandel2018a}
Mandel I.,  2018, \mn@doi [The Astrophysical Journal]
  {10.3847/2041-8213/aaa6c1}, 853, L12

\bibitem[\protect\citeauthoryear{{Mangano} et~al.,}{{Mangano}
  et~al.}{2007}]{Mangano2007a}
{Mangano} V.,  et~al., 2007, \mn@doi [\aap] {10.1051/0004-6361:20077232}, \href
  {https://ui.adsabs.harvard.edu/abs/2007A&A...470..105M} {470, 105}

\bibitem[\protect\citeauthoryear{{Margutti} et~al.,}{{Margutti}
  et~al.}{2012}]{Margutti2012a}
{Margutti} R.,  et~al., 2012, \mn@doi [\apj] {10.1088/0004-637X/756/1/63},
  \href {https://ui.adsabs.harvard.edu/abs/2012ApJ...756...63M} {756, 63}

\bibitem[\protect\citeauthoryear{{Margutti} et~al.,}{{Margutti}
  et~al.}{2018}]{Margutti2018a}
{Margutti} R.,  et~al., 2018, \mn@doi [\apjl] {10.3847/2041-8213/aab2ad}, \href
  {http://adsabs.harvard.edu/abs/2018ApJ...856L..18M} {856, L18}

\bibitem[\protect\citeauthoryear{{Metzger}}{{Metzger}}{2017}]{Metzger2017a}
{Metzger} B.~D.,  2017, \mn@doi [Living Reviews in Relativity]
  {10.1007/s41114-017-0006-z}, \href
  {http://adsabs.harvard.edu/abs/2017LRR....20....3M} {20, 3}

\bibitem[\protect\citeauthoryear{{Metzger} \& {Piro}}{{Metzger} \&
  {Piro}}{2014}]{MetzgerPiro2014a}
{Metzger} B.~D.,  {Piro} A.~L.,  2014, \mn@doi [\mnras] {10.1093/mnras/stu247},
  \href {http://adsabs.harvard.edu/abs/2014MNRAS.439.3916M} {439, 3916}

\bibitem[\protect\citeauthoryear{{Metzger} et~al.,}{{Metzger}
  et~al.}{2010}]{Metzger2010a}
{Metzger} B.~D.,  et~al., 2010, \mn@doi [\mnras]
  {10.1111/j.1365-2966.2010.16864.x}, \href
  {https://ui.adsabs.harvard.edu/abs/2010MNRAS.406.2650M} {406, 2650}

\bibitem[\protect\citeauthoryear{{Metzger}, {Thompson}  \&
  {Quataert}}{{Metzger} et~al.}{2018}]{Metzger2018a}
{Metzger} B.~D.,  {Thompson} T.~A.,   {Quataert} E.,  2018, \mn@doi [\apj]
  {10.3847/1538-4357/aab095}, \href
  {http://adsabs.harvard.edu/abs/2018ApJ...856..101M} {856, 101}

\bibitem[\protect\citeauthoryear{{Mooley} et~al.,}{{Mooley}
  et~al.}{2018a}]{Mooley2018a}
{Mooley} K.~P.,  et~al., 2018a, \mn@doi [\nat] {10.1038/nature25452}, \href
  {http://adsabs.harvard.edu/abs/2018Natur.554..207M} {554, 207}

\bibitem[\protect\citeauthoryear{{Mooley} et~al.,}{{Mooley}
  et~al.}{2018b}]{Mooley2018b}
{Mooley} K.~P.,  et~al., 2018b, \mn@doi [\nat] {10.1038/s41586-018-0486-3},
  \href {http://adsabs.harvard.edu/abs/2018Natur.561..355M} {561, 355}

\bibitem[\protect\citeauthoryear{{Nicuesa Guelbenzu} et~al.,}{{Nicuesa
  Guelbenzu} et~al.}{2011}]{Nicuesa2011a}
{Nicuesa Guelbenzu} A.,  et~al., 2011, \mn@doi [\aap]
  {10.1051/0004-6361/201116657}, \href
  {http://adsabs.harvard.edu/abs/2011A%26A...531L...6N} {531, L6}

\bibitem[\protect\citeauthoryear{{Nicuesa Guelbenzu} et~al.,}{{Nicuesa
  Guelbenzu} et~al.}{2012a}]{Nicuesa2012b}
{Nicuesa Guelbenzu} A.,  et~al., 2012a, \mn@doi [\aap]
  {10.1051/0004-6361/201118416}, \href
  {http://adsabs.harvard.edu/abs/2012A%26A...538L...7N} {538, L7}

\bibitem[\protect\citeauthoryear{{Nicuesa Guelbenzu} et~al.,}{{Nicuesa
  Guelbenzu} et~al.}{2012b}]{Nicuesa2012a}
{Nicuesa Guelbenzu} A.,  et~al., 2012b, \mn@doi [\aap]
  {10.1051/0004-6361/201219551}, \href
  {http://adsabs.harvard.edu/abs/2012A%26A...548A.101N} {548, A101}

\bibitem[\protect\citeauthoryear{{Perego}, {Radice}  \& {Bernuzzi}}{{Perego}
  et~al.}{2017}]{Perego2017a}
{Perego} A.,  {Radice} D.,   {Bernuzzi} S.,  2017, \mn@doi [\apjl]
  {10.3847/2041-8213/aa9ab9}, \href
  {http://adsabs.harvard.edu/abs/2017ApJ...850L..37P} {850, L37}

\bibitem[\protect\citeauthoryear{{P{\'e}rez-Ram{\'{\i}}rez}
  et~al.,}{{P{\'e}rez-Ram{\'{\i}}rez} et~al.}{2013}]{PerezRamirez2013a}
{P{\'e}rez-Ram{\'{\i}}rez} D.,  et~al., 2013, in {Castro-Tirado} A.~J.,
  {Gorosabel} J.,   {Park} I.~H.,  eds,  EAS Publications Series Vol. 61, EAS
  Publications Series. pp 345--349, \mn@doi{10.1051/eas/1361055}

\bibitem[\protect\citeauthoryear{{Perley}}{{Perley}}{2015}]{Perley2015a}
{Perley} D.~A.,  2015, GRB Coordinates Network, Circular Service, No.~17744,
  \#1 (2015), \href {http://adsabs.harvard.edu/abs/2015GCN.17744....1P} {17744}

\bibitem[\protect\citeauthoryear{{Perley} \& {Cenko}}{{Perley} \&
  {Cenko}}{2015}]{Perley2015b}
{Perley} D.~A.,  {Cenko} S.~B.,  2015, GRB Coordinates Network, \href
  {https://ui.adsabs.harvard.edu/abs/2015GCN.17312....1P} {17312, 1}

\bibitem[\protect\citeauthoryear{{Perley} et~al.,}{{Perley}
  et~al.}{2009}]{Perley2009a}
{Perley} D.~A.,  et~al., 2009, \mn@doi [\apj] {10.1088/0004-637X/696/2/1871},
  \href {http://adsabs.harvard.edu/abs/2009ApJ...696.1871P} {696, 1871}

\bibitem[\protect\citeauthoryear{{Perley}, {Modjaz}, {Morgan}, {Cenko},
  {Bloom}, {Butler}, {Filippenko}  \& {Miller}}{{Perley}
  et~al.}{2012}]{Perley2012a}
{Perley} D.~A.,  {Modjaz} M.,  {Morgan} A.~N.,  {Cenko} S.~B.,  {Bloom} J.~S.,
  {Butler} N.~R.,  {Filippenko} A.~V.,   {Miller} A.~A.,  2012, \mn@doi [\apj]
  {10.1088/0004-637X/758/2/122}, \href
  {https://ui.adsabs.harvard.edu/abs/2012ApJ...758..122P} {758, 122}

\bibitem[\protect\citeauthoryear{{Pian} et~al.,}{{Pian}
  et~al.}{2017}]{Pian2017a}
{Pian} E.,  et~al., 2017, \mn@doi [\nat] {10.1038/nature24298}, \href
  {http://adsabs.harvard.edu/abs/2017Natur.551...67P} {551, 67}

\bibitem[\protect\citeauthoryear{{Piro} et~al.,}{{Piro}
  et~al.}{2019}]{Piro2019a}
{Piro} L.,  et~al., 2019, \mn@doi [\mnras] {10.1093/mnras/sty3047}, \href
  {http://adsabs.harvard.edu/abs/2019MNRAS.483.1912P} {483, 1912}

\bibitem[\protect\citeauthoryear{{Planck Collaboration}:~{Ade} et~al.,}{{Planck
  Collaboration}:~{Ade} et~al.}{2016}]{Planck2016a}
{Planck Collaboration}:~{Ade} P.~A.~R.,  et~al., 2016, \mn@doi [\aap]
  {10.1051/0004-6361/201525830}, \href
  {http://adsabs.harvard.edu/abs/2016A%26A...594A..13P} {594, A13}

\bibitem[\protect\citeauthoryear{{Price}, {Berger}  \& {Fox}}{{Price}
  et~al.}{2006}]{Price2006a}
{Price} P.~A.,  {Berger} E.,   {Fox} D.~B.,  2006, GRB Coordinates Network,
  \href {http://adsabs.harvard.edu/abs/2006GCN..5275....1P} {5275}

\bibitem[\protect\citeauthoryear{{Radice}, {Perego}, {Hotokezaka}, {Fromm},
  {Bernuzzi}  \& {Roberts}}{{Radice} et~al.}{2018}]{Radice2018a}
{Radice} D.,  {Perego} A.,  {Hotokezaka} K.,  {Fromm} S.~A.,  {Bernuzzi} S.,
  {Roberts} L.~F.,  2018, \mn@doi [\apj] {10.3847/1538-4357/aaf054}, \href
  {http://adsabs.harvard.edu/abs/2018ApJ...869..130R} {869, 130}

\bibitem[\protect\citeauthoryear{{Rossi}, {Stratta}, {Maiorano}, {Amati},
  {Nicastro}  \& {Palazzi}}{{Rossi} et~al.}{2018a}]{Rossi2018a}
{Rossi} A.,  {Stratta} G.,  {Maiorano} E.,  {Amati} L.,  {Nicastro} L.,
  {Palazzi} E.,  2018a, \memsai, \href
  {https://ui.adsabs.harvard.edu/abs/2018MmSAI..89..254R} {89, 254}

\bibitem[\protect\citeauthoryear{Rossi et~al.,}{Rossi
  et~al.}{2018b}]{Rossi2018gcn}
Rossi A.,  et~al., 2018b, GRB Coordinates Network, Circular Service, 22763

\bibitem[\protect\citeauthoryear{{Rowlinson} et~al.,}{{Rowlinson}
  et~al.}{2010}]{Rowlinson2010a}
{Rowlinson} A.,  et~al., 2010, \mn@doi [\mnras]
  {10.1111/j.1365-2966.2010.17354.x}, \href
  {http://adsabs.harvard.edu/abs/2010MNRAS.409..531R} {409, 531}

\bibitem[\protect\citeauthoryear{{Sakamoto} et~al.,}{{Sakamoto}
  et~al.}{2013}]{Sakamoto2013a}
{Sakamoto} T.,  et~al., 2013, \mn@doi [\apj] {10.1088/0004-637X/766/1/41},
  \href {https://ui.adsabs.harvard.edu/abs/2013ApJ...766...41S} {766, 41}

\bibitem[\protect\citeauthoryear{{Salafia}, {Ghirlanda}, {Ascenzi}  \&
  {Ghisellini}}{{Salafia} et~al.}{2019}]{Salafia2019a}
{Salafia} O.~S.,  {Ghirlanda} G.,  {Ascenzi} S.,   {Ghisellini} G.,  2019,
  \mn@doi [\aap] {10.1051/0004-6361/201935831}, \href
  {https://ui.adsabs.harvard.edu/abs/2019A&A...628A..18S} {628, A18}

\bibitem[\protect\citeauthoryear{{Sari}, {Piran}  \& {Narayan}}{{Sari}
  et~al.}{1998}]{Sari1998a}
{Sari} R.,  {Piran} T.,   {Narayan} R.,  1998, \mn@doi [\apjl]
  {10.1086/311269}, \href {http://adsabs.harvard.edu/abs/1998ApJ...497L..17S}
  {497, L17}

\bibitem[\protect\citeauthoryear{{Sari}, {Piran}  \& {Halpern}}{{Sari}
  et~al.}{1999}]{Sari1999a}
{Sari} R.,  {Piran} T.,   {Halpern} J.~P.,  1999, \apjl, 519, L17

\bibitem[\protect\citeauthoryear{{Sathyaprakash} et~al.,}{{Sathyaprakash}
  et~al.}{2012}]{ET2012a}
{Sathyaprakash} B.,  et~al., 2012, \mn@doi [Classical and Quantum Gravity]
  {10.1088/0264-9381/29/12/124013}, \href
  {https://ui.adsabs.harvard.edu/\#abs/2012CQGra..29l4013S} {29, 124013}

\bibitem[\protect\citeauthoryear{{Savchenko} et~al.,}{{Savchenko}
  et~al.}{2017}]{Savchenko2017a}
{Savchenko} V.,  et~al., 2017, \mn@doi [\apjl] {10.3847/2041-8213/aa8f94},
  \href {http://adsabs.harvard.edu/abs/2017ApJ...848L..15S} {848, L15}

\bibitem[\protect\citeauthoryear{{Schlafly} \& {Finkbeiner}}{{Schlafly} \&
  {Finkbeiner}}{2011}]{SchlaflyFinkbeiner2011a}
{Schlafly} E.~F.,  {Finkbeiner} D.~P.,  2011, \mn@doi [\apj]
  {10.1088/0004-637X/737/2/103}, \href
  {http://adsabs.harvard.edu/abs/2011ApJ...737..103S} {737, 103}

\bibitem[\protect\citeauthoryear{{Selsing} et~al.,}{{Selsing}
  et~al.}{2018}]{Selsing2018a}
{Selsing} J.,  et~al., 2018, \mn@doi [\aap] {10.1051/0004-6361/201731475},
  \href {https://ui.adsabs.harvard.edu/\#abs/2018A&A...616A..48S} {616, A48}

\bibitem[\protect\citeauthoryear{{Selsing} et~al.,}{{Selsing}
  et~al.}{2019}]{Selsing2019a}
{Selsing} J.,  et~al., 2019, \mn@doi [\aap] {10.1051/0004-6361/201832835},
  \href {http://adsabs.harvard.edu/abs/2019A%26A...623A..92S} {623, A92}

\bibitem[\protect\citeauthoryear{{Siegel} \& {Ciolfi}}{{Siegel} \&
  {Ciolfi}}{2016a}]{Siegel2016a}
{Siegel} D.~M.,  {Ciolfi} R.,  2016a, \mn@doi [\apj]
  {10.3847/0004-637X/819/1/14}, \href
  {http://adsabs.harvard.edu/abs/2016ApJ...819...14S} {819, 14}

\bibitem[\protect\citeauthoryear{{Siegel} \& {Ciolfi}}{{Siegel} \&
  {Ciolfi}}{2016b}]{Siegel2016b}
{Siegel} D.~M.,  {Ciolfi} R.,  2016b, \mn@doi [\apj]
  {10.3847/0004-637X/819/1/15}, \href
  {http://adsabs.harvard.edu/abs/2016ApJ...819...15S} {819, 15}

\bibitem[\protect\citeauthoryear{{Smartt} et~al.,}{{Smartt}
  et~al.}{2017}]{Smartt2017a}
{Smartt} S.~J.,  et~al., 2017, \mn@doi [\nat] {10.1038/nature24303}, \href
  {http://adsabs.harvard.edu/abs/2017Natur.551...75S} {551, 75}

\bibitem[\protect\citeauthoryear{{Spyromilio}, {Comer{\'o}n}, {D'Odorico},
  {Kissler-Patig}  \& {Gilmozzi}}{{Spyromilio} et~al.}{2008}]{Spyromilio2008a}
{Spyromilio} J.,  {Comer{\'o}n} F.,  {D'Odorico} S.,  {Kissler-Patig} M.,
  {Gilmozzi} R.,  2008, The Messenger, \href
  {https://ui.adsabs.harvard.edu/abs/2008Msngr.133....2S} {133, 2}

\bibitem[\protect\citeauthoryear{{Stratta} et~al.,}{{Stratta}
  et~al.}{2007}]{Stratta2007a}
{Stratta} G.,  et~al., 2007, \mn@doi [\aap] {10.1051/0004-6361:20078006}, \href
  {http://adsabs.harvard.edu/abs/2007A%26A...474..827S} {474, 827}

\bibitem[\protect\citeauthoryear{{Stratta} et~al.,}{{Stratta}
  et~al.}{2018a}]{Stratta2018b}
{Stratta} G.,  et~al., 2018a, \mn@doi [Advances in Space Research]
  {10.1016/j.asr.2018.04.013}, \href
  {https://ui.adsabs.harvard.edu/abs/2018AdSpR..62..662S} {62, 662}

\bibitem[\protect\citeauthoryear{{Stratta}, {Dainotti}, {Dall'Osso},
  {Hernandez}  \& {De Cesare}}{{Stratta} et~al.}{2018b}]{Stratta2018a}
{Stratta} G.,  {Dainotti} M.~G.,  {Dall'Osso} S.,  {Hernandez} X.,   {De
  Cesare} G.,  2018b, \mn@doi [\apj] {10.3847/1538-4357/aadd8f}, \href
  {http://adsabs.harvard.edu/abs/2018ApJ...869..155S} {869, 155}

\bibitem[\protect\citeauthoryear{{Tanaka} et~al.,}{{Tanaka}
  et~al.}{2018}]{Tanaka2018a}
{Tanaka} M.,  et~al., 2018, \mn@doi [\apj] {10.3847/1538-4357/aaa0cb}, \href
  {http://adsabs.harvard.edu/abs/2018ApJ...852..109T} {852, 109}

\bibitem[\protect\citeauthoryear{{Tanvir} et~al.,}{{Tanvir}
  et~al.}{2010}]{Tanvir2010a}
{Tanvir} N.~R.,  et~al., 2010, GRB Coordinates Network, Circular Service,
  No.~11123, \#1 (2010), \href
  {http://adsabs.harvard.edu/abs/2010GCN.11123....1T} {11123}

\bibitem[\protect\citeauthoryear{{Tanvir}, {Levan}, {Fruchter}, {Hjorth},
  {Hounsell}, {Wiersema}  \& {Tunnicliffe}}{{Tanvir}
  et~al.}{2013}]{Tanvir2013a}
{Tanvir} N.~R.,  {Levan} A.~J.,  {Fruchter} A.~S.,  {Hjorth} J.,  {Hounsell}
  R.~A.,  {Wiersema} K.,   {Tunnicliffe} R.~L.,  2013, \mn@doi [\nat]
  {10.1038/nature12505}, \href
  {http://adsabs.harvard.edu/abs/2013Natur.500..547T} {500, 547}

\bibitem[\protect\citeauthoryear{{Tanvir} et~al.,}{{Tanvir}
  et~al.}{2015}]{Tanvir2015a}
{Tanvir} N.~R.,  et~al., 2015, GRB Coordinates Network, Circular Service,
  No.~18100, \#1 (2015), \href
  {http://adsabs.harvard.edu/abs/2015GCN.18100....1T} {18100}

\bibitem[\protect\citeauthoryear{{Tanvir} et~al.,}{{Tanvir}
  et~al.}{2017}]{Tanvir2017a}
{Tanvir} N.~R.,  et~al., 2017, \mn@doi [\apjl] {10.3847/2041-8213/aa90b6},
  \href {http://adsabs.harvard.edu/abs/2017ApJ...848L..27T} {848, L27}

\bibitem[\protect\citeauthoryear{{Troja} et~al.,}{{Troja}
  et~al.}{2016}]{Troja2016a}
{Troja} E.,  et~al., 2016, \mn@doi [\apj] {10.3847/0004-637X/827/2/102}, \href
  {https://ui.adsabs.harvard.edu/\#abs/2016ApJ...827..102T} {827, 102}

\bibitem[\protect\citeauthoryear{{Troja} et~al.,}{{Troja}
  et~al.}{2017}]{Troja2017a}
{Troja} E.,  et~al., 2017, \mn@doi [\nat] {10.1038/nature24290}, \href
  {http://adsabs.harvard.edu/abs/2017Natur.551...71T} {551, 71}

\bibitem[\protect\citeauthoryear{{Troja} et~al.,}{{Troja}
  et~al.}{2018a}]{Troja2018b}
{Troja} E.,  et~al., 2018a, \mn@doi [Nature Communications]
  {10.1038/s41467-018-06558-7}, \href
  {http://adsabs.harvard.edu/abs/2018NatCo...9.4089T} {9, 4089}

\bibitem[\protect\citeauthoryear{{Troja} et~al.,}{{Troja}
  et~al.}{2018b}]{Troja2018a}
{Troja} E.,  et~al., 2018b, \mn@doi [\mnras] {10.1093/mnrasl/sly061}, \href
  {http://adsabs.harvard.edu/abs/2018MNRAS.478L..18T} {478, L18}

\bibitem[\protect\citeauthoryear{{Troja} et~al.,}{{Troja}
  et~al.}{2019}]{Troja2019a}
{Troja} E.,  et~al., 2019, \mn@doi [\mnras] {10.1093/mnras/stz2255}, \href
  {https://ui.adsabs.harvard.edu/abs/2019MNRAS.489.2104T} {489, 2104}

\bibitem[\protect\citeauthoryear{{Varela}, {Knust}  \& {Greiner}}{{Varela}
  et~al.}{2015}]{Varela2015a}
{Varela} K.,  {Knust} F.,   {Greiner} J.,  2015, GRB Coordinates Network, \href
  {https://ui.adsabs.harvard.edu/abs/2015GCN.17732....1V} {17732, 1}

\bibitem[\protect\citeauthoryear{{Villar} et~al.,}{{Villar}
  et~al.}{2017}]{Villar2017b}
{Villar} V.~A.,  et~al., 2017, \mn@doi [\apjl] {10.3847/2041-8213/aa9c84},
  \href {http://adsabs.harvard.edu/abs/2017ApJ...851L..21V} {851, L21}

\bibitem[\protect\citeauthoryear{{Watson}, {Hjorth}, {Jakobsson}, {Xu},
  {Fynbo}, {Sollerman}, {Th{\"o}ne}  \& {Pedersen}}{{Watson}
  et~al.}{2006}]{Watson2006a}
{Watson} D.,  {Hjorth} J.,  {Jakobsson} P.,  {Xu} D.,  {Fynbo} J.~P.~U.,
  {Sollerman} J.,  {Th{\"o}ne} C.~C.,   {Pedersen} K.,  2006, \mn@doi [\aap]
  {10.1051/0004-6361:20065380}, \href
  {http://adsabs.harvard.edu/abs/2006A%26A...454L.123W} {454, L123}

\bibitem[\protect\citeauthoryear{{Xu} et~al.,}{{Xu} et~al.}{2009}]{Xu2009a}
{Xu} D.,  et~al., 2009, \mn@doi [\apj] {10.1088/0004-637X/696/1/971}, \href
  {https://ui.adsabs.harvard.edu/abs/2009ApJ...696..971X} {696, 971}

\bibitem[\protect\citeauthoryear{{Yang} et~al.,}{{Yang}
  et~al.}{2015}]{Yang2015a}
{Yang} B.,  et~al., 2015, \mn@doi [Nature Communications] {10.1038/ncomms8323},
  \href {http://adsabs.harvard.edu/abs/2015NatCo...6E7323Y} {6, 7323}

\bibitem[\protect\citeauthoryear{{Yaron} \& {Gal-Yam}}{{Yaron} \&
  {Gal-Yam}}{2012}]{Yaron2012a}
{Yaron} O.,  {Gal-Yam} A.,  2012, \mn@doi [\pasp] {10.1086/666656}, \href
  {http://adsabs.harvard.edu/abs/2012PASP..124..668Y} {124, 668}

\bibitem[\protect\citeauthoryear{{Yonetoku}, {Murakami}, {Nakamura},
  {Yamazaki}, {Inoue}  \& {Ioka}}{{Yonetoku} et~al.}{2004}]{Yonetoku2004a}
{Yonetoku} D.,  {Murakami} T.,  {Nakamura} T.,  {Yamazaki} R.,  {Inoue} A.~K.,
   {Ioka} K.,  2004, \mn@doi [\apj] {10.1086/421285}, \href
  {http://adsabs.harvard.edu/abs/2004ApJ...609..935Y} {609, 935}

\bibitem[\protect\citeauthoryear{{Yu}, {Zhang}  \& {Gao}}{{Yu}
  et~al.}{2013}]{Yu2013a}
{Yu} Y.-W.,  {Zhang} B.,   {Gao} H.,  2013, \mn@doi [\apjl]
  {10.1088/2041-8205/776/2/L40}, \href
  {http://adsabs.harvard.edu/abs/2013ApJ...776L..40Y} {776, L40}

\bibitem[\protect\citeauthoryear{{Zhang} \& {M{\'e}sz{\'a}ros}}{{Zhang} \&
  {M{\'e}sz{\'a}ros}}{2001}]{ZhangMeszaros2001a}
{Zhang} B.,  {M{\'e}sz{\'a}ros} P.,  2001, \mn@doi [\apjl] {10.1086/320255},
  \href {http://adsabs.harvard.edu/abs/2001ApJ...552L..35Z} {552, L35}

\bibitem[\protect\citeauthoryear{{Zhang} \& {M{\'e}sz{\'a}ros}}{{Zhang} \&
  {M{\'e}sz{\'a}ros}}{2004}]{Zhang2004a}
{Zhang} B.,  {M{\'e}sz{\'a}ros} P.,  2004, \mn@doi [International Journal of
  Modern Physics A] {10.1142/S0217751X0401746X}, \href
  {http://adsabs.harvard.edu/abs/2004IJMPA..19.2385Z} {19, 2385}

\bibitem[\protect\citeauthoryear{{Zhang}, {Fan}, {Dyks}, {Kobayashi},
  {M{\'e}sz{\'a}ros}, {Burrows}, {Nousek}  \& {Gehrels}}{{Zhang}
  et~al.}{2006}]{Zhang2006c}
{Zhang} B.,  {Fan} Y.~Z.,  {Dyks} J.,  {Kobayashi} S.,  {M{\'e}sz{\'a}ros} P.,
  {Burrows} D.~N.,  {Nousek} J.~A.,   {Gehrels} N.,  2006, \mn@doi [\apj]
  {10.1086/500723}, \href {http://adsabs.harvard.edu/abs/2006ApJ...642..354Z}
  {642, 354}

\bibitem[\protect\citeauthoryear{{de Ugarte Postigo} et~al.,}{{de Ugarte
  Postigo} et~al.}{2014}]{deUgartePostigo2014a}
{de Ugarte Postigo} A.,  et~al., 2014, \mn@doi [\aap]
  {10.1051/0004-6361/201322985}, \href
  {https://ui.adsabs.harvard.edu/abs/2014A&A...563A..62D} {563, A62}

\makeatother
\end{thebibliography}




\appendix

\section{Additional material}

\onecolumn

\begin{longtable}{l c      c c       c       c       c      c     c     c     c}
\caption{Table with GRB and AT2017gfo luminosity ratios}\label{tab:ratiolum}\\
GRB  &  time  &  Band  &  $L_{GRB}$  &  $L_{KN}$  &  $L_{GRB}/L_{kn}$  &  z  &  $\lambda_{rest}$  &  $\lambda_{obs}$ \\
     &   hr   &    eff.           &  $10^{26}$ erg/s Hz$^{-1}$  &  $10^{26}$ erg/s Hz$^{-1}$  &                   &     &  nm  &  nm \\
\hline
\hline
\endfirsthead

\multicolumn{9}{c}{{\bfseries \tablename\ \thetable{} -- continued from previous page}} \\
\midrule
\endhead

\midrule
\multicolumn{9}{r}{{Continued on next page}} \\
\endfoot

\bottomrule
\endlastfoot
\footnotetext

050509B  &  40.9  &  g*  &  < 3.21  &  1.24  &   < 2.59  &  0.225  &  391.7  &  479.8 \\
050509B  &  77.1  &  V  &  < 0.50  &  0.36  &   < 1.38  &  0.225  &  454.9  &  557.1 \\
050509B  &  155.6  &  V  &  < 5.01  &  0.08  &   < 59.43  &  0.225  &  454.9  &  557.1 \\
050509B  &  233.0  &  V  &  < 10.50  &  0.08  &   < 123.82  &  0.225  &  454.9  &  557.1 \\
050509B  &  16.1  &  R  &  < 60.60  &  10.50  &   < 5.77  &  0.225  &  523.5  &  641.2 \\
050509B  &  21.2  &  R  &  < 2.62  &  8.58  &   < 0.31  &  0.225  &  523.5  &  641.2 \\
050509B  &  36.6  &  R  &  < 0.96  &  4.50  &   < 0.21  &  0.225  &  523.5  &  641.2 \\
050509B  &  40.9  &  R  &  < 7.22  &  3.91  &   < 1.85  &  0.225  &  523.5  &  641.2 \\
\end{longtable}

\clearpage
\normalsize
\twocolumn


\onecolumn


\begin{longtable}{l c      c       c | c c c c}
\caption{Rest frame light-curves of {\bf AT2017gfo}}\label{tab:lckn}\\
time  & Luminosity & M$_{\rm abs}$ &  filter  & time  & Luminosity & M$_{\rm abs}$ &  filter\\
days  & [erg/s/Hz] & AB    &       & days  & [erg/s/Hz] & AB    &       \\
\hline
\hline
\endfirsthead

\multicolumn{8}{c}{{\bfseries \tablename\ \thetable{} -- continued from previous page}} \\
\midrule
\endhead

\midrule 
\multicolumn{8}{r}{{Continued on next page}} \\
\endfoot

\bottomrule
\endlastfoot


0.50  & --             &  --     & U & 0.50  & 1.29E$+27 $ & $-16.16  $ & r \\
0.64  & 7.18E$+26 $ & $-15.52  $ & U & 0.64  & 1.15E$+27 $ & $-16.03  $ & r \\    
0.99  & 3.81E$+26 $ & $-14.83  $ & U & 0.99  & 8.50E$+26 $ & $-15.70  $ & r \\    
1.49  & 1.02E$+26 $ & $-13.40  $ & U & 1.49  & 6.22E$+26 $ & $-15.36  $ & r \\    
2.43  & 1.65E$+25 $ & $-11.42  $ & U & 2.43  & 2.43E$+26 $ & $-14.35  $ & r \\    
3.47  & 5.35E$+24 $ & $-10.20  $ & U & 3.47  & 1.18E$+26 $ & $-13.56  $ & r \\    
4.41  & 4.51E$+24 $ & $-10.01  $ & U & 4.41  & 6.90E$+25 $ & $-12.98  $ & r \\    
5.45  & 3.18E$+24 $ & $-9.64   $ & U & 5.45  & 4.64E$+25 $ & $-12.54  $ & r \\
\bottomrule
\end{longtable}


\normalsize
\twocolumn



\begin{figure*}
\begin{center}
\begin{tabular}{  c  c  }
\includegraphics[scale=0.45]{figaft/GRB050509B_KN_mag.pdf}&
\includegraphics[scale=0.45]{figaft/GRB050709_KN_mag.pdf}\\
\includegraphics[scale=0.45]{figaft/GRB061201_KN_mag.pdf}&
\includegraphics[scale=0.45]{figaft/GRB080905A_KN_mag.pdf}\\
\includegraphics[scale=0.45]{figaft/GRB090515_KN_mag.pdf}&
\includegraphics[scale=0.45]{figaft/GRB100206A_KN_mag.pdf}\\
\includegraphics[scale=0.45]{figaft/GRB160821B_KN_mag.pdf}&
\end{tabular}
\end{center}
\caption{Short GRBs for which the optical counterpart luminosity ({\bf circles mark the detections and triangles the upper limits}) are fainter than AT2017gfo luminosity (dotted lines with crosses) in at least one effective rest-frame  filter (see \S 3.2). Note that GRB 050709, GRB 090515 and GRB160821B {\bf show evidence of a temporal decay index lower than the shallowest index predicted by the fireball model (i.e. $\alpha<0.75$, see \S 4.2, black dashed line). 
}
}
\label{fig:constraints}
\end{figure*}

\begin{figure*}
\begin{center}
\begin{tabular}{  c  c  }
\includegraphics[scale=0.45]{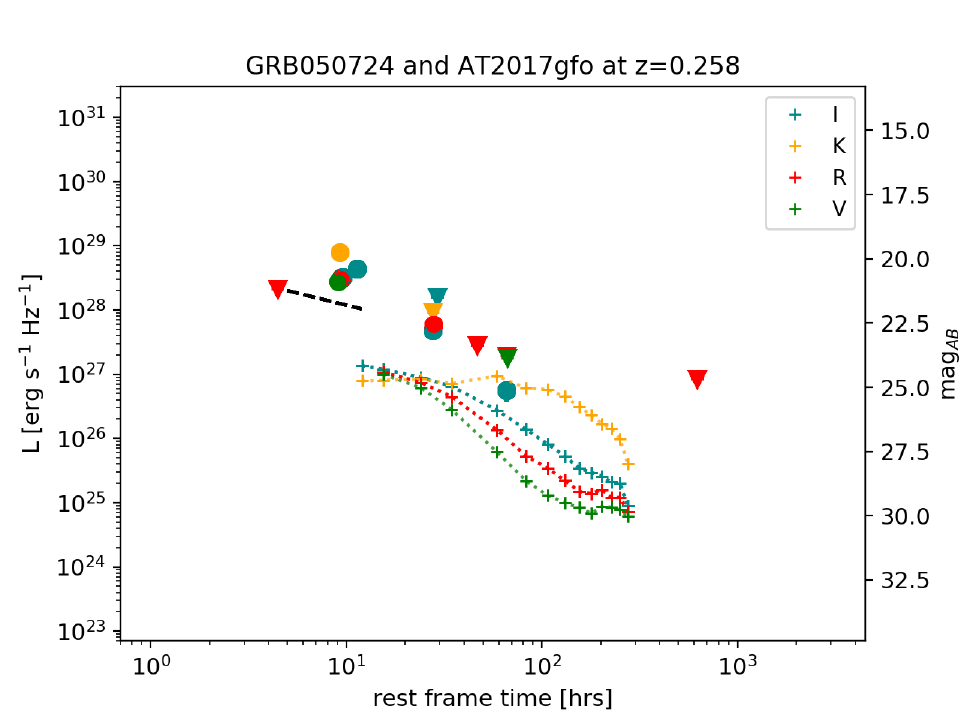}&
\includegraphics[scale=0.45]{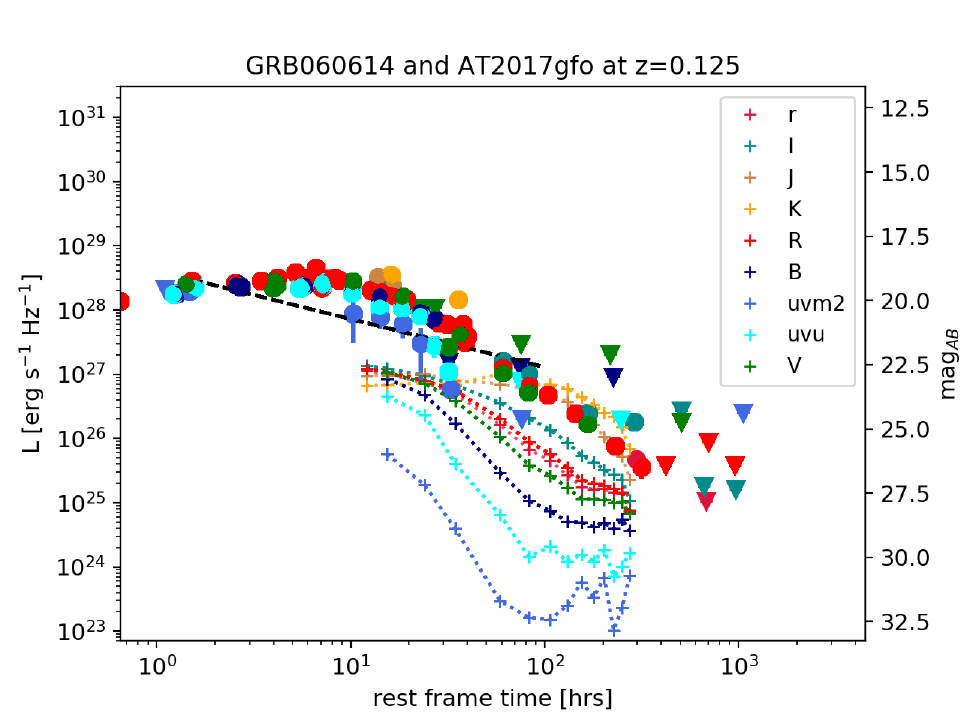}\\
\includegraphics[scale=0.45]{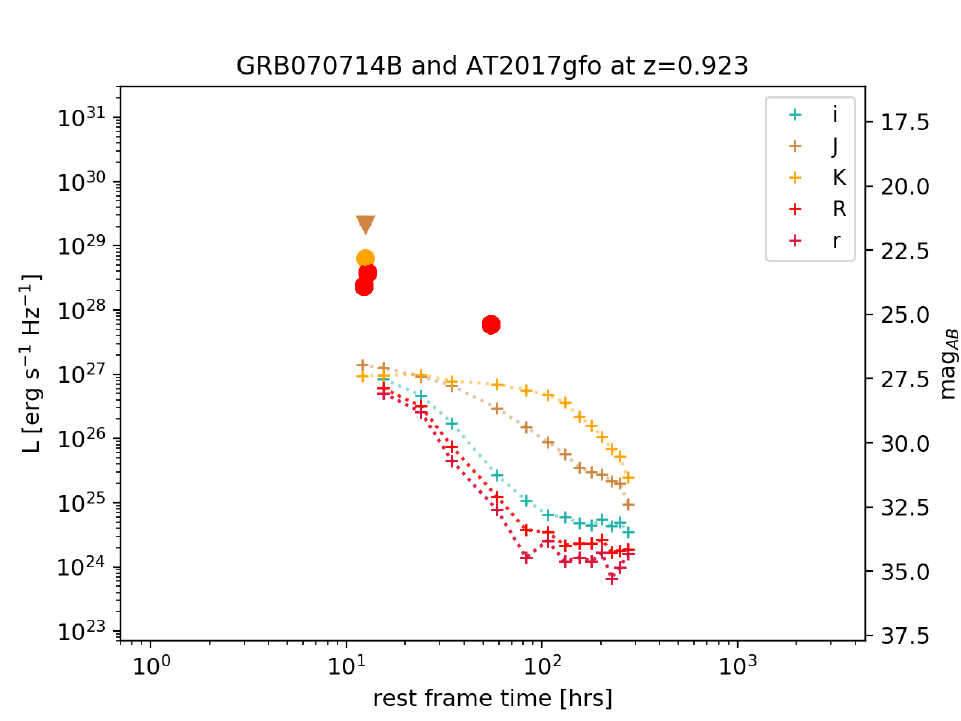}&
\includegraphics[scale=0.45]{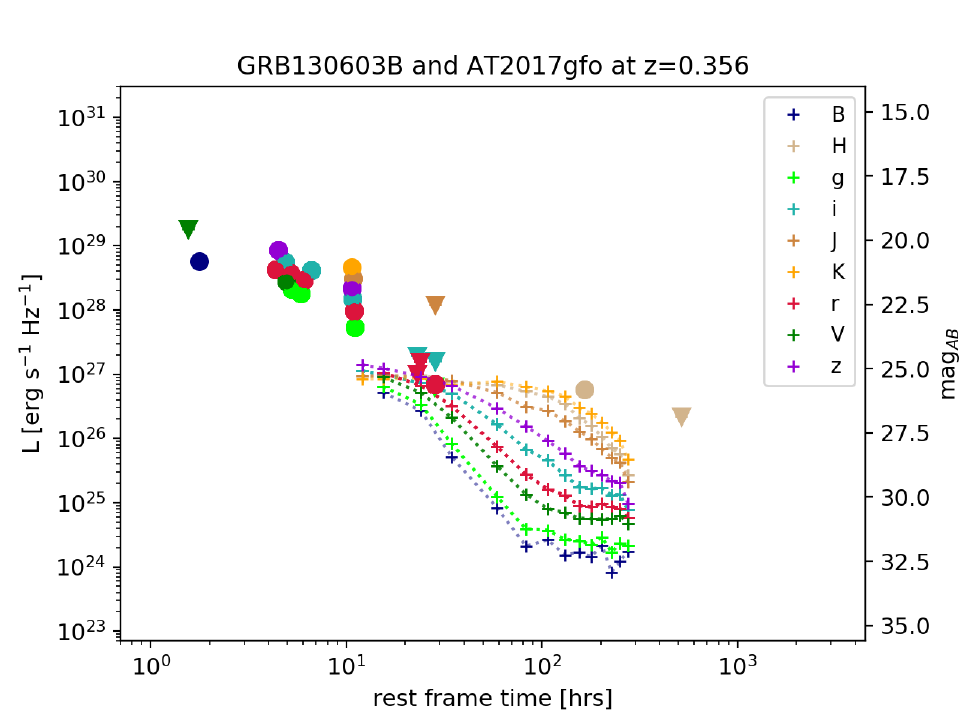}\\ 
\includegraphics[scale=0.45]{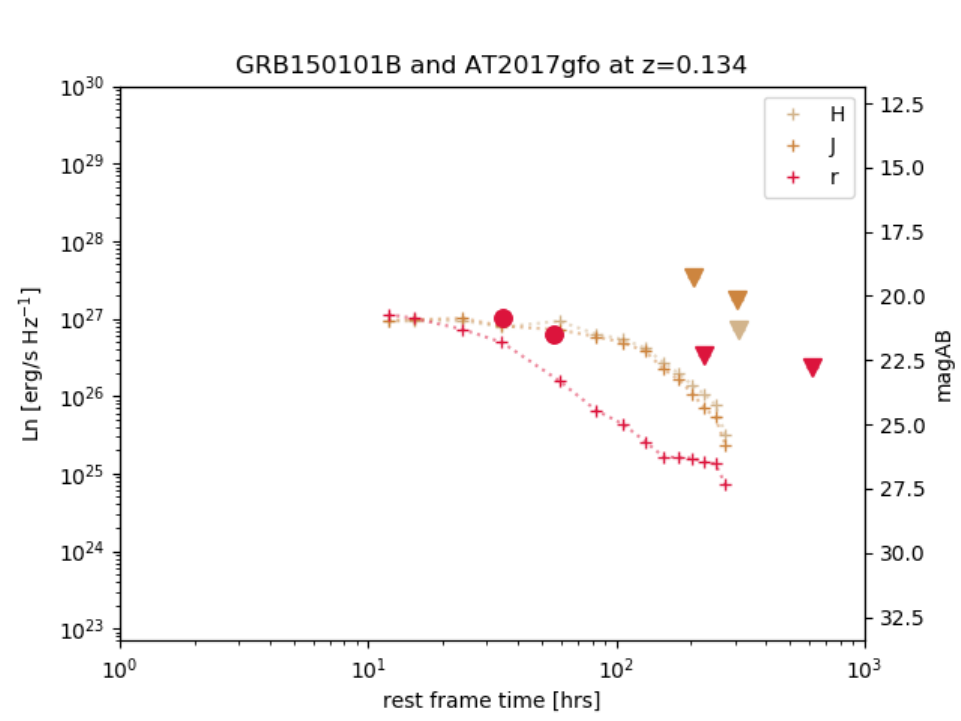}&
\includegraphics[scale=0.45]{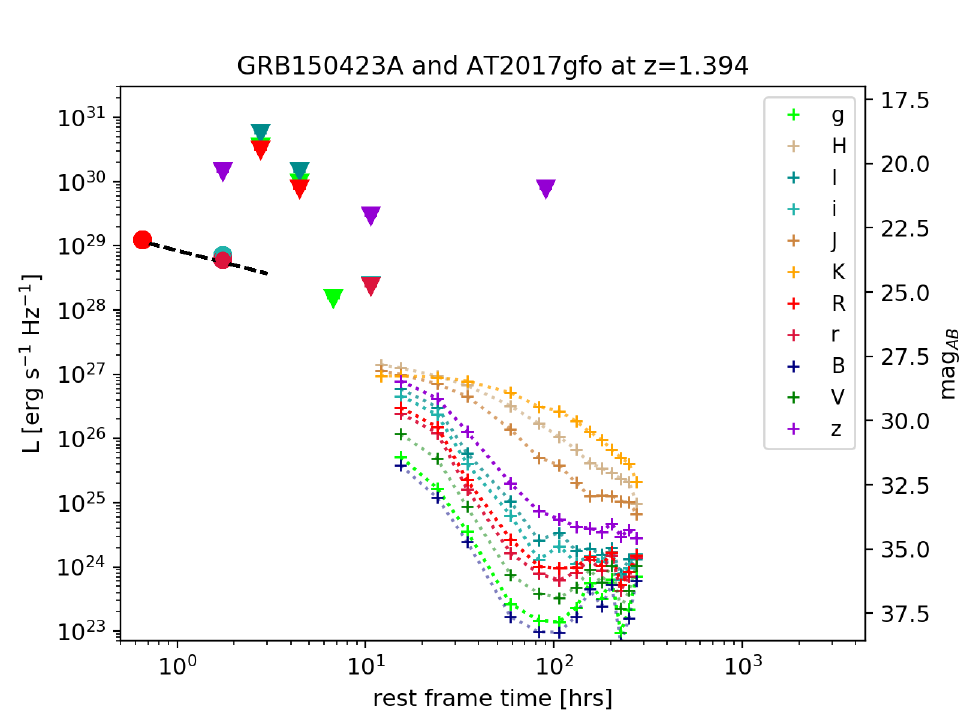}\\
\includegraphics[scale=0.45]{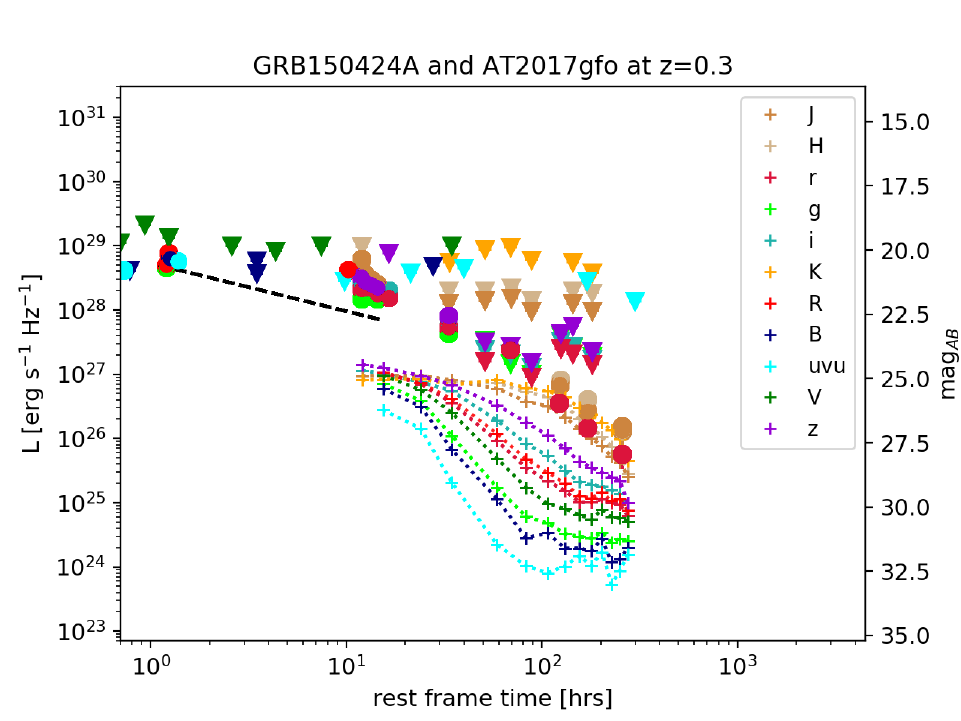}&
\end{tabular}
\end{center}
\caption{Same as Figure \ref{fig:constraints} but for GRB optical counterparts for which the luminosity light curves are above the AT2017gfo luminosity in any filter and for which we found evidence of an anomalous shallow decay.  If this feature is due to an emerging kilonova emission, from these short GRBs we can infer the upper range of possible kilonova luminosity values (i.e. not just upper limits).
We include here also the light curves of GRBs 070714B, 130603B, and 150101B for which a kilonova is claimed in the literature.
}
\label{fig:lcplot}
\end{figure*}

\begin{figure*}
\begin{center}
\begin{tabular}{  c  c  }
\includegraphics[scale=0.45]{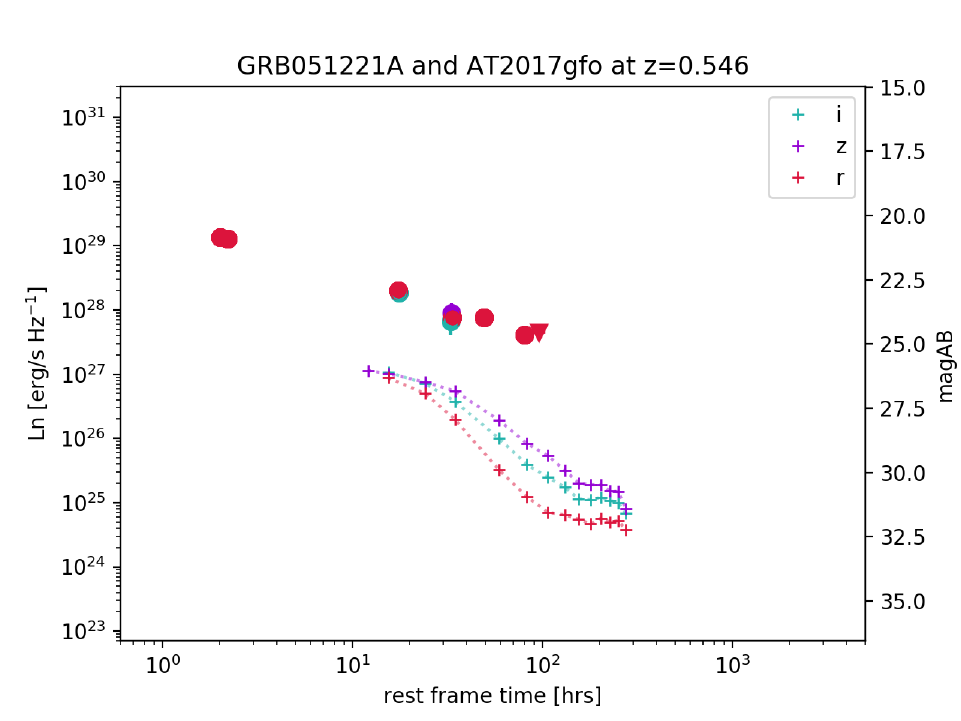}&
\includegraphics[scale=0.45]{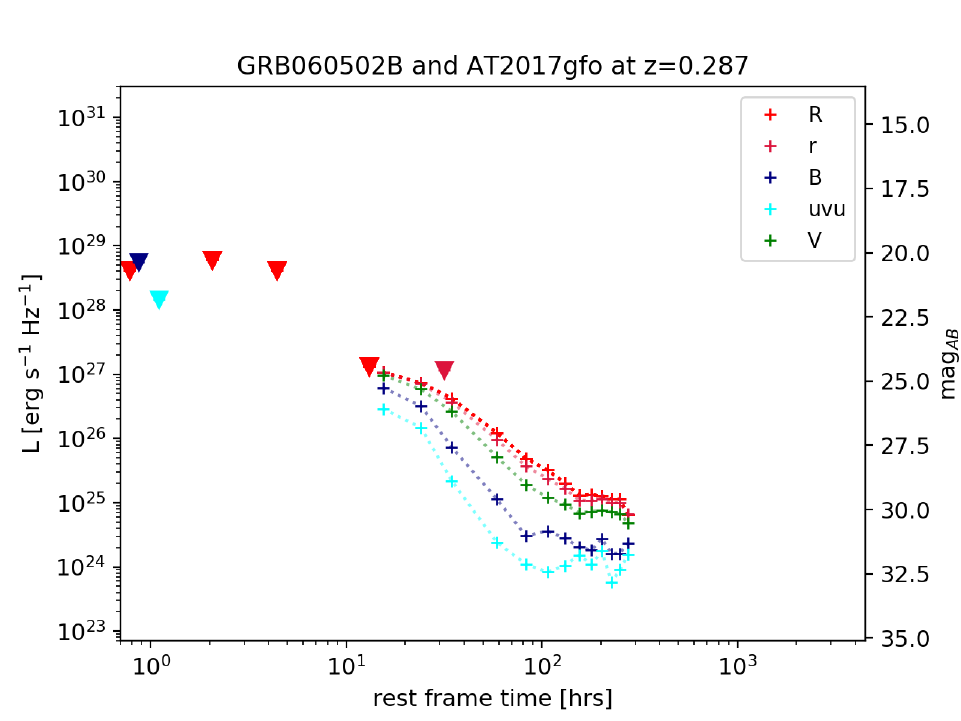}\\
\includegraphics[scale=0.45]{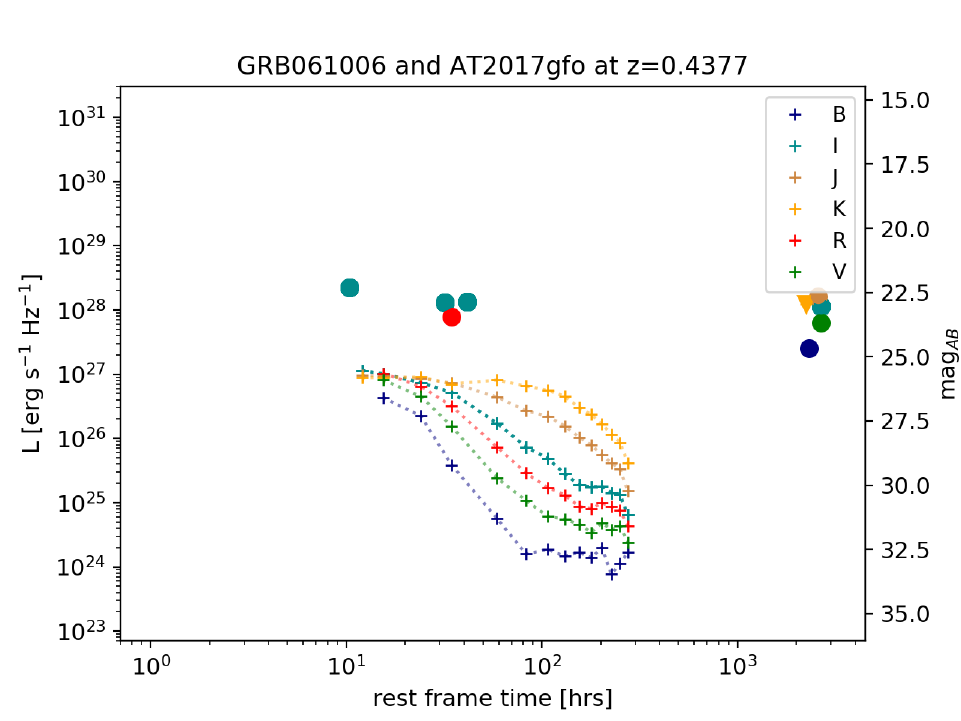}&
\includegraphics[scale=0.45]{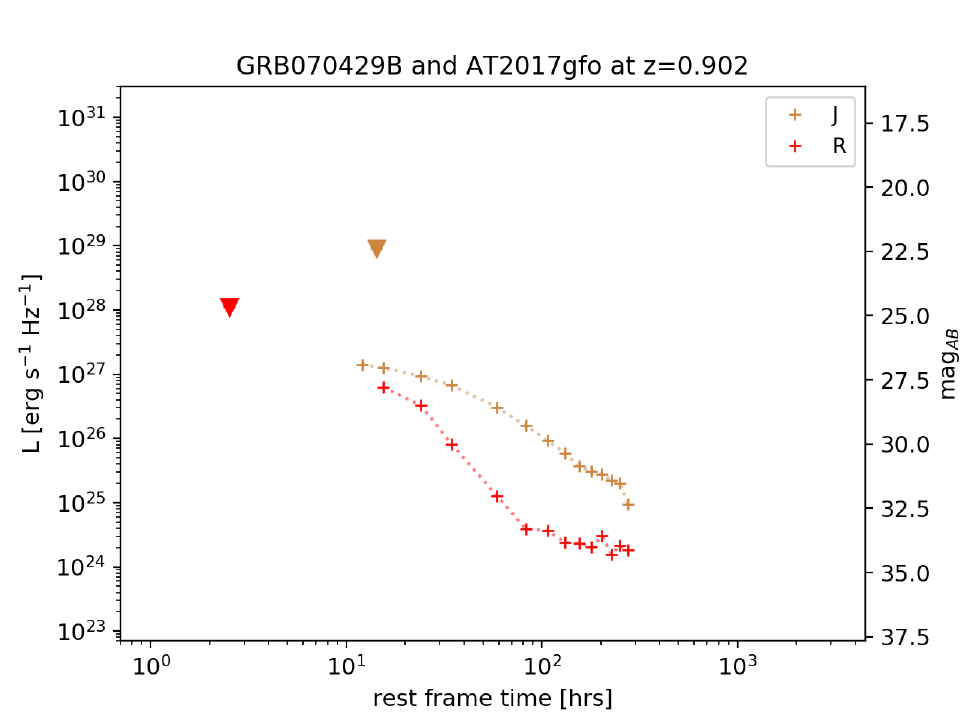}\\
\includegraphics[scale=0.45]{figaft/GRB070714B_KN_mag.pdf}&
\includegraphics[scale=0.45]{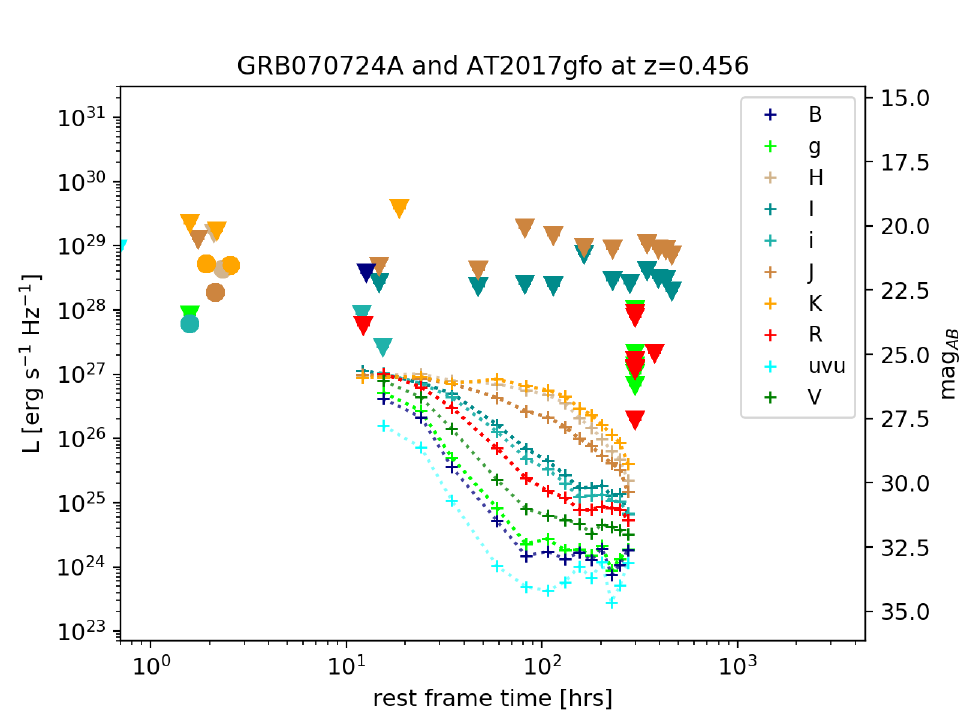}\\
\includegraphics[scale=0.45]{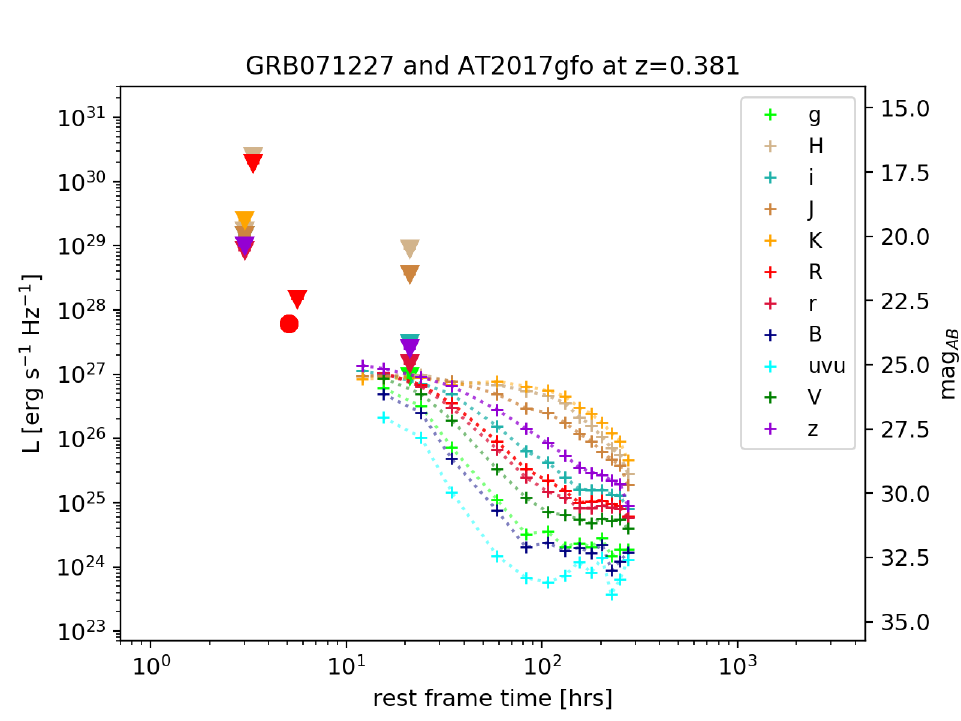}&
\includegraphics[scale=0.45]{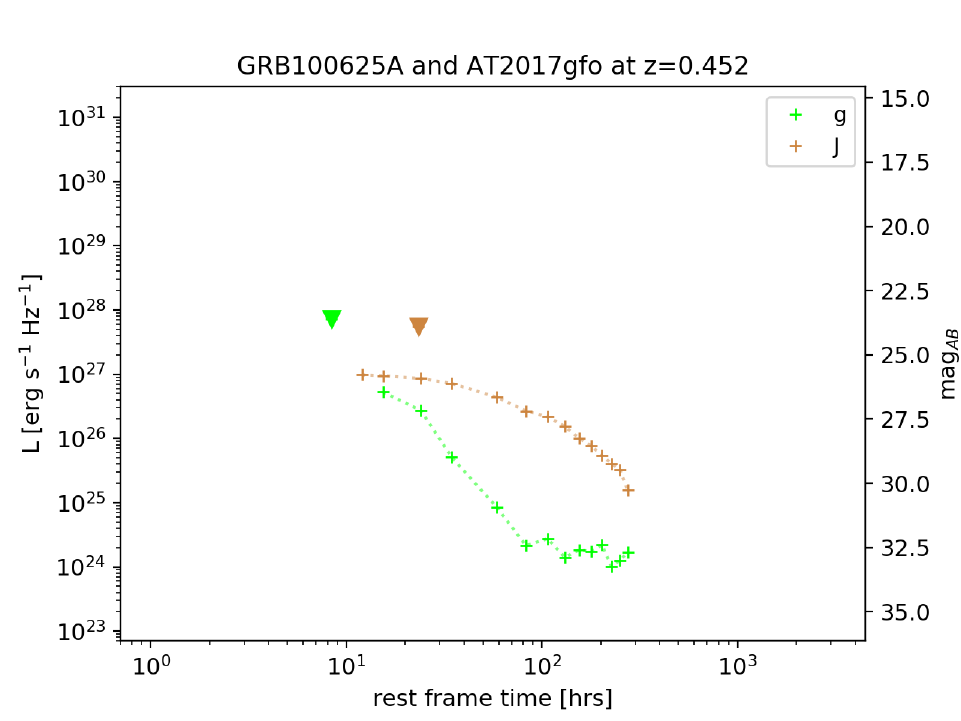}\\
\end{tabular}
\end{center}
\caption{Short GRB optical counterparts for which the luminosity light curves are
above the AT2017gfo luminosity in any filter. Circles mark the detections and triangles mark the upper limits, AT2017gfo luminosity is indicated with dotted lines with crosses.}
\label{fig:lcplot3}
\end{figure*}

\begin{figure*}
\begin{center}
\begin{tabular}{  c  c  }
\includegraphics[scale=0.45]{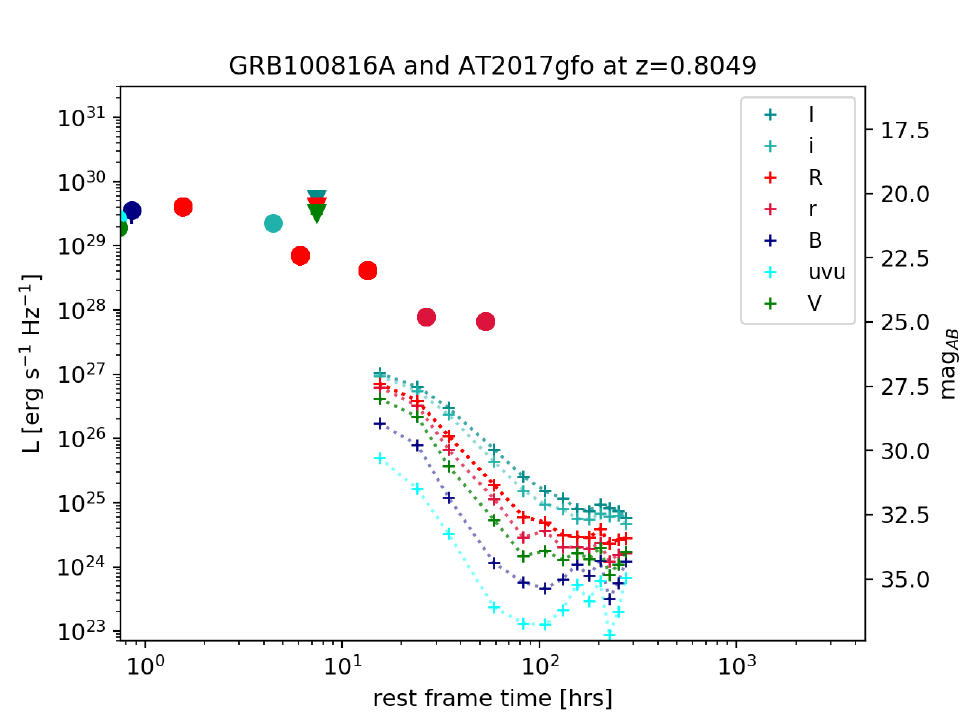}&
\includegraphics[scale=0.45]{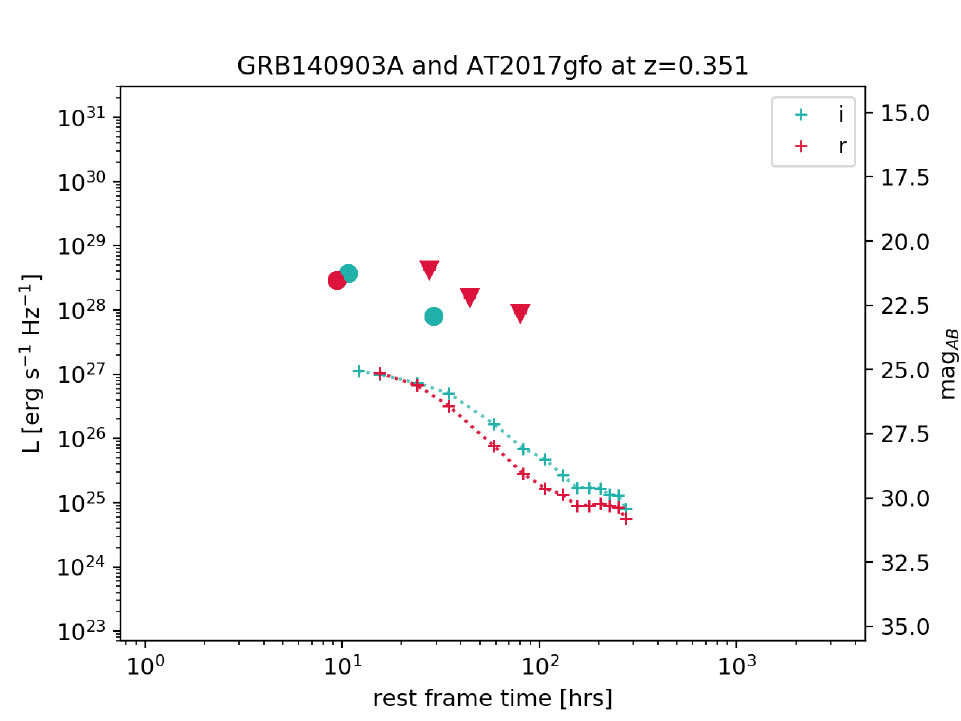}\\
\includegraphics[scale=0.45]{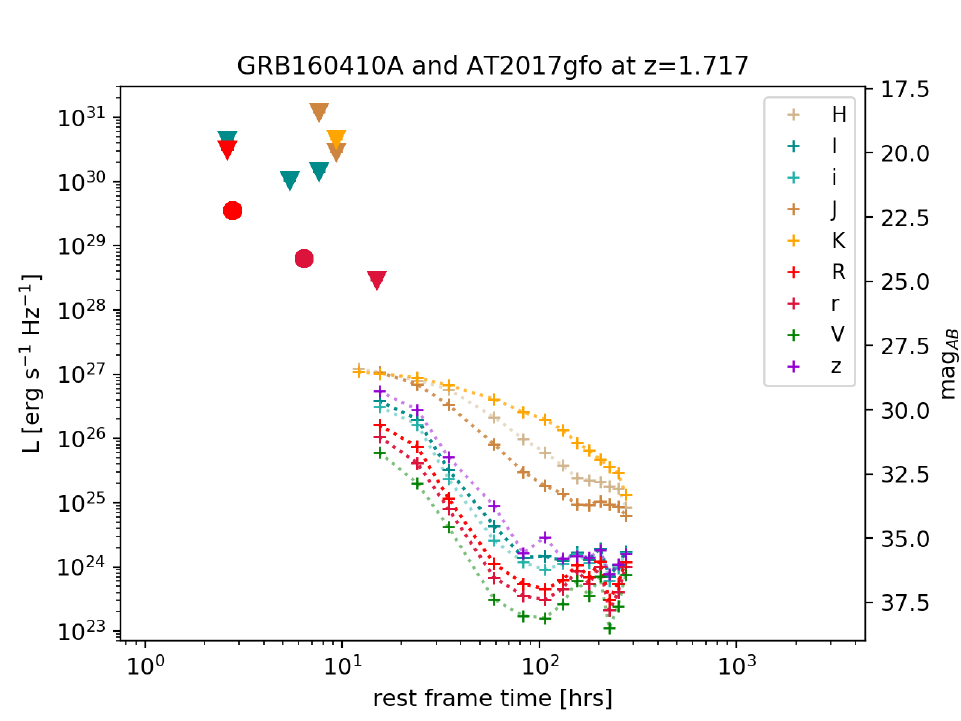}&
\includegraphics[scale=0.45]{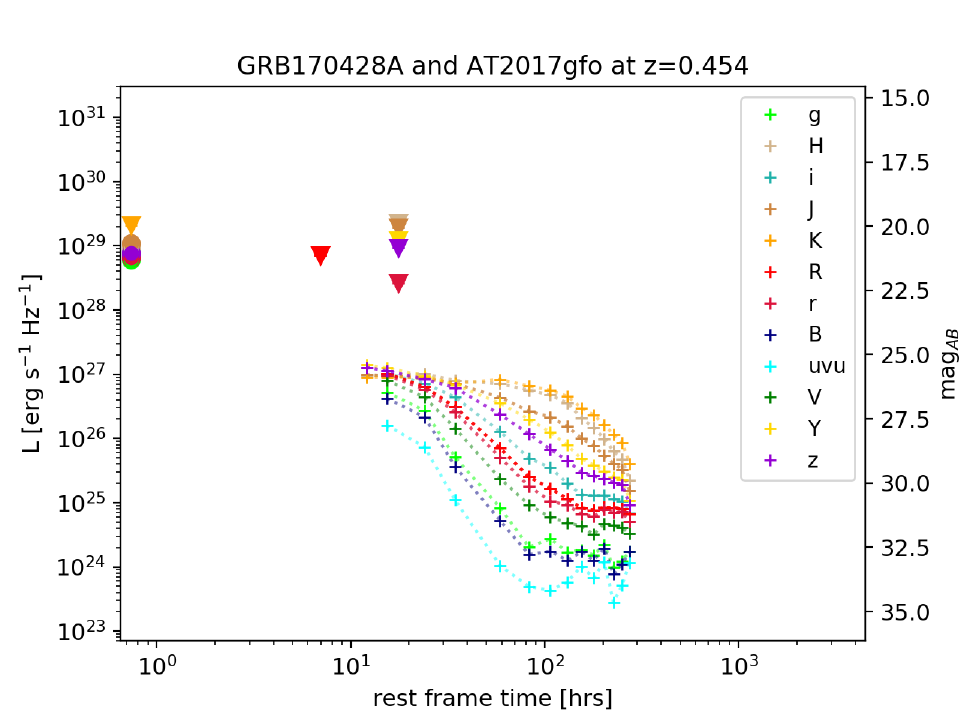}\\
\end{tabular}
\end{center}
\contcaption{} 
\label{fig:lcplot4}
\end{figure*}

\begin{figure*}
\begin{center}
\begin{tabular}{  c  c  }
\includegraphics[scale=0.45]{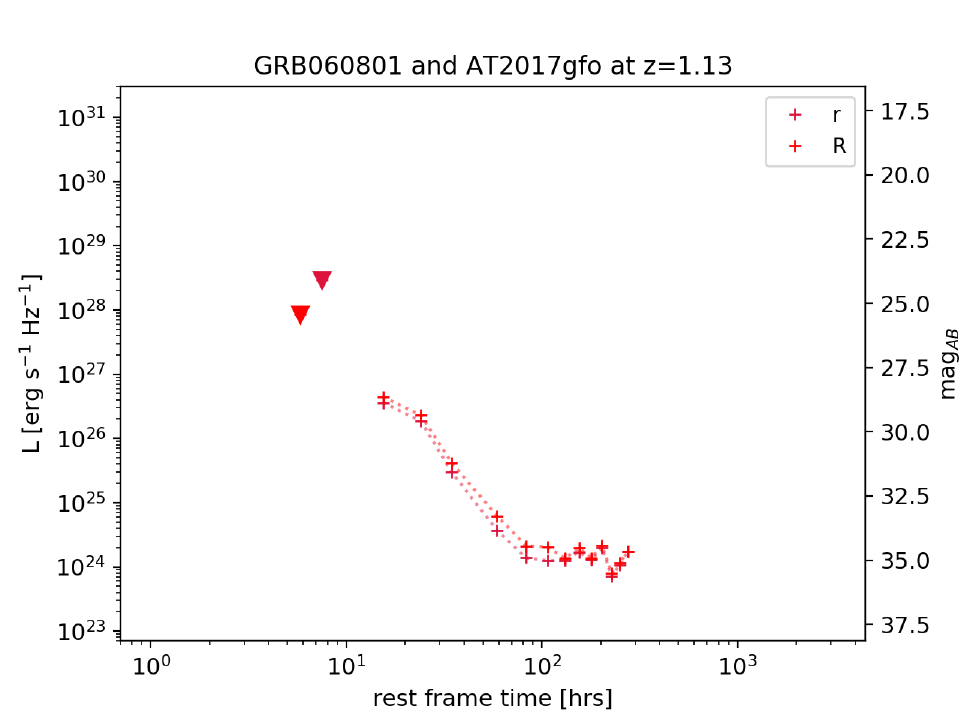}&
\includegraphics[scale=0.45]{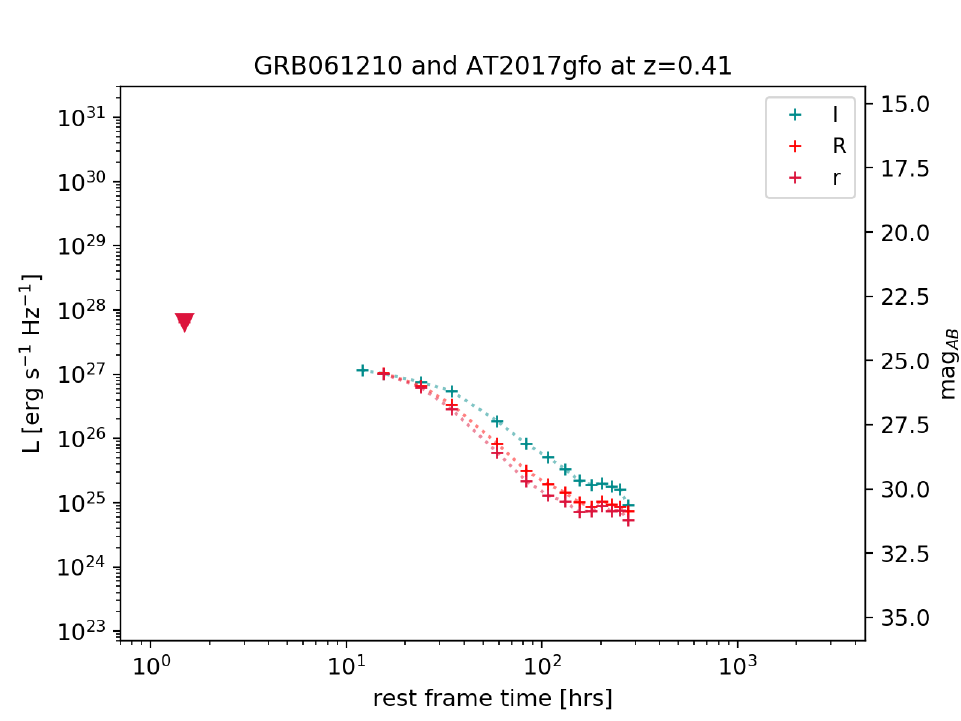}\\
\includegraphics[scale=0.45]{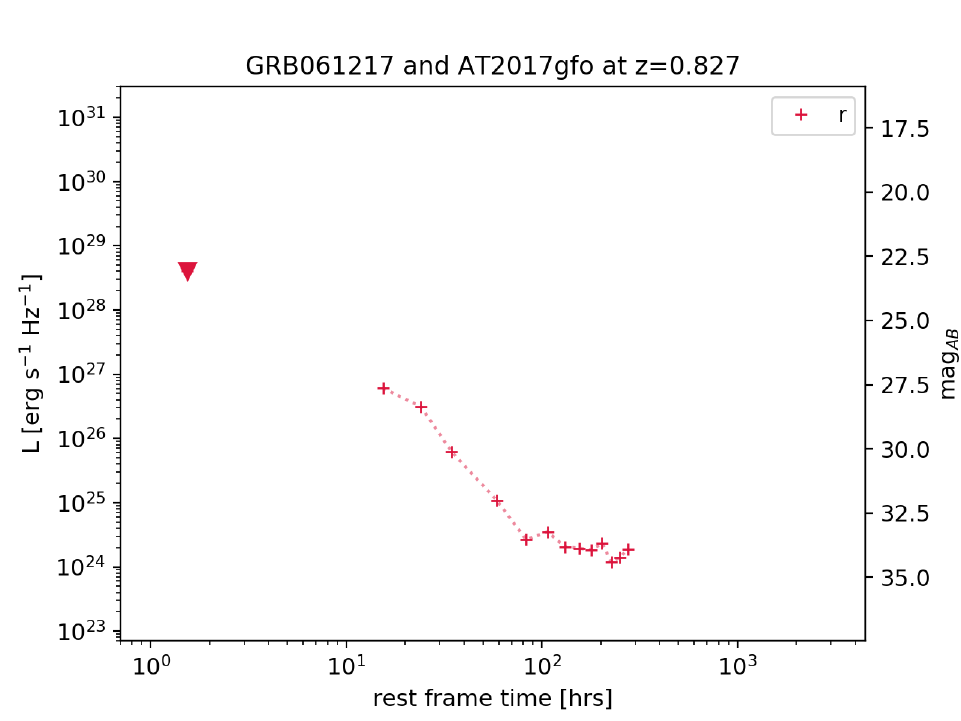}& 
\includegraphics[scale=0.45]{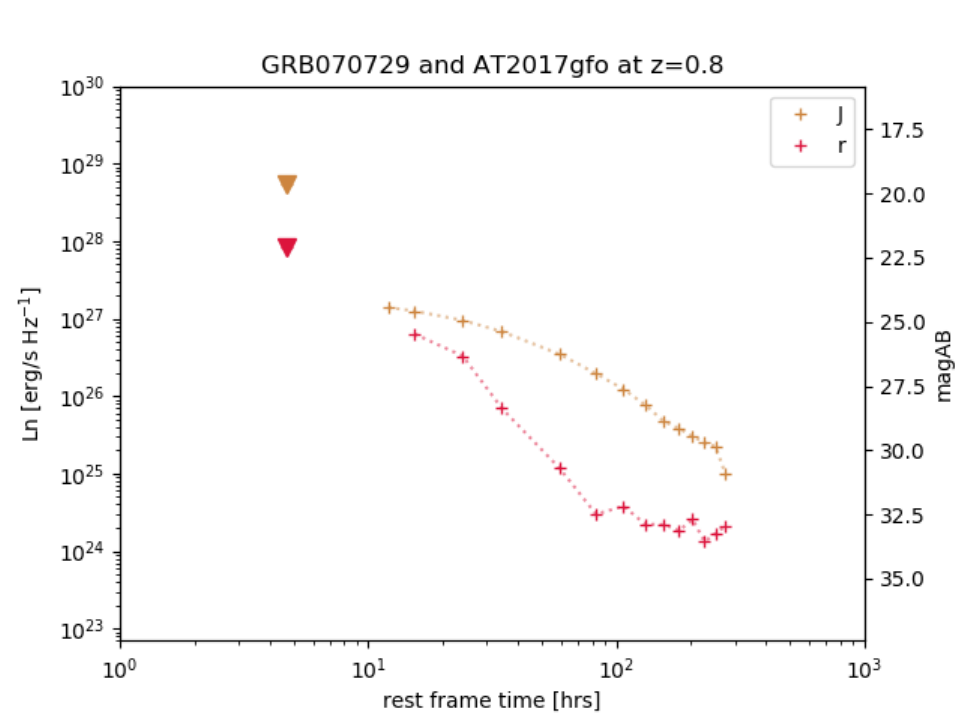}\\
\includegraphics[scale=0.45]{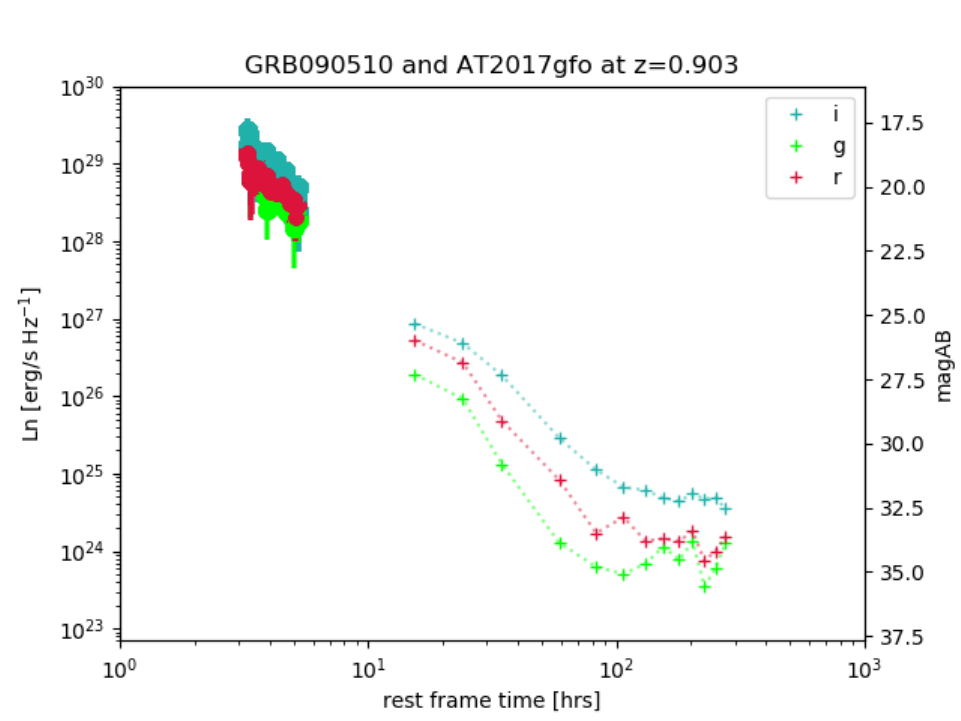}&
\includegraphics[scale=0.45]{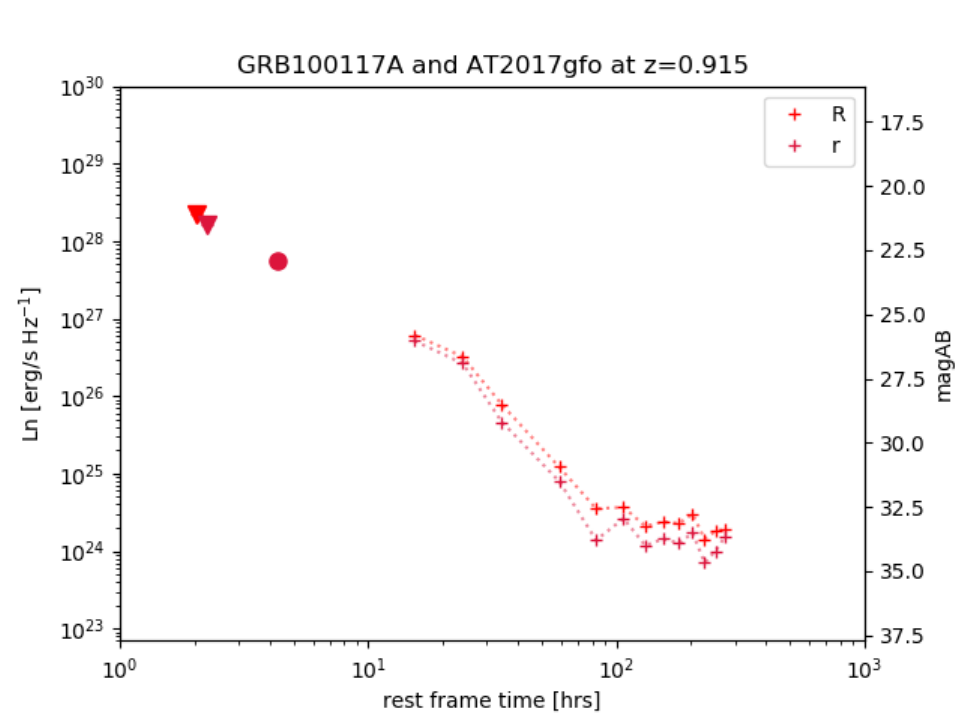}\\
\includegraphics[scale=0.45]{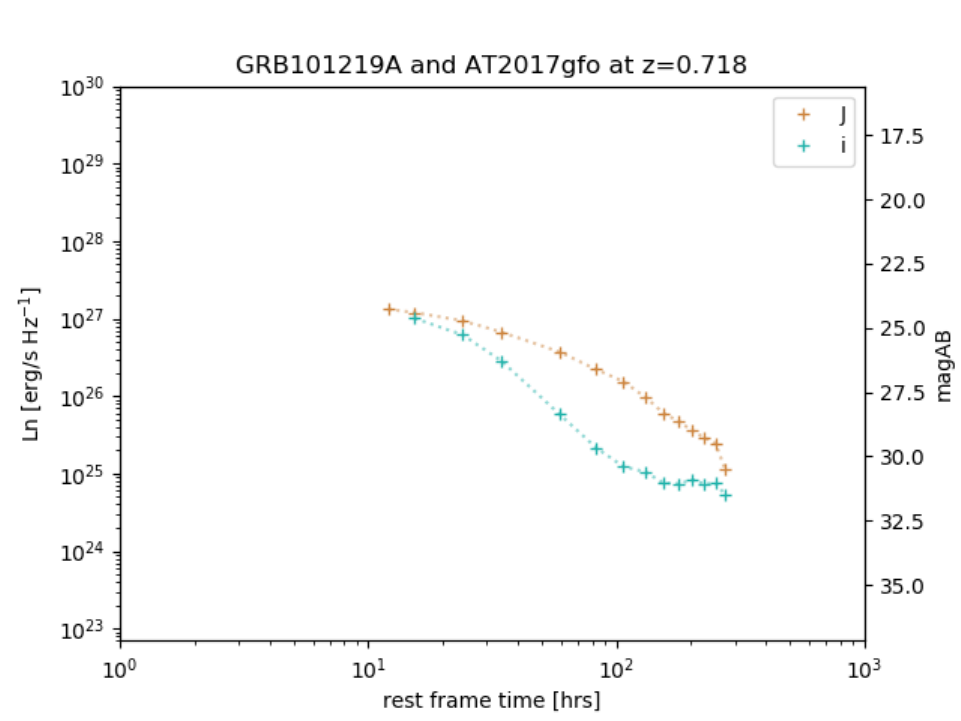}&
\includegraphics[scale=0.45]{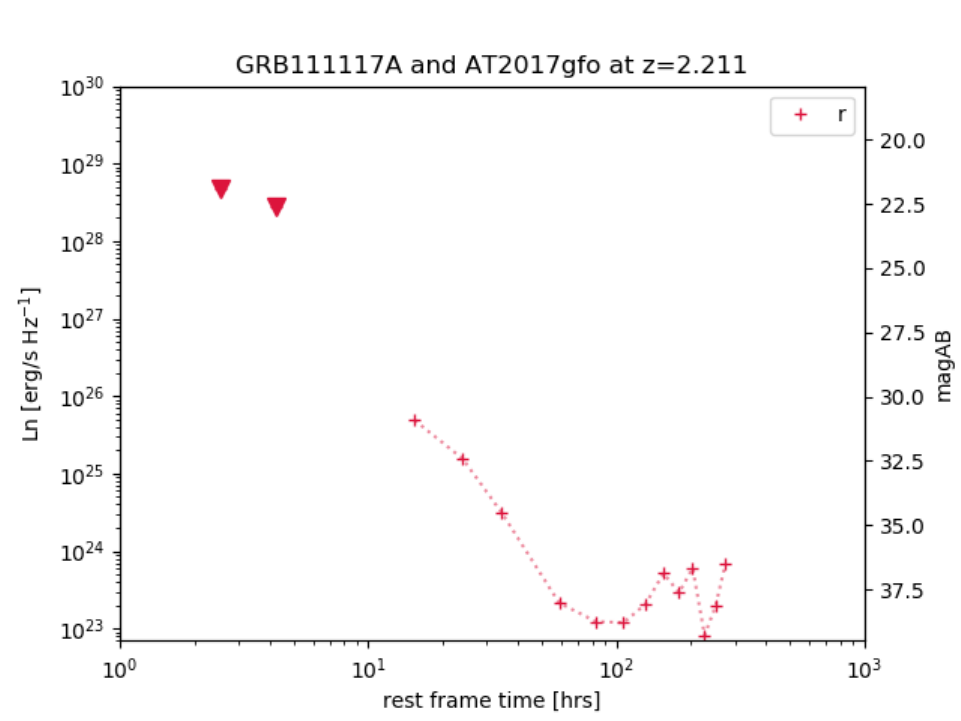}\\
\end{tabular}
\end{center}
\caption{Short GRBs for which no {\bf optical} data fall within the {\bf AT2017gfo} sampled
temporal window.}
\label{fig:noconstraints}
\end{figure*}

\begin{figure*}
\begin{center}
\begin{tabular}{  c  c  }
\includegraphics[scale=0.45]{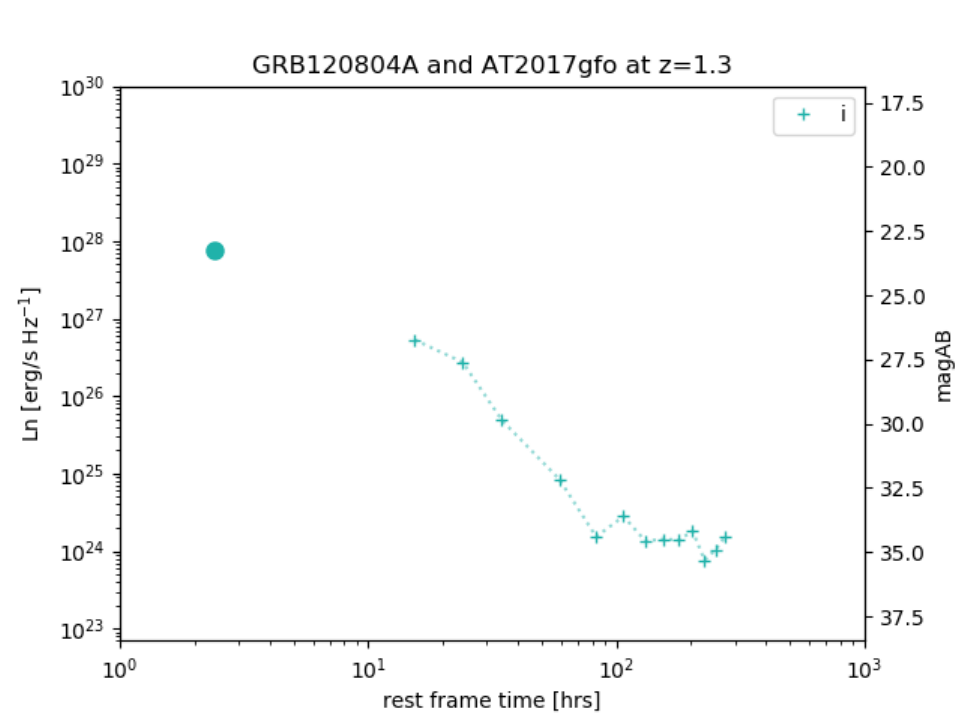}&
\includegraphics[scale=0.45]{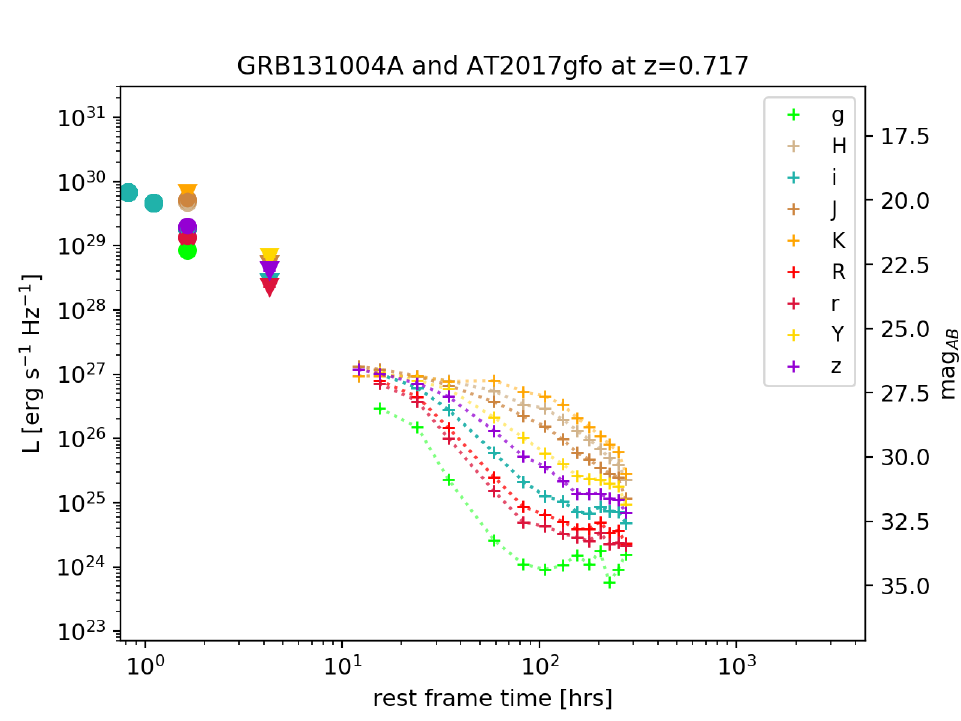}\\
\includegraphics[scale=0.45]{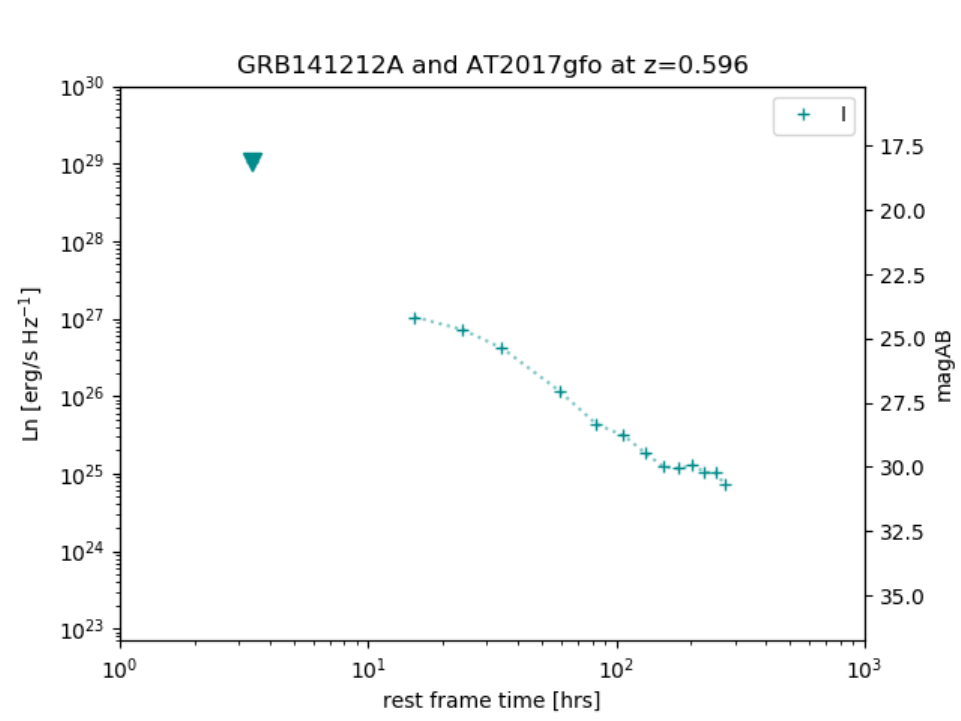}&
\includegraphics[scale=0.45]{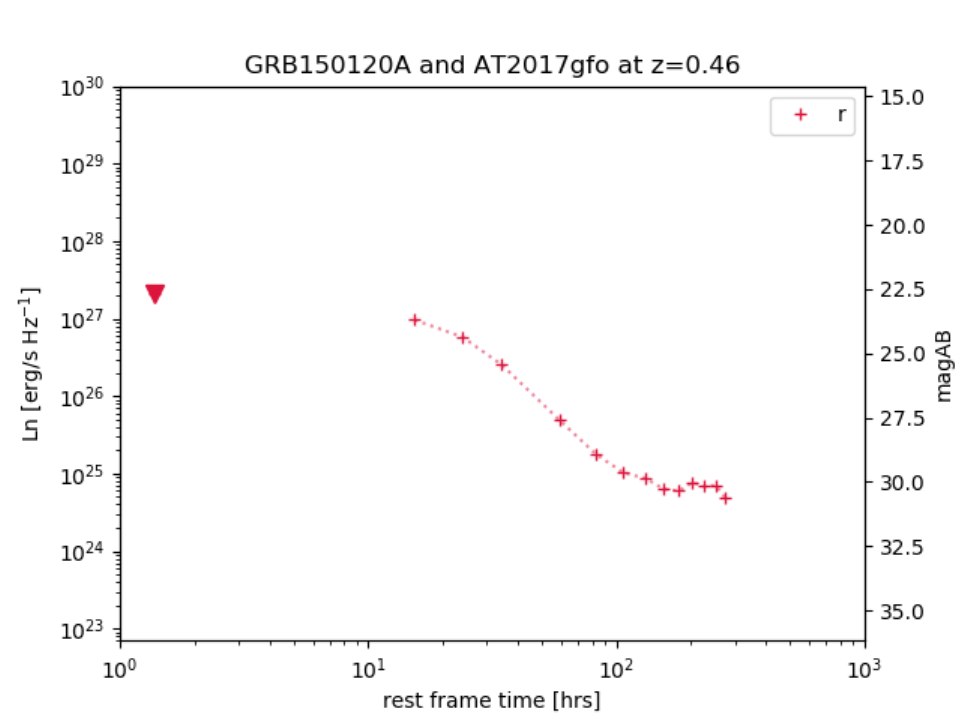}\\
\includegraphics[scale=0.45]{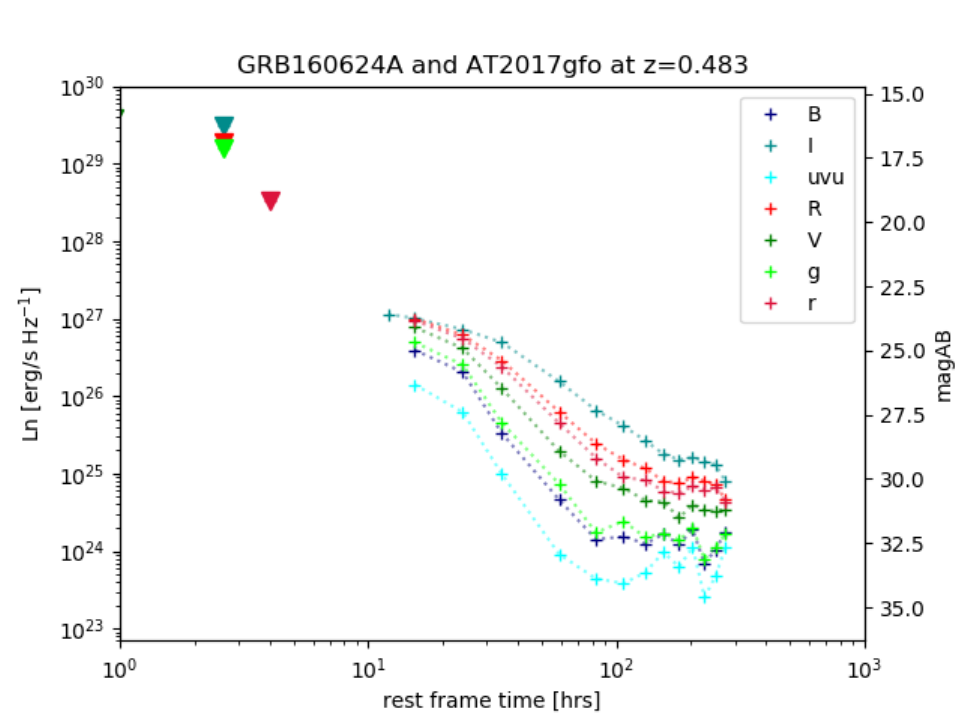}&
\end{tabular}
\end{center}
\contcaption{} 
\label{fig:noconstraints2}
\end{figure*}

\bsp	
\label{lastpage}
\end{document}